\documentclass{aa}  
\usepackage[utf8]{inputenc}

\usepackage{graphicx}
\usepackage{siunitx}
\usepackage{amsmath}
\usepackage{textcomp}
\usepackage{multirow}                
\usepackage{booktabs}                              
\usepackage{multicol,lipsum}
\usepackage{colortbl}
\usepackage{fixltx2e} 
\usepackage[table, dvipsnames]{xcolor}
\usepackage{txfonts}
\usepackage[]{hyperref}
\usepackage{hyperref}
 \hypersetup{
     colorlinks=true,
     linkcolor=blue,
     filecolor=blue,
     citecolor =blue,      
     urlcolor=blue,
     }

\DeclareRobustCommand{\ion}[2]{%
\relax\ifmmode
\ifx\testbx\f@series
{\mathbf{#1\,\mathsc{#2}}}\else
{\mathrm{#1\,\mathsc{#2}}}\fi
\else\textup{#1\,{\mdseries\textsc{#2}}}%
\fi}

\usepackage{comment}
\usepackage{balance}
\usepackage{subfig}

\begin{document}

   \title{Analysis of the planetary mass uncertainties on the accuracy of atmospherical
retrieval }

   \author{C. Di Maio
          \inst{1,2}
          \and
          Q. Changeat\inst{3}
          \and
          S. Benatti\inst{2}
          \and
          G. Micela\inst{2}
          } 

   \institute{Università degli Studi di Palermo, Dipartimento di Fisica e Chimica, via Archirafi 36, Palermo, Italy.\\
              \email{claudia.dimaio@inaf.it}
         \and
             INAF – Osservatorio Astronomico di Palermo, Piazza del Parlamento, 1, 90134 Palermo, Italy
        \and 
        Department of Physics and Astronomy, University College London, Gower Street, London WC1E 6BT, UK
        }

   \date{Received; Accepted}

\titlerunning{Analysis of planetary mass uncertainties on atmospherical retrieval accuracy}
\authorrunning{Di Maio et al.}
 
\abstract
  {Characterising the properties of exoplanet atmospheres relies on several interconnected parameters, which makes it difficult to determine them independently. Planetary mass plays a role in determining the scale height of atmospheres, similarly to the contribution from the average molecular weight of the gas. Analogously, the clouds masking the real atmospheric scale height make it difficult to correctly derive the atmospheric properties.}
  {We investigate the relevance of planetary mass knowledge in spectral retrievals, identifying cases where mass measurements are needed for clear or cloudy and primary or secondary atmospheres, along with the relevant precision, in the context of the ESA M4 Ariel Mission.}
  {We used TauREx to simulate the Ariel transmission spectra of representative targets of the Ariel mission reference sample, assuming different scenarios: a primordial cloudy atmosphere of a hot Jupiter and a hot Neptune, as well as the secondary atmosphere of a super-Earth that also exhibits a cloud presence. We extracted information on the various properties of the atmospheres for the cases of unknown mass or mass with different uncertainties. We also tested how the signal-to-noise ratio impacts atmospheric retrieval for different wavelength ranges.}
  {We accurately retrieved the primordial atmospheric composition independently from mass uncertainties for clear atmospheres, while we found that the uncertainties increased for high altitude clouds. We highlight the importance of the signal-to-noise ratio in the Rayleigh scattering region of the spectrum, which is crucial to retrieving the cloud pressure and to accurately retrieving all other relevant parameters. For the secondary atmosphere cases, a mass uncertainty no larger than 50\% is sufficient to retrieve the atmospheric parameters, even in the presence of clouds.}
   {Our analysis suggests that even in the worst-case scenario, a 50\% mass precision level is enough for producing reliable retrievals, while an atmospheric retrieval without any knowledge of a planetary mass could lead to biases in cloudy primary atmospheres as well as in secondary atmospheres.}


   \maketitle
%
\section{Introduction}
In the last decade, our knowledge of exoplanet atmospheres has been revolutionised. The majority of planets for which detailed atmospheric information is available have been shown to transit their parent star. The atmospheres of about sixty exoplanets have been observed using transmission spectroscopy. By modelling the transmission spectra of exoplanets, we are able to  extract information about various properties and processes in the atmosphere \citep{Charbonneau2002ApJ...568..377C, Tinetti2007Natur.448..169T, Swain2008Natur.452..329S, Kreidberg2014Natur.505...69K, Schwarz2015A&A...576A.111S, Sing2016Natur.529...59S, Hoeijmakers2018Natur.560..453H, deWit2018NatAs...2..214D, Tsiaras2019NatAs...3.1086T,Brogi2019AJ....157..114B,Welbanks2019ApJ...887L..20W,Edwards2020AJ....160....8E, Changeat2021ApJ...907L..22C, Changeat2022ApJS..260....3C, Roudier2021AJ....162...37R, Yip2021AJ....161....4Y}. This is commonly done through a forward model, which generates a spectrum from atmospheric parameters, and a parameter estimation scheme, which samples the parameter space to calculate the probability distribution of the set of parameters. This method, called atmospheric retrieval, has become a fundamental tool for explaining individual observations from transit, eclipse, and phase curve spectroscopy at both low and high resolution.

With NASA's Kepler \citep{Borucki2010DPS....42.4703B} and Transiting Exoplanet Survey Satellite (TESS, \citet{Ricker2015JATIS...1a4003R}), we have already been able to identify a large number of targets suitable for atmospheric characterisation with the Hubble Space Telescope (HST) as well as the James Webb Space Telescope (JWST, \citet{Greene2016ApJ...817...17G}). A new generation of observatories from space and the ground and dedicated missions will come online, offering a broader spectral coverage and higher signal-to-noise ratios (S/N), allowing us to study a significantly larger number of targets. The ESA-Ariel mission alone was designed for this purpose: it will provide transit, eclipse, and phase-curve spectra for hundreds of planets. It is expected to revolutionise our understanding of the physical and chemical properties of a large and diverse sample of extrasolar worlds. To maximise the science return of Ariel, the observations will be performed in four tiers \citep{Tinetti2021}, each one with different binning of the spectra in order to reach the required S/N and for a decreasing number of targets aiming to obtain both an unprecedented statistics of planetary atmospheres and their full characterisation for a number of benchmark cases.  


Most of the planets with mass measurements, mainly coming from radial velocity follow-up confirmations, have typical error bars of the order of 10\%, in particular for a planet with M > 0.1 M$_J$. Planets smaller than Neptune have larger mass errors, often greater than 40-50\%. This uncertainty may contribute to the degeneracy in retrieving the mean molecular weight of the atmosphere, especially when clouds are present \citep{Batalha2019ApJ...885L..25B}. 
In addition, \citet{deWit2013Sci...342.1473D} showed that the next-generation of transmission spectra would contain the information necessary to independently constrain the mass of an exoplanet based on its temperature, pressure, and composition profile.

In a previous work, using a set of simulations,
 \citet{Changeat2020ApJ...896..107C} performed an atmospheric retrieval to study the influence of the knowledge of the planetary mass on the retrieved parameters. In particular, these authors found that for clear-sky gaseous atmospheres, the results obtained when the mass is known or retrieved as a free parameters are the same. In the case of a secondary atmospheres, the retrievals are more challenging due to the higher degree of freedom for the atmospheric main components. In cases where clouds are added, the mass uncertainties may  substantially impact the retrieval due to the degeneracy with the mean molecular weight. 

\begin{figure}[h!]
    \centering
    \includegraphics[scale=0.18]{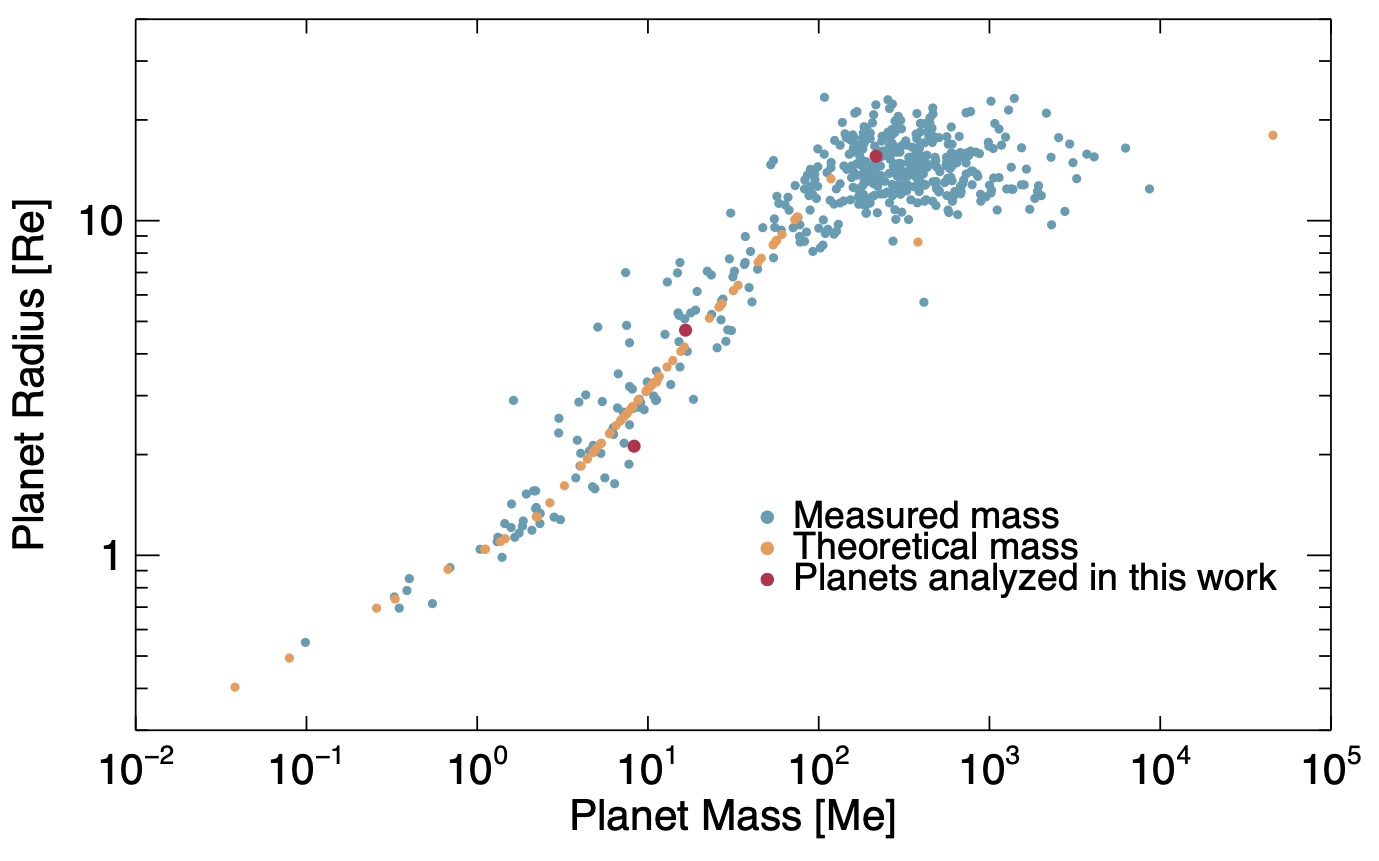}
    \caption{Mass-radius distribution of the planets of the mission reference sample (MRS) for which we have an estimation of the mass. In red, we highlighted the targets analysed in this work (data courtesy by Edwards).}
    \label{fig:MassaRaggio}
\end{figure}

In this context, we aim to understand how precisely we ought to estimate the planetary mass in order to robustly characterise the atmosphere. According to \citet{Edwards2022AJ....164...15E}, the mission reference sample (MRS) of Ariel will contain a selection of planets that could be observed in the prime mission lifetime. About 2000 will be included in the MRS and half of them will be actually investigated by Ariel. Today, about 570 of them are confirmed planets and for $\sim$ 500 of them, we have an estimates of the mass. 
In Fig. \ref{fig:MassaRaggio} we reported the mass-radius relation for the planets that have a mass estimates and we highlighted  the targets analysed in this work with red dots.

The results of this work could provide an input for radial velocity campaigns that are expected to prioritise the most impacted planets. This paper is organised as follows. We describe the methodology used for the retrieval analysis in Sect. \ref{sec:methodology}. In Sect. \ref{Sec:Primordial}, we present the analysis performed for the primordial atmosphere cases. The impact of S/N on atmospheric retrieval is discussed in Sect. \ref{Sec:S/N}, while Sects. \ref{Sec:Secondary} and \ref{Sec:ClodySecondary} present the retrieval analysis of the clear and cloudy secondary atmosphere cases. Our conclusions are given in Sect. \ref{sec:conclusions}.

\section{Retrieval analysis}
\subsection{Methodology}\label{sec:methodology}
In order to analyse the atmospheric retrieval accuracy and how it depends on the planetary mass uncertainties, we used the open-source TauREx 3.1, the new version of TauREx \citep{Waldmann_2015b,Waldmann_2015a}. This fully Bayesian inverse atmospheric retrieval framework \citep{Al_Refaie_2021} is useful in simulating different atmospheric configuration with different star-planet systems and perform retrievals. It uses the highly accurate line lists from the ExoMol \citep{TENNYSON2016_73}, HITEMP \citep{Rothman2014}, and HITRAN \citep{Gordon2016} databases to build forward and retrieval models. In out study, the molecular cross sections were taken
from ExoMol (H$_2$O, \citet{Polyansky2018MNRAS.480.2597P}; CO, \citet{Li2015ApJS..216...15L}; and CH$_4$, \citet{Yurchenko2017A&A...605A..95Y}).

For each tested case, we used TauREx in forward mode to generate a high-resolution theoretical spectrum. We focused only on transit spectra. We specified the main properties of the star and the planet and the main constituents of the atmosphere using their relative abundances. Then, by convolving the high-resolution spectrum through the instrument model (ArielRad v. 2.4.6, \citet{Mugnai2020ExA....50..303M}, Ariel Payload v. 0.0.5, ExoRad v. 2.1.94), we simulated a spectrum as observed by ARIEL and used it as the input of the retrieval. 
The instrument model was obtained for each target and to simulate the Ariel Tier-2 performance, we took into account the number of transit required for the Tier-2 to obtain the adequate S/N. We investigated the parameter space with the nested sampling algorithm MultiNest \citep{Feroz2009MNRAS.398.1601F} with 500 live points and an evidence tolerance of 0.5.

In Sect. \ref{Sec:Primordial}, we tested the case of a hypothetical hot-Jupiter, with parameters based on HD 209458b (see Table \ref{tab:Primordialcases}). In order to investigate the benefits on an increased accuracy in the planetary mass estimation, we performed the retrieval when the mass was totally unknown and, thus, retrieved as a free parameter -- that is, when we know the mass with an uncertainty of 40\% and 10\%.
Also, in order to test the atmospheric retrieval for a smaller planet, we performed a retrieval for a hot-Neptune around a G star, with parameters based on HD 219666b (see Table \ref{tab:Primordialcases}), considering (even in this case) a mass uncertainty of 40\% and 10\%. 

In Sect. \ref{Sec:S/N}, we also discuss the importance of guaranteeing an adequate S/N value by performing the retrieval for the same cases, but considering 10.5th magnitude stars. Furthermore, we compared the retrieval performed on the same object considering different uncertainties at different wavelength ranges to investigate whether the retrieval is more sensitive to a specific range of the spectrum.

\begin{table}[h!]\small
    \centering
    \caption{Planetary and stellar parameters used to produce the forward models and the boundary used in our retrieval analyses for the primordial atmosphere of the hot Jupiter and the Neptunian planet.}
    \begin{tabular}{lcc}
    \toprule[0.05cm]
    \multicolumn{3}{c}{Stellar Parameters} \\
    \toprule
         & HD 209458 & HD 219666 \\
         \midrule
         Sp. type & G0 V & G5 V \\ 
         R$_s$ (R$_\odot$)& 1.19 \tablefootmark{(a)} & 1.03 \tablefootmark{(c)}\\
         M$_s$ (M$_\odot$)& 1.23 \tablefootmark{(a)} & 0.92 \tablefootmark{(c)}\\
         T$_s$ (K) & 6091 \tablefootmark{(a)} & 5527 \tablefootmark{(c)}\\
         d (pc) & 48 \tablefootmark{(a)} & 94 \tablefootmark{(c)}\\
         m$_v$ & 7.65 \tablefootmark{(b)} & 9.81 \tablefootmark{(d)}\\
    \toprule[0.05cm]
    \multicolumn{3}{c}{Hot Jupiter - HD 209458b} \\
    \toprule
          & Input & Boundary \\
        \midrule
        R$_p$ (R$_J$) & 1.39 &  (0.9,1.5)\\
        M$_p$ (M$_J$) & 0.73 & (0.5,1)\\
        T$_p$ (K) & 1450 & (100,4000)\\
    \toprule[0.05cm]
    \multicolumn{3}{c}{Neptunian planet - HD 219666b} \\
    \toprule
          & Input & Boundary \\
        \midrule
        R$_p$ (R$_J$) & 0.42 &  (0.4,0.44)\\
        M$_p$ (M$_J$) & 0.05 & (0.02,0.07)\\
        T$_p$ (K) & 1041 & (100,4000)\\
    \bottomrule[0.05cm]
    
    \end{tabular}
    \tablefoot{\tablefoottext{a}{\citet{Stassun2017AJ....153..136S}}
    \tablefoottext{b}{\citet{delBurgo2016MNRAS.463.1400D}}
    \tablefoottext{c}{\citet{Esposito2019A&A...623A.165E}}
    \tablefoottext{d}{\citet{Hog2000A&A...355L..27H}}.}
    \label{tab:Primordialcases}
\end{table}
In Section \ref{Sec:Secondary}, we investigate the case of a hypothetical super-Earth, with parameters based on HD 97658b (see Table \ref{tab:Secondary}), one of the targets on the ARIEL Target List \citep{Edwards2019AJ....157..242E}. We tested three different atmospheric configurations by considering a heavy atmosphere containing a significant fraction of H$_2$O, CO, and N$_2$, respectively. Also, in order to test the impact of the mass uncertainties onto the retrieval of the atmospheric properties we considered three different mass uncertainties (10\%, 30\%, and 50\%).
\begin{table}[h!]
    \centering
    \caption{Planetary and stellar parameters used to produce the forward models for the secondary atmosphere of a Super-Earth planet. }
    \begin{tabular}{lc}
    \toprule[0.05cm]
         Parameters & HD 97658 \\
         \midrule
         Sp. type & K1 V \\ 
         R$_s$ (R$_\odot$)& 0.73 \tablefootmark{(a)}  \\
         M$_s$ (M$_\odot$)& 0.85 \tablefootmark{(a)} \\
         T$_s$ (K) & 5212 \tablefootmark{(b)}\\
         d (pc) & 21.546 \tablefootmark{(b)} \\
         m$_v$ & 7.78 \tablefootmark{(b)} \\
         \midrule
         R$_p$ (R$_J$) & 0.189 \\
         M$_p$ (M$_J$) & 0.02611\\
         T$_p$ (K) & 720.33 \\
         
    \bottomrule[0.05cm]
    \end{tabular}
    \tablefoot{\tablefoottext{a}{\citet{Howard2011ApJ...730...10H}}
    \tablefoottext{b}{\citet{Ellis2021AJ....162..118E}}.}
    
    \label{tab:Secondary}
\end{table}
In Section \ref{Sec:ClodySecondary}, we investigated the case of cloudy N$_2$-dominated secondary atmospheres. In order to test the difference in the retrieval of a atmosphere dominated by active gases, which are characterised by traceable molecular features directly observable in the spectrum, we analysed two other different scenarios where we considered a H$_2$O- and a CO-dominated atmosphere.

For all the tested cases, we assumed a planetary atmosphere constituted by 100 layers in a plane-parallel geometry, uniformly distributed in log space between $10^{-1}$ and $10^6$ Pa. The temperature structure was modelled with a isothermal $T-p$ profile. The trace gases considered here were allowed to vary freely between $10^{-12}$ and $10^{-2}$ in volume mixing ratio. 

Regarding the processes in the atmosphere that contribute to the optical depth to be considered, we set the molecular profile of each species to be constant at each atmospheric layer. Also, we took into account the collision-induced absorption (CIA) from H$_2$-H$_2$ \citep{Abel2011JPCA..115.6805A, Fletcher2018ApJS..235...24F} and  H$_2$-He \citep{Abel2012JChPh.136d4319A}, as well as Rayleigh scattering for all molecules.

\subsection{Primordial atmosphere}\label{Sec:Primordial}
To investigate the contribution of the planetary mass uncertainties onto the retrieval of a primary atmosphere, we simulated a spectrum of a hot Jupiter based on HD 209458b and its parent star.

In a previous work, \citet{Changeat2020ApJ...896..107C}  already performed a retrieval on this object, comparing the case where the planetary mass is assumed to be known to one where it is retrieved as a free parameter. In particular, they found that for a clear sky atmosphere, the knowledge of the mass does not impact the results. However, if clouds are modelled, some discrepancies appear only in the retrieval of the radius when the cloud pressure gets closer to $10^{-3}$ bar and the retrieved mass also becomes less accurate.

Here, we want to investigate the benefits of increased accuracy in the planetary mass estimation on the atmospheric retrieval.
We adopted the same parameters used by \citet{Changeat2020ApJ...896..107C}. In particular, for trace gases, we included H$_2$O, CH$_4$, and CO, with mixing ratios of $10^{-5}$, $5 \times 10^{-6}$, and $10^{-4}$, respectively. 
We first simulated a clear sky atmosphere case and then we tested the behaviour of the retrievals when clouds are present with four different configurations (P$_{clouds}$ = $10^{-1}$, $10^{-2}$, $5 \times 10^{-2}$, and $10^{-3}$ bar for our worst-case scenario). For each scenario, we performed the retrieval for three cases: in the first case, the planetary mass is retrieved as a free parameter (we used a large boundary range, by supposing a mass uncertainty of about 100\%, so we can assume the mass as totally unknown); in the second, we supposed that we know the mass with an uncertainty of 40\%; and in the third case, we applied an uncertainty of 10\%. We also performed the retrieval for a Neptunian planet around a G star to investigate how the retrieval depends from the planet characteristic. The parameters used to generate the forward model and the prior bounds employed for each fitted parameter are reported in Table \ref{tab:Primordialcases}. 

\begin{figure*}[ht!]
 \centering
        {\subfloat[Radius\label{fig:Radius_Gstar}]{\includegraphics[scale=0.41]{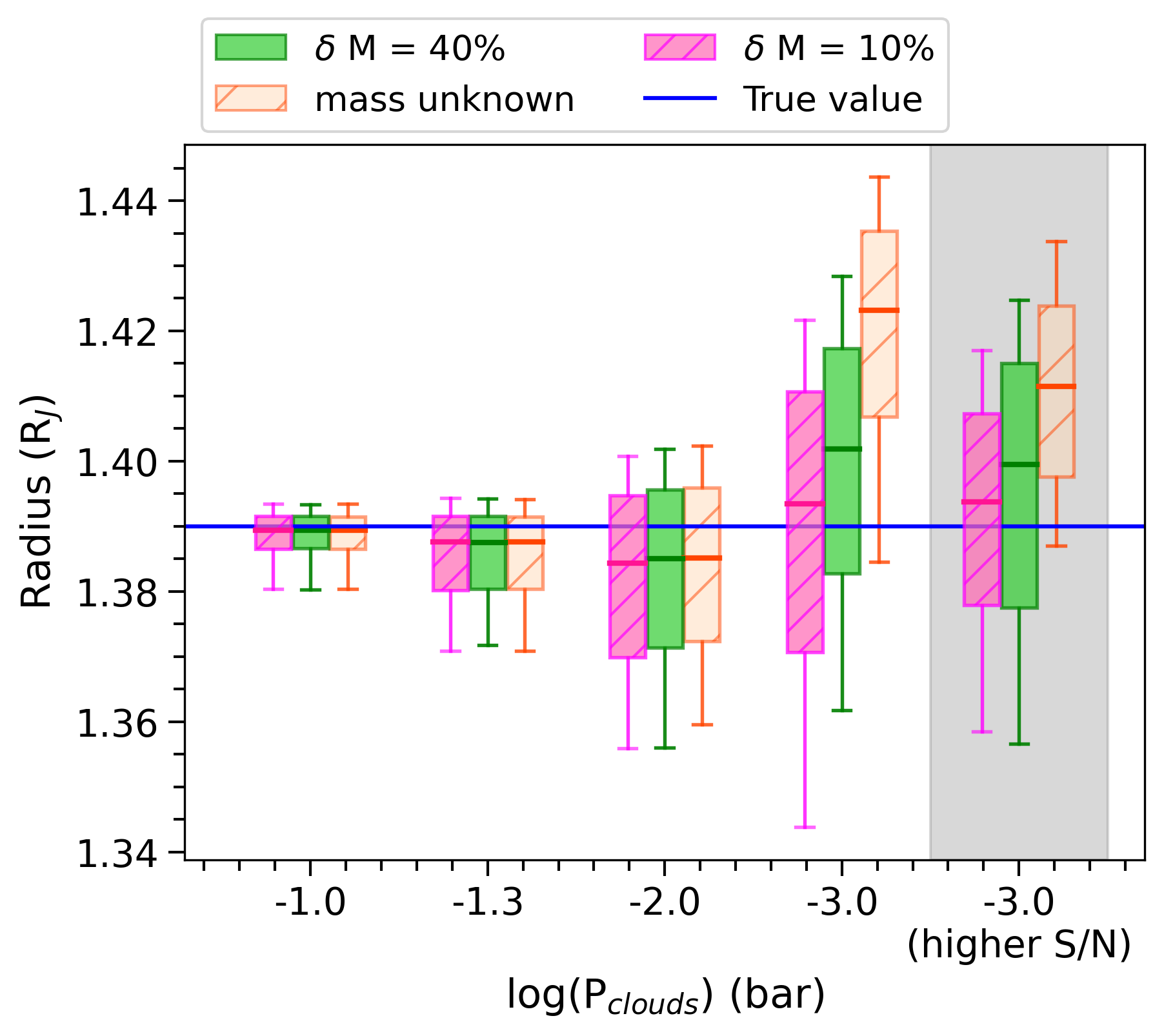}}
        \subfloat[Temperature\label{fig:Temp_Gstar}]{\includegraphics[scale=0.41]{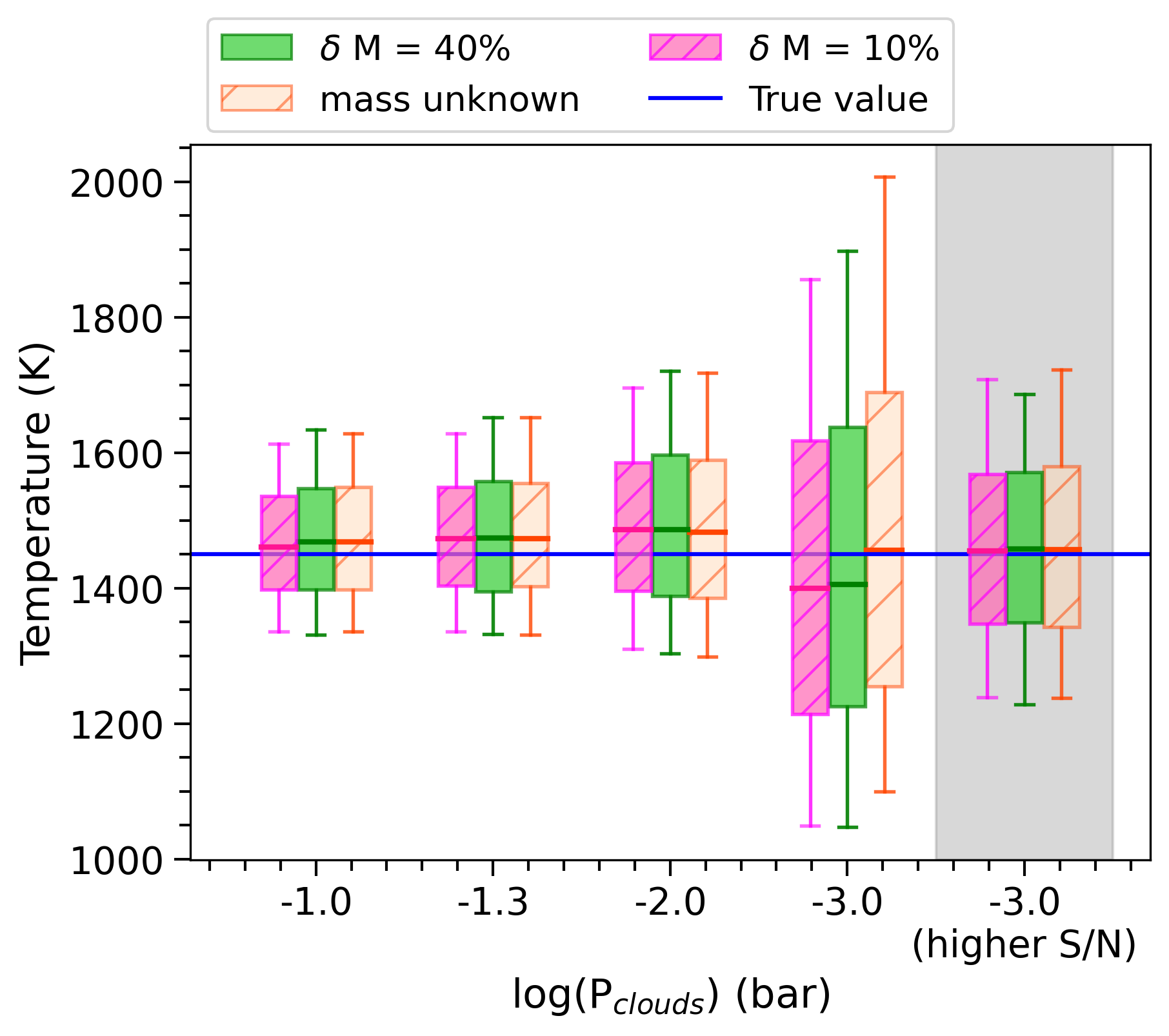}}
        \subfloat[Clouds\label{fig:Clouds_Gstar}]{\includegraphics[scale=0.41]{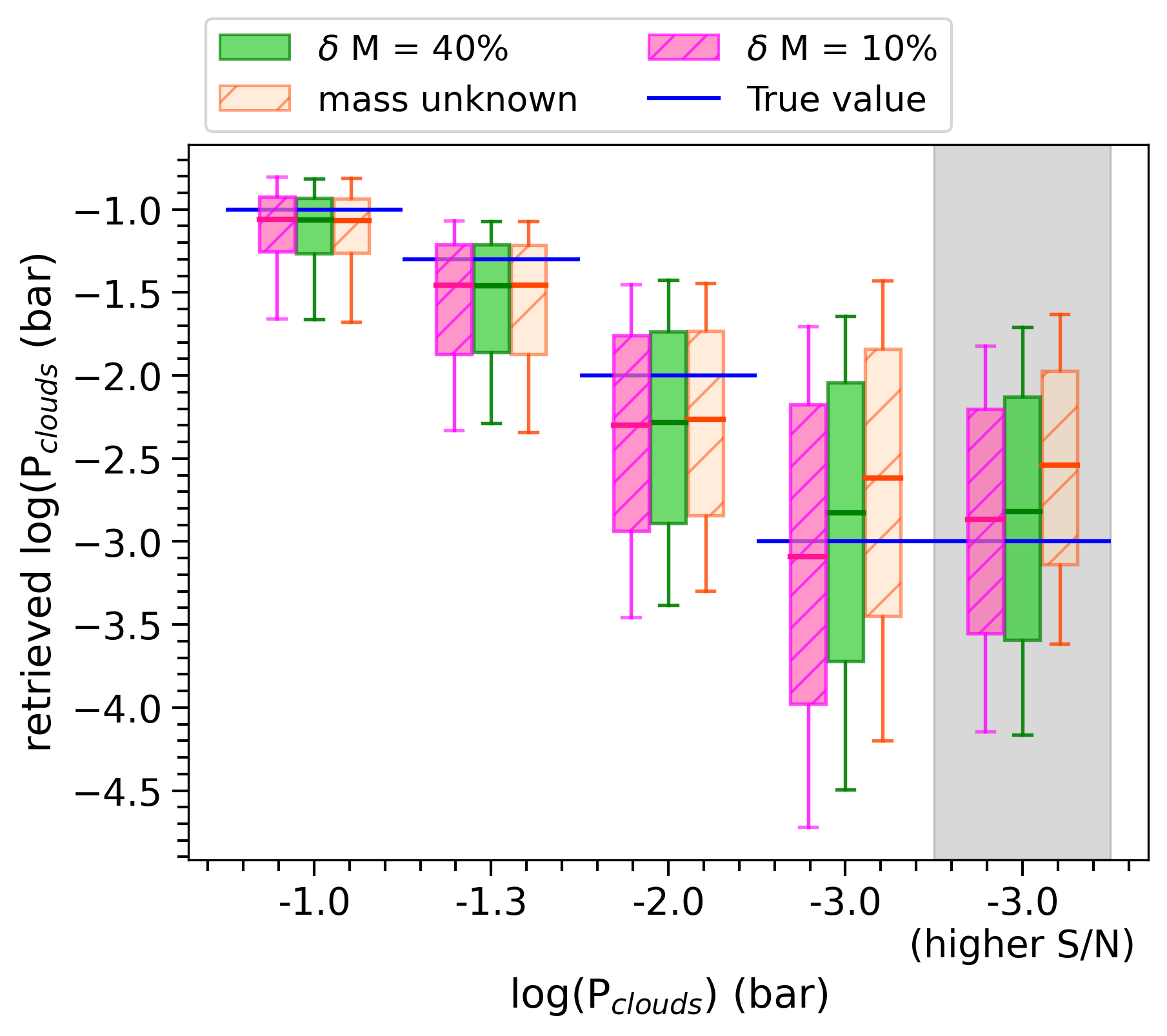}} \\
        \subfloat[H$_2$O mixing ratio \label{fig:H2O_Gstar}]{\includegraphics[scale=0.41]{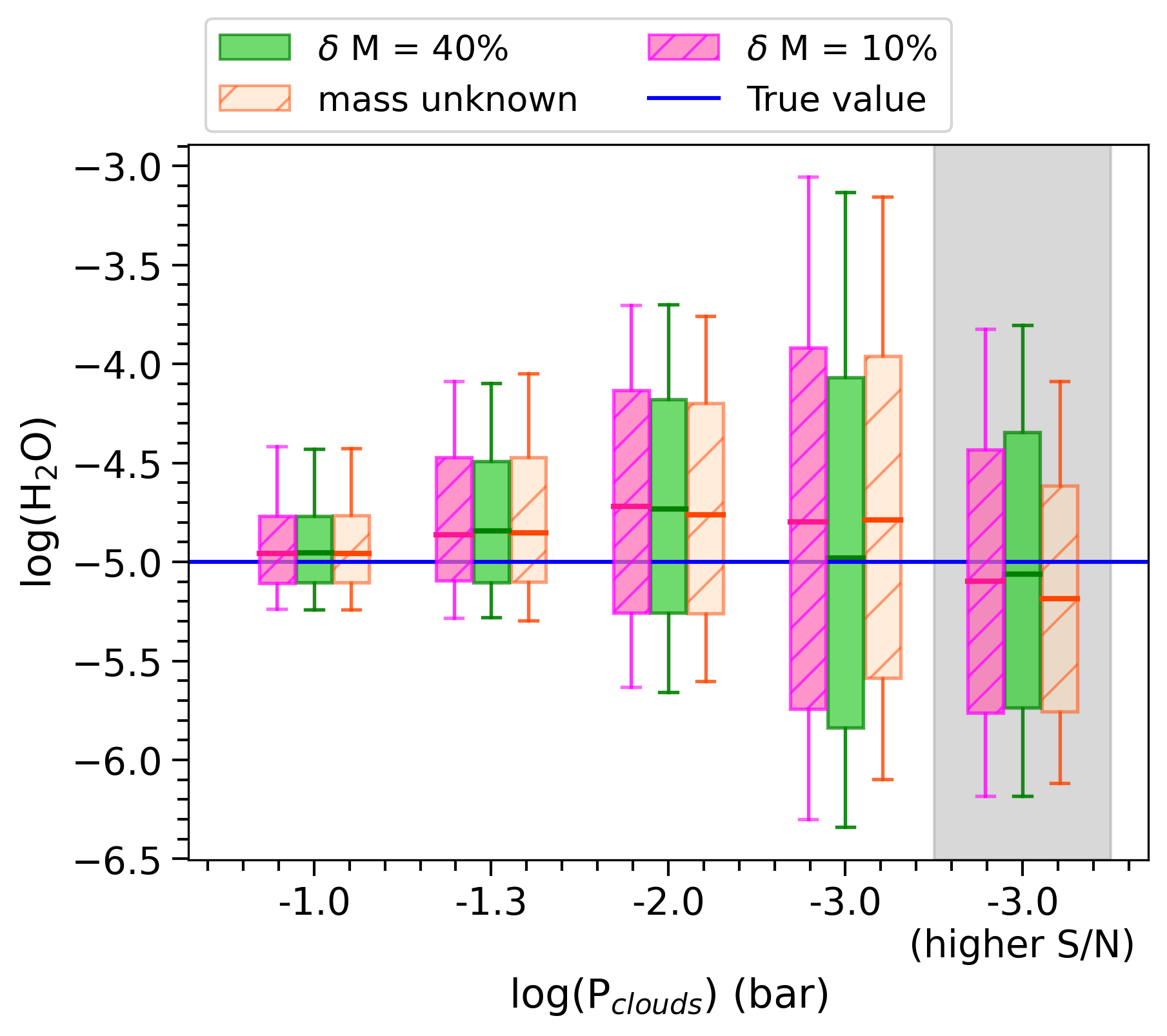}}
        \subfloat[CH$_4$ mixing ratio\label{fig:CH4_Gstar}]{\includegraphics[scale=0.41]{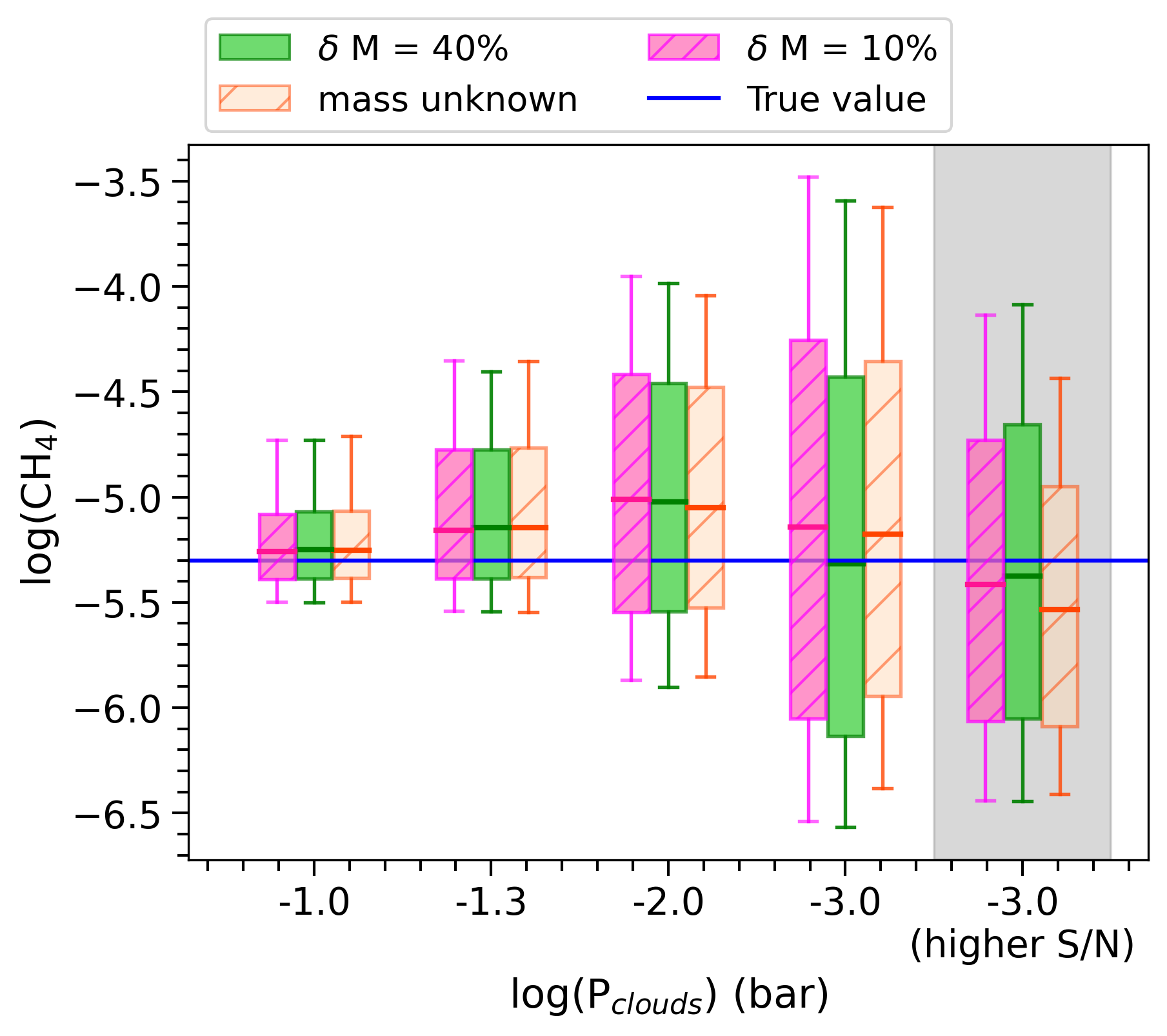}}
        \subfloat[CO mixing ratio\label{fig:CO_Gstar}]{\includegraphics[scale=0.41]{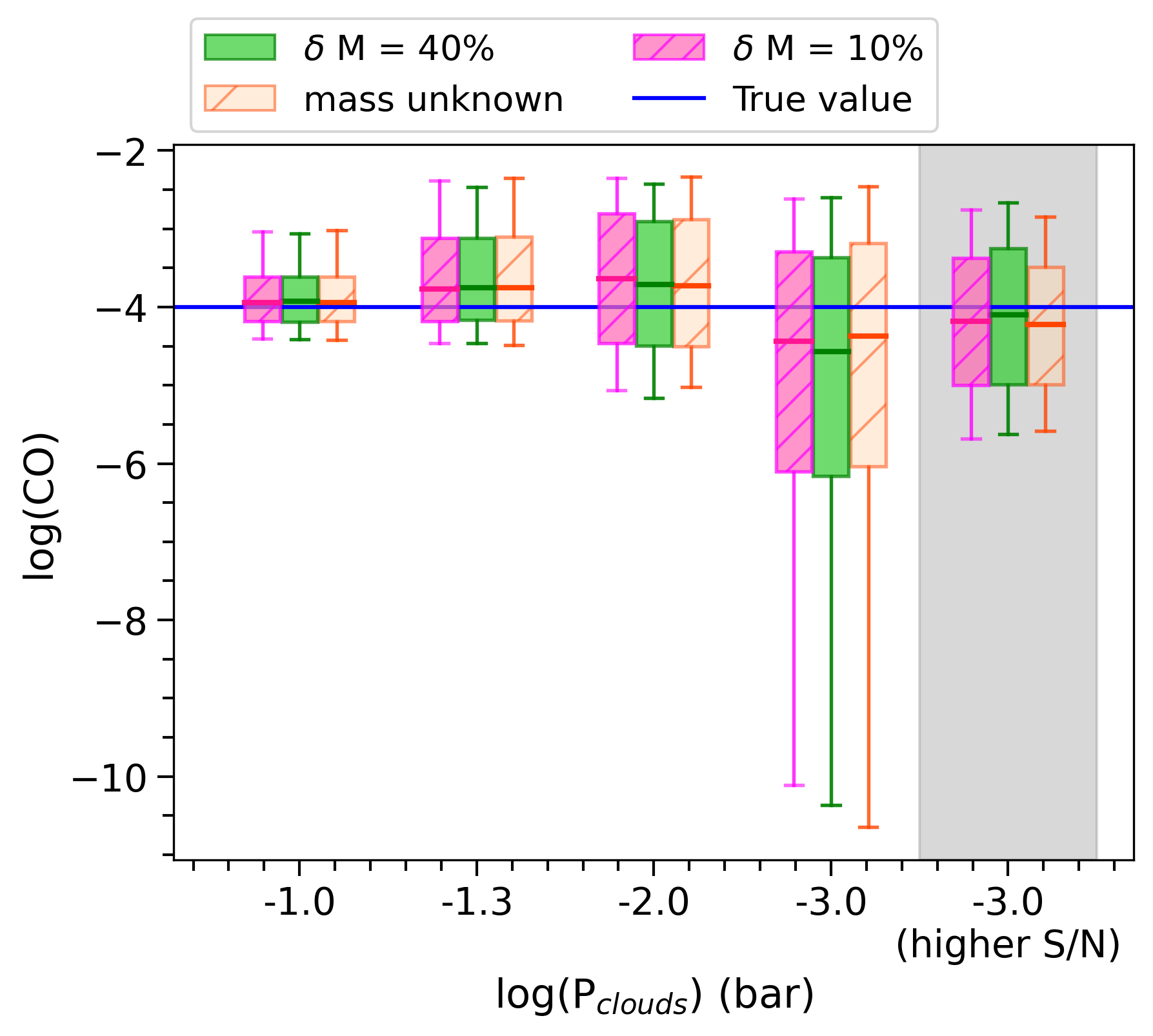}}
    }
    \caption{Comparison between the results obtained from the retrieval performed for the case of a hot-Jupiter around a G star when the mass is known with an uncertainty of 40\% (in green) and of 10\% (in magenta) as well as when is totally unknown (in orange) as a function of cloud pressure. In the gray area, we report the results obtained for a P$_{clouds} = 10^{-3}$ bar, assuming noise decreased by a factor of two.
    The size of the box and the error bar represent the points within 1$\sigma$ and 2$\sigma$ of the median of the distribution (highlighted with solid lines), respectively. Blue line is the real value.}
    \label{fig:Gstar}
 \end{figure*}

In Fig. \ref{fig:Gstar}, we compare the results obtained for a hot-Jupiter orbiting around a G star, as a function of cloud pressure, in the case where we know the mass with an uncertainty of 40\% (in green) and of 10\% (in magenta), as well as the case in which the mass is totally unknown (in orange). The discrepancies in the retrieval of the radius, which appear when the clouds pressure gets closer to $10^{-3}$ bar, and which are the same as obtained by \citet{Changeat2020ApJ...896..107C}, disappear when we performed the retrieval while considering a mass uncertainty of about 40\% or less. In these cases, the retrieved radius for high altitude clouds is within 1$\sigma$ of the true value. Also, for all the parameters, we obtained a more accurate and precise retrieval when we know the mass with an uncertainty of 40\% in case with high-altitude clouds as well. 

\begin{figure}
    \centering
    \includegraphics[scale=0.55]{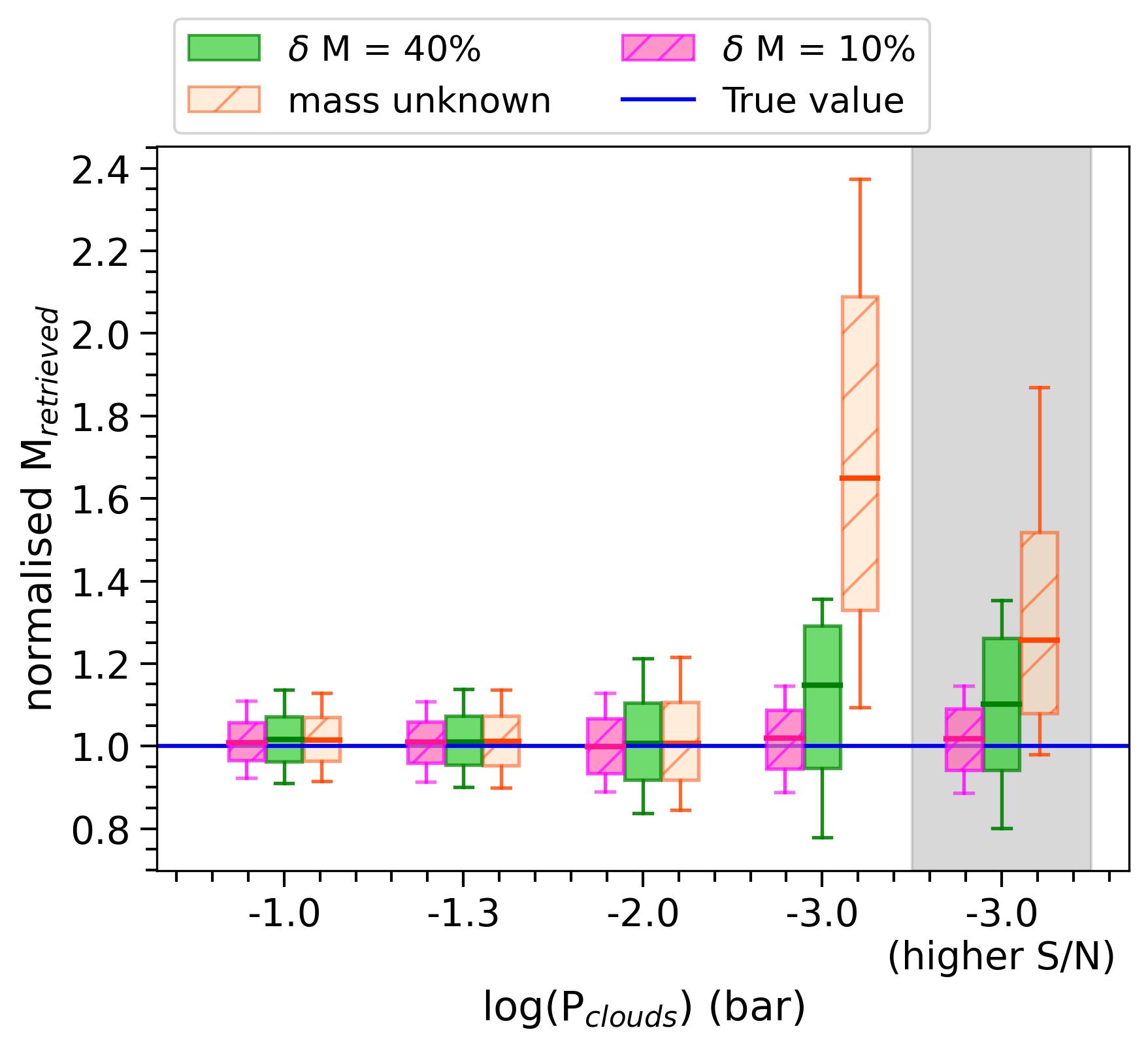}
    \caption{Comparison between the normalised retrieved mass in the case of a hot-Jupiter around a G star when the mass is estimated with an uncertainty of 40\% (in green), of 10\% (in magenta) and when is totally unknown (in orange) as a function of clouds pressure. In the grey area we reported the results obtained for a P$_{clouds} = 10^{-3}$ bar considering a noise decreased by a factor of two. The size of the box and the error bar represent the points within 1$\sigma$ and 2$\sigma$ of the median of the distribution (highlighted with solid-lines), respectively. The blue line is the real value.}
    \label{fig:Gstar_Mass}
\end{figure}

Focusing on the retrieval of the mass, in Fig. \ref{fig:Gstar_Mass}, we compare the results of the normalised retrieved mass of each tested case obtained for the mass as totally unknown and for the mass with an uncertainty of 40\% and 10\%. The mass is well retrieved for all cases with clouds at low altitudes even when we totally unknown the mass. The retrieved mass becomes less accurate when the clouds pressure is lower than 10$^{-2}$ bar. With a mass uncertainty of 40\%, we significantly increase the accuracy and precision in the normalised retrieved mass. Indeed, in this case the mass is well retrieved even for high altitude clouds and the retrieved values are within 1$\sigma$ with the true values. 
Additionally, we note that while a better estimation of the mass (mass uncertainty of 10\%), could allow us to retrieve the mass and the radius with more precision also in the cases with high altitude clouds, we do not observe significant difference between the results obtained with a mass uncertainty of 40\% and 10\% of all the other parameters.

Furthermore, we performed the retrieval for the worst cloudy case (P$_{clouds}=10^{-3}$ bar) considering an increased S/N (see the results in the grey part of the plots in Fig. \ref{fig:Gstar}). To this purpose we considered four times as many observations, so we decreased the noise by a factor of 2. 
From this test, we note that for the worst scenario, in which the contribution of the clouds determines a less accurate estimate of the retrieved parameters, an increased S/N could help to better estimate the parameters and, in particular, the trace composition of the atmosphere. 

From these results, we note that we could use TauREx as a tool to estimate the mass of hot Jupiters with more precision than what we already know. In particular, if we know the mass with an uncertainty of 40\%, we could be able to retrieve the mass with an uncertainty of about 15\%. The results obtained from the retrieval of the analysed cases are summarised in Table \ref{tab:primordial-G} in Appendix \ref{app:tabelle}.

To test the atmospheric retrieval in the case of smaller planet, we simulated the spectrum of a Neptunian planet, based on HD 219666b around its host star (Table \ref{tab:Primordialcases}). 
We used the same parameters of the hot Jupiter case for the atmospheric composition. 
In Table \ref{tab:neptunian} in Appendix \ref{app:tabelle}, we summarise the results obtained from this test. In this case, some discrepancies appear in the atmospheric retrieval when the cloud pressure gets closer to 10$^{-3}$ bar. All the other parameters are well retrieved, even for high altitude clouds. 
Focusing on the retrieval of the mass, we note that the mass is well retrieved for all the cases at lower altitudes, while the retrieved mass becomes less accurate when the clouds pressure decreases. In all tested cases for neptunian planets, we can refine the mass to within 20\% -- provided the initial mass uncertainty is $\leq$ 40\%.

 \begin{figure*}[h!]
 \centering
        {\subfloat[Radius\label{fig:Radius_mv8vsmv10.5}]{\includegraphics[scale=0.42]{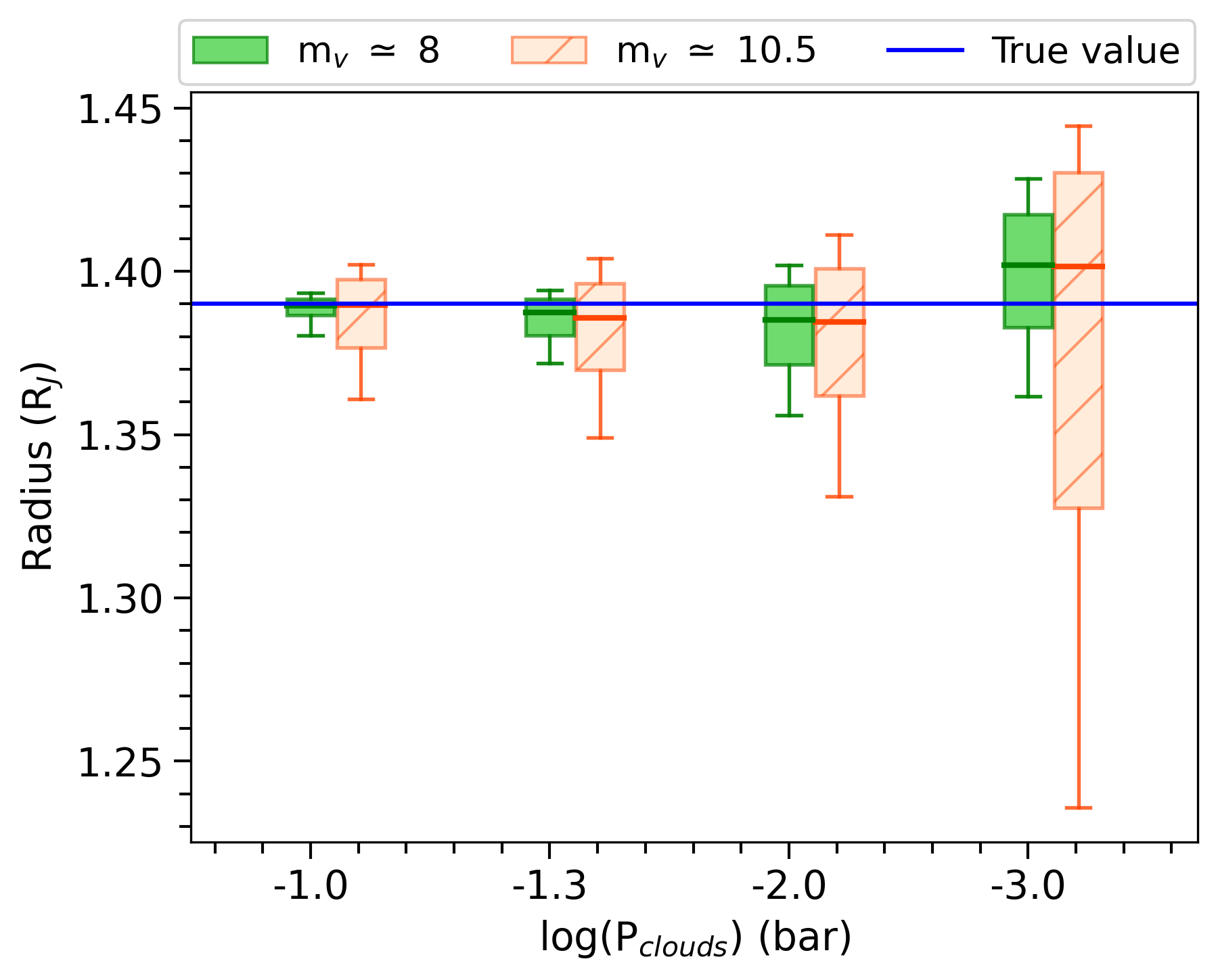}}
        \subfloat[Temperature\label{fig:Temp_mv8vsmv10.5}]{\includegraphics[scale=0.42]{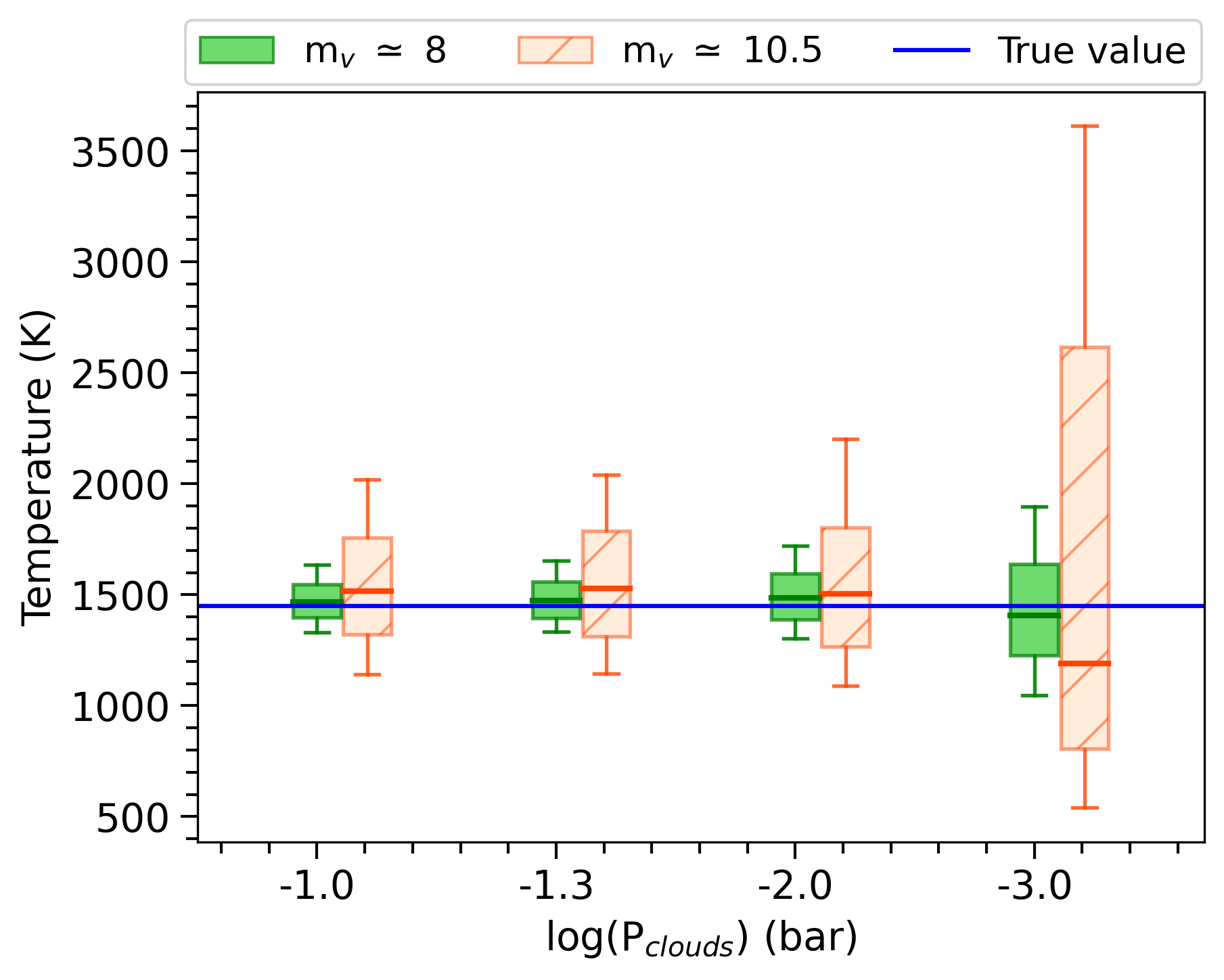}}
        \subfloat[Clouds\label{fig:Clouds_mv8vsmv10.5}]{\includegraphics[scale=0.42]{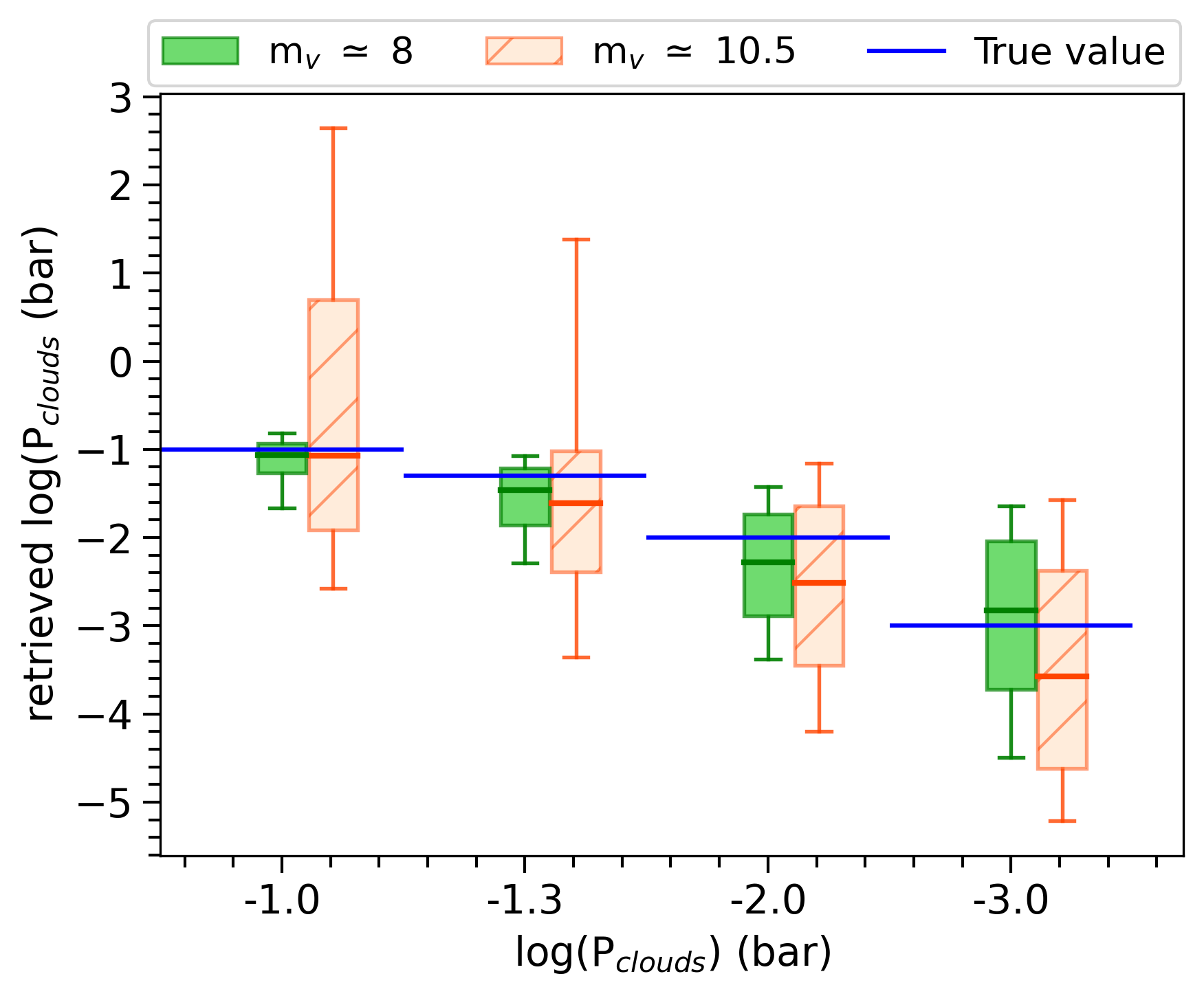}} \\
        \subfloat[H$_2$O mixing ratio \label{fig:H2O_mv8vsmv10.5}]{\includegraphics[scale=0.42]{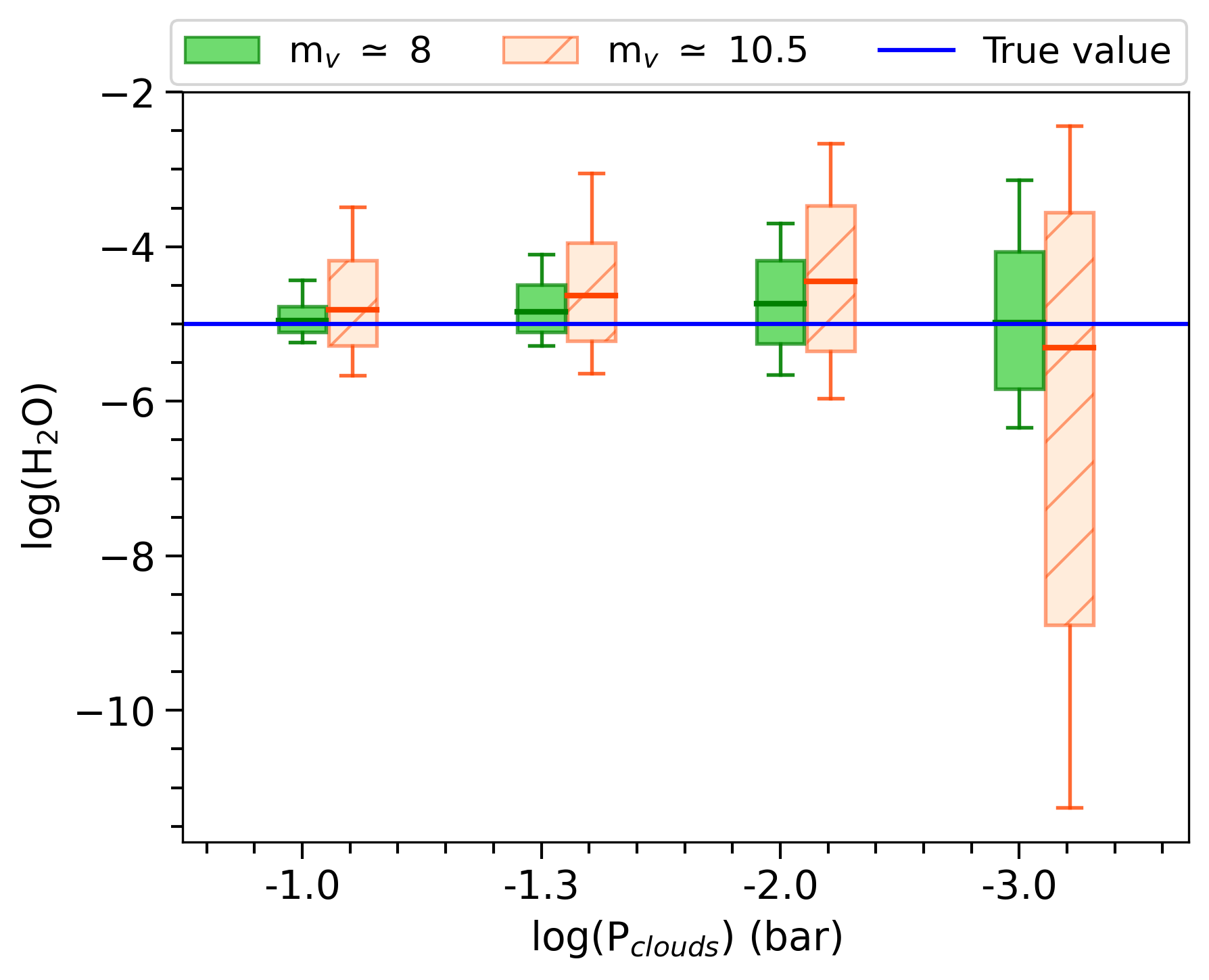}}
        \subfloat[CH$_4$ mixing ratio\label{fig:CH4_mv8vsmv10.5}]{\includegraphics[scale=0.42]{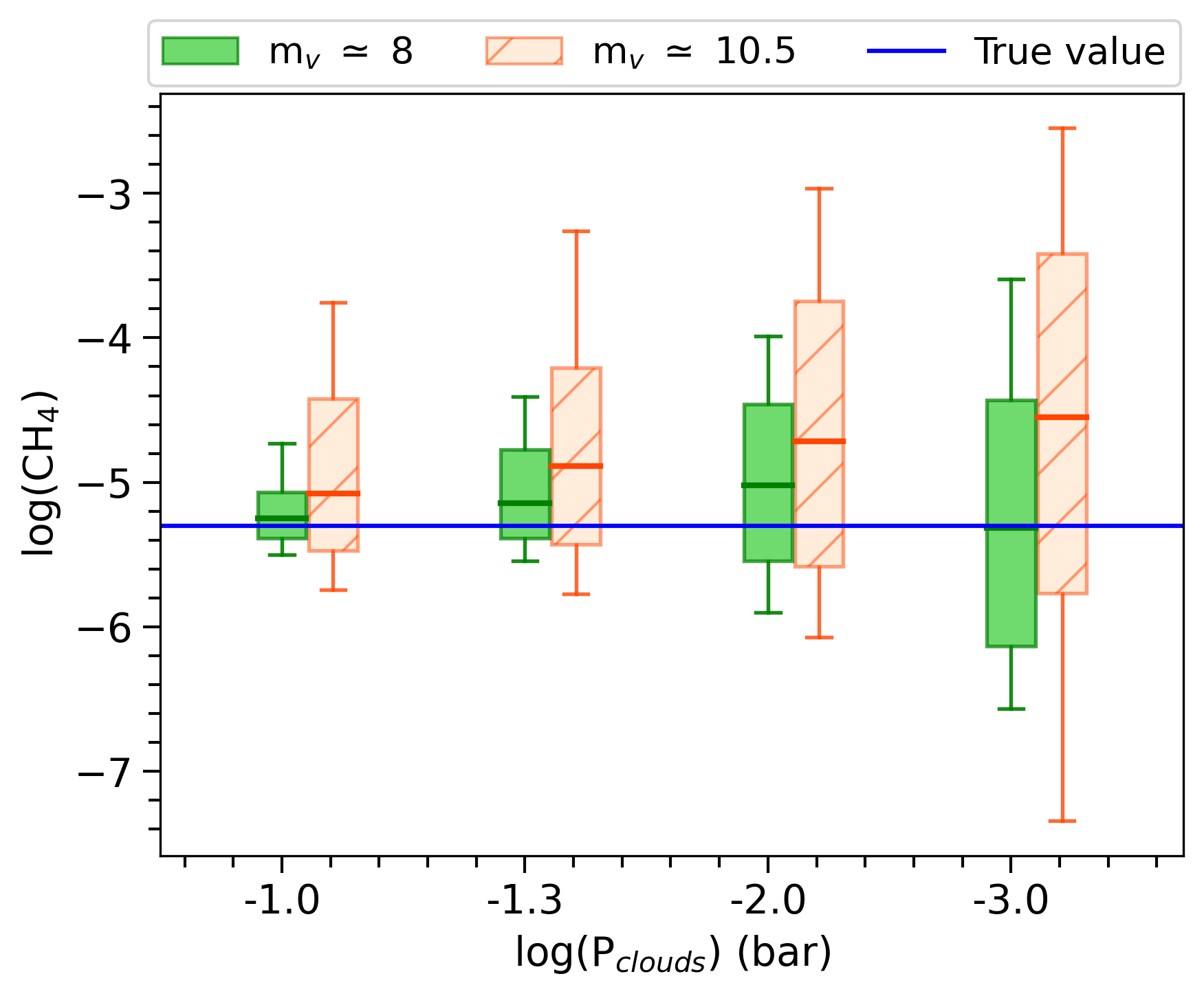}}
        \subfloat[CO mixing ratio\label{fig:CO_mv8vsmv10.5}]{\includegraphics[scale=0.42]{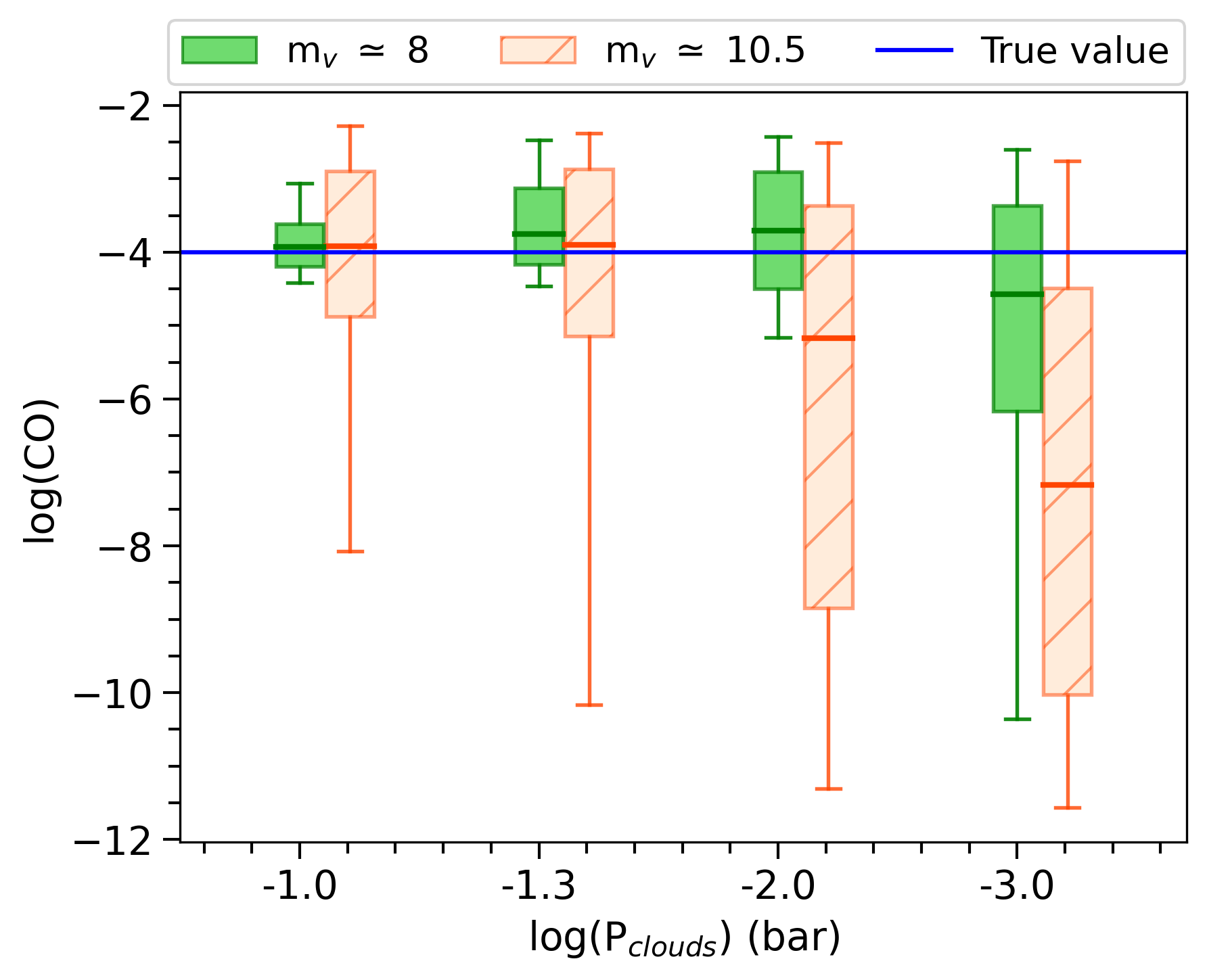}}
    }
    \caption{Comparison between the results obtained from the retrieval performed for the case of a hot Jupiter around an 8th magnitude G star (in green) and around a 10.5th magnitude G star (in orange) as a function of clouds pressure,  assuming a mass uncertainty of about 40\%. The size of the box and the error bar represent the points within 1$\sigma$ and 2$\sigma$ of the median of the distribution (highlighted with a red and an orange solid line), respectively. Blue line is the real value.}
    \label{fig:mv8vsmv10.5}
 \end{figure*}
 
We note that the discrepancies obtained in the atmospheric retrieval at high altitude clouds does not disappear when the mass in known with an uncertainty of 10\%. In this case, while the retrieval of the mass increase in precision, as expected, the mass does not seem to impact the retrieval of the atmospheric composition.

\subsection{Signal-to-noise ratio impact on the atmospheric retrieval at different wavelength ranges}\label{Sec:S/N}
\citet{Changeat2020ApJ...896..107C} highlighted the importance of guaranteeing the adequate S/N when we observe heavy secondary atmosphere, by suggesting that an adequate S/N is necessary to correctly estimate  the mass through transit spectroscopy. 

Here, we test the importance of the S/N for the primary atmosphere.
To this purpose, in Fig. \ref{fig:mv8vsmv10.5}, we compare the results obtained in the previous section, where we consider  an 8th magnitude G star, with the results obtained for the same planet supposed in the previous section, orbiting around a 10.5th magnitude star. Of course, a higher magnitude for the star implies a lower S/N. The results of this test are summarised in Table \ref{tab:mv8mv10} in Appendix \ref{app:tabelle}.

As expected, in the case of 10.5th magnitude star the uncertainties of all the fitted parameters increase with respect to the uncertainties obtained for the 8th magnitude star case and also increase when the clouds pressure decreases. The retrieved values are within 1$\sigma$ of the true values, except for the CO mixing ratio where the accuracy decreases at high altitude clouds and is not compatible (at 1$\sigma)$ with the true value.

\begin{figure}[hb!]
    \centering    \includegraphics[scale=0.50]{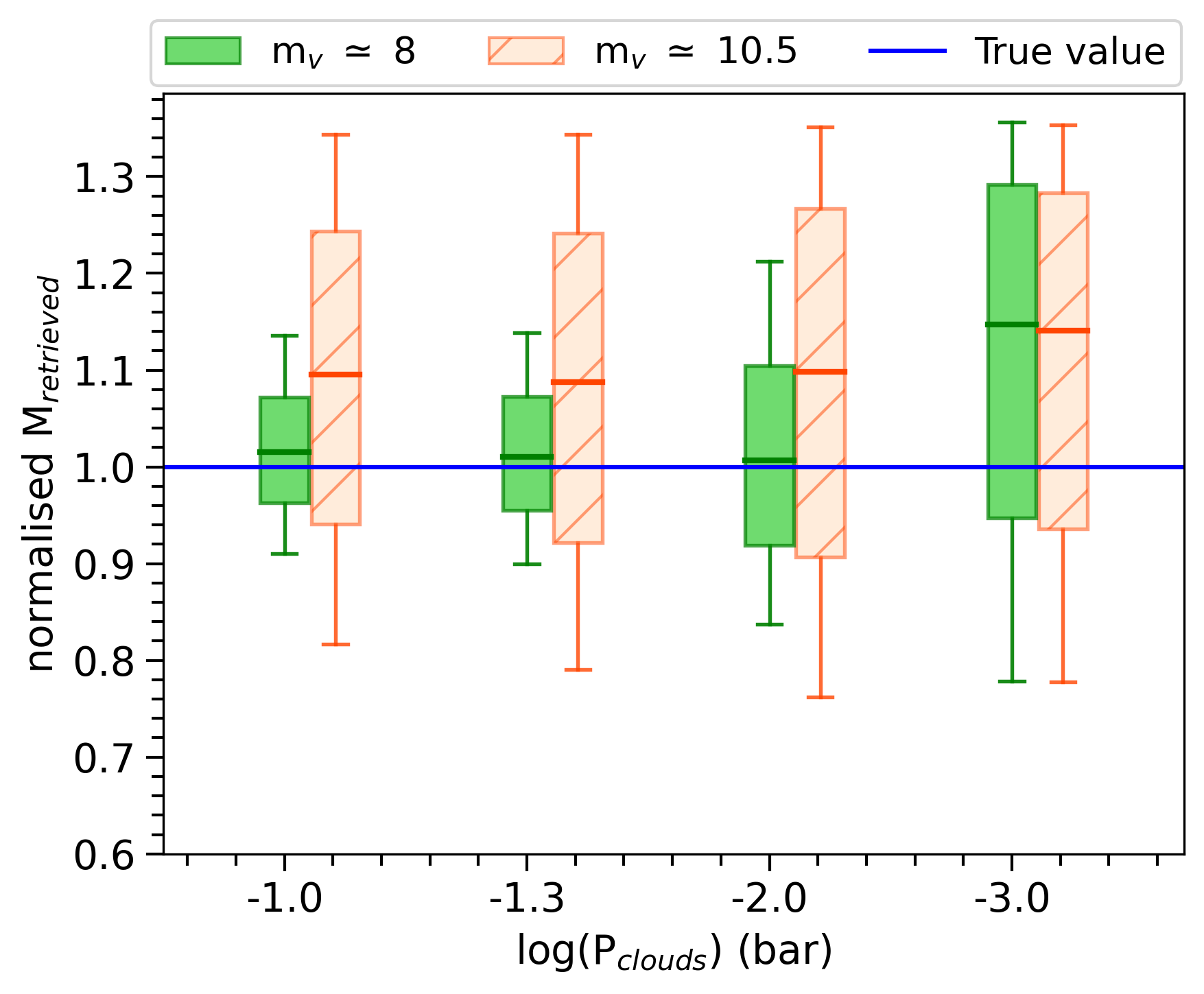}
    \caption{Comparison between the retrieved mass for the case of a hot-Jupiter around an eighth-magnitude G star (in green) and around a 10.5th magnitude G star (in orange) as a function of cloud pressure (mass uncertainty of about 40\%). Colour scale is the same as in Fig. \ref{fig:mv8vsmv10.5}.}
    \label{fig:mv8vsmv10MASS}%
 \end{figure}

Focusing on the retrieval of the mass, see Fig. \ref{fig:mv8vsmv10MASS}, we note that the mass is retrieved for all cases but with less accuracy and precision, and still within 1$\sigma$ of the true value, in the case of 10.5th magnitude star, regardless of the height of the clouds.

\begin{figure}[h!]
    \centering
    \includegraphics[scale=0.32]{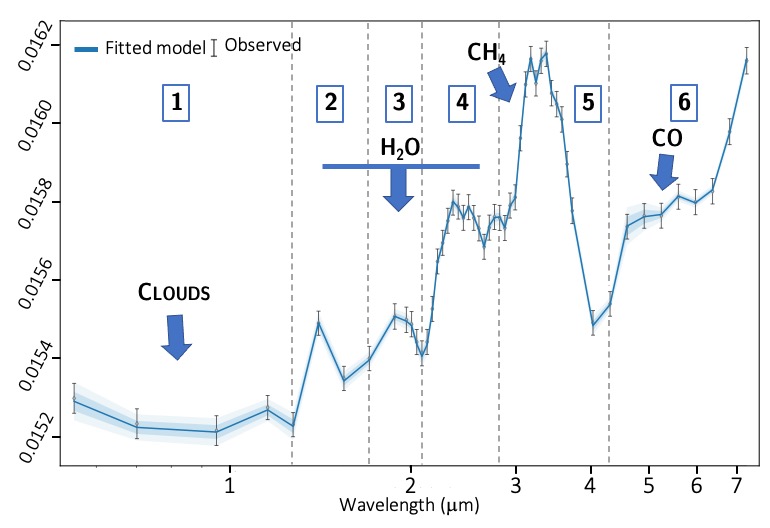}
    \caption{Example of spectrum obtained for a primordial atmosphere case with a cloud pressure of 10$^{-1}$ bar. We highlighted the range of the spectrum in which we expected the main contribution of H$_2$O, CH$_4$, and CO. We selected six different ranges of the spectrum and we increased the S/N in each of them to investigate the contribution of each range on the retrieval (the points at the edge of the ranges belong to both of the adjacent selected sections).}
    \label{fig:split}%
 \end{figure}

In addition, we investigated the impact of the S/N of specific range of the spectrum onto the atmospheric retrieval. We performed a retrieval analysis of a Jovian planet around a G star, considering a cloud pressure of 10$^{-1}$ bar. We decided to split the spectrum in six different ranges (see Fig. \ref{fig:split}), each of which is dominated by a different atmospheric features, to understand which range of the spectrum provides the main contribution to the retrieval.  

 \begin{figure*}[ht!]
 \centering
        {\subfloat[Radius\label{fig:Radius_SN}]{\includegraphics[scale=0.41]{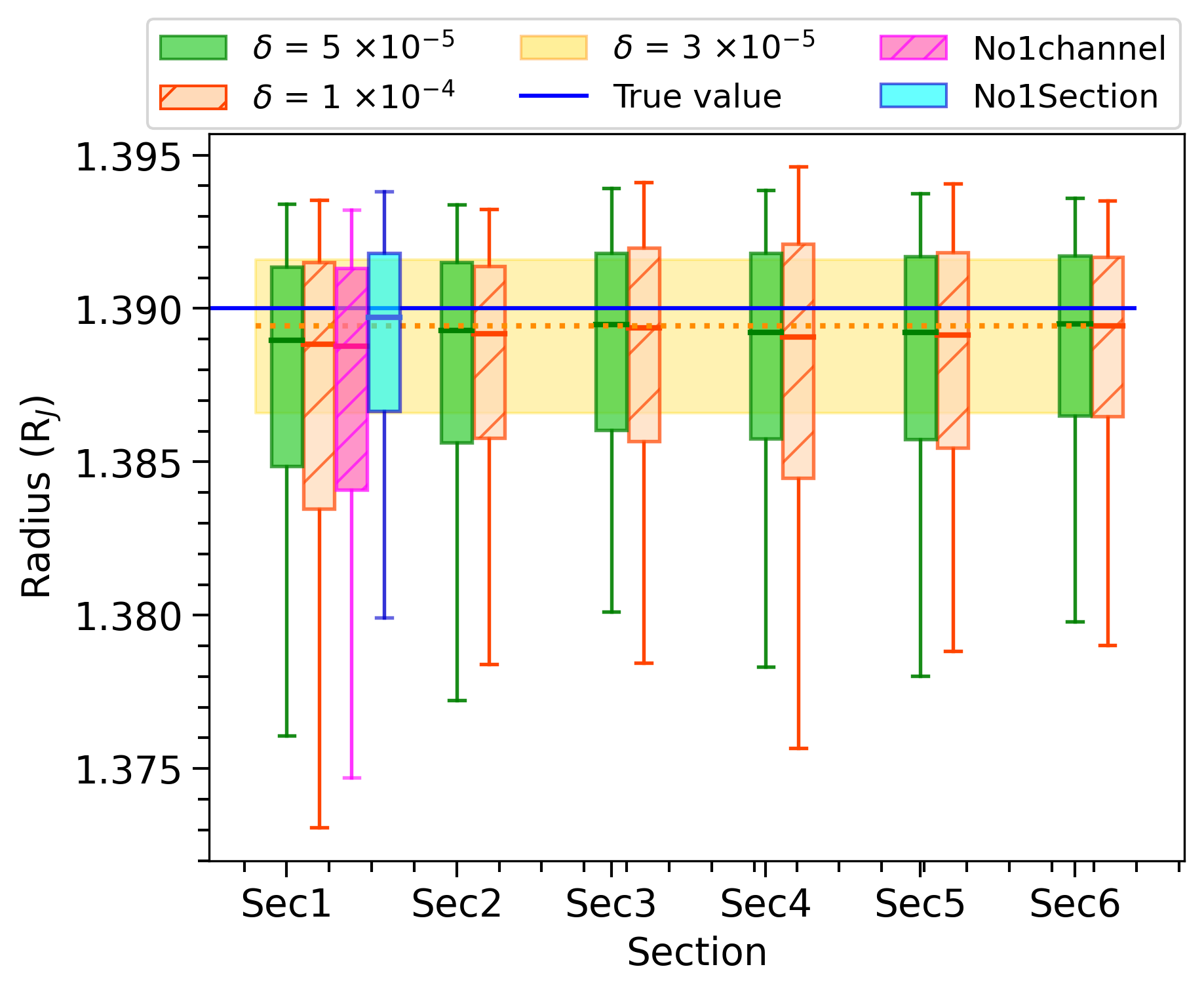}}
        \subfloat[Temperature\label{fig:Temp_SN}]{\includegraphics[scale=0.41]{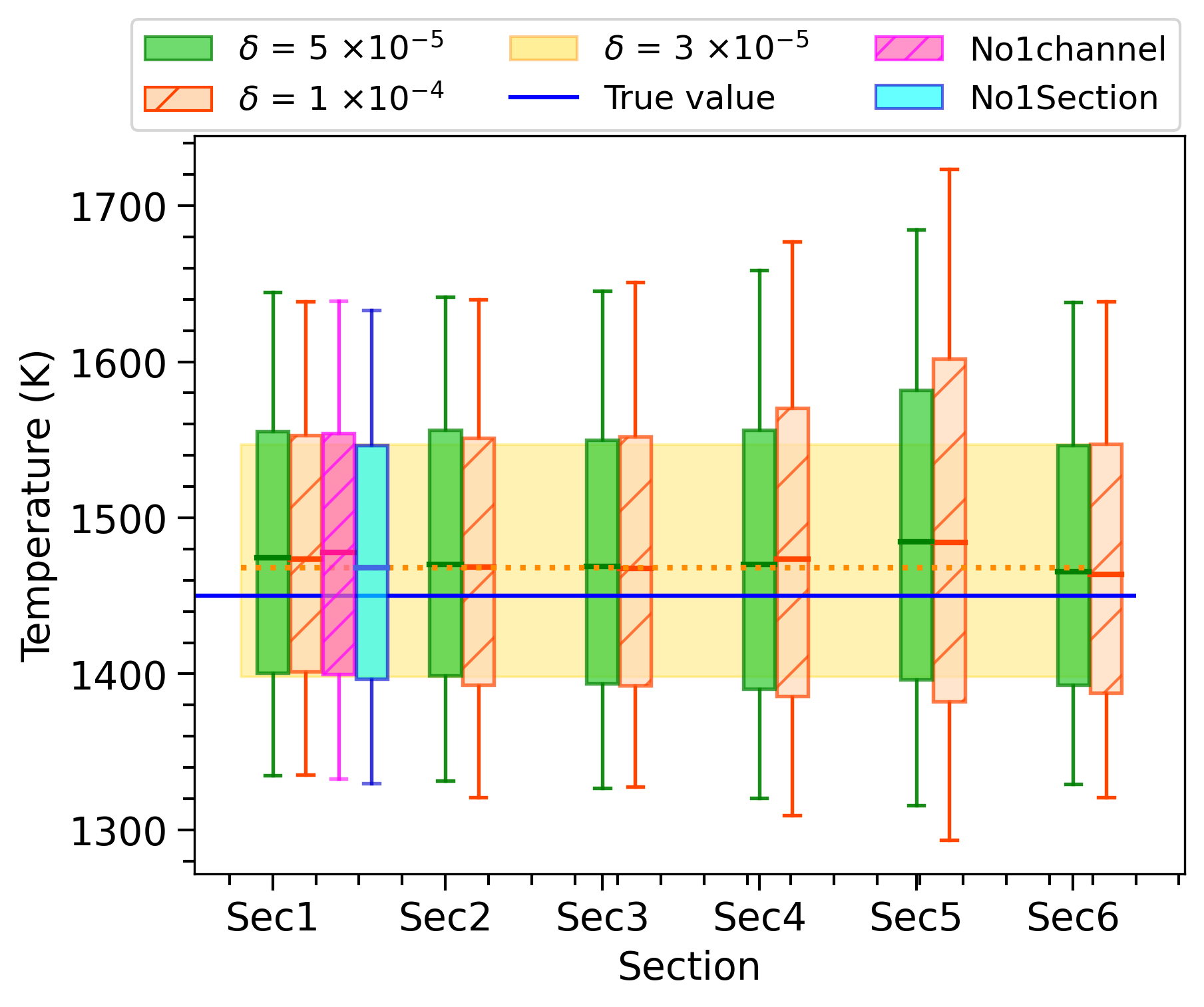}}
        \subfloat[Mass\label{fig:Mass_SN}]{\includegraphics[scale=0.41]{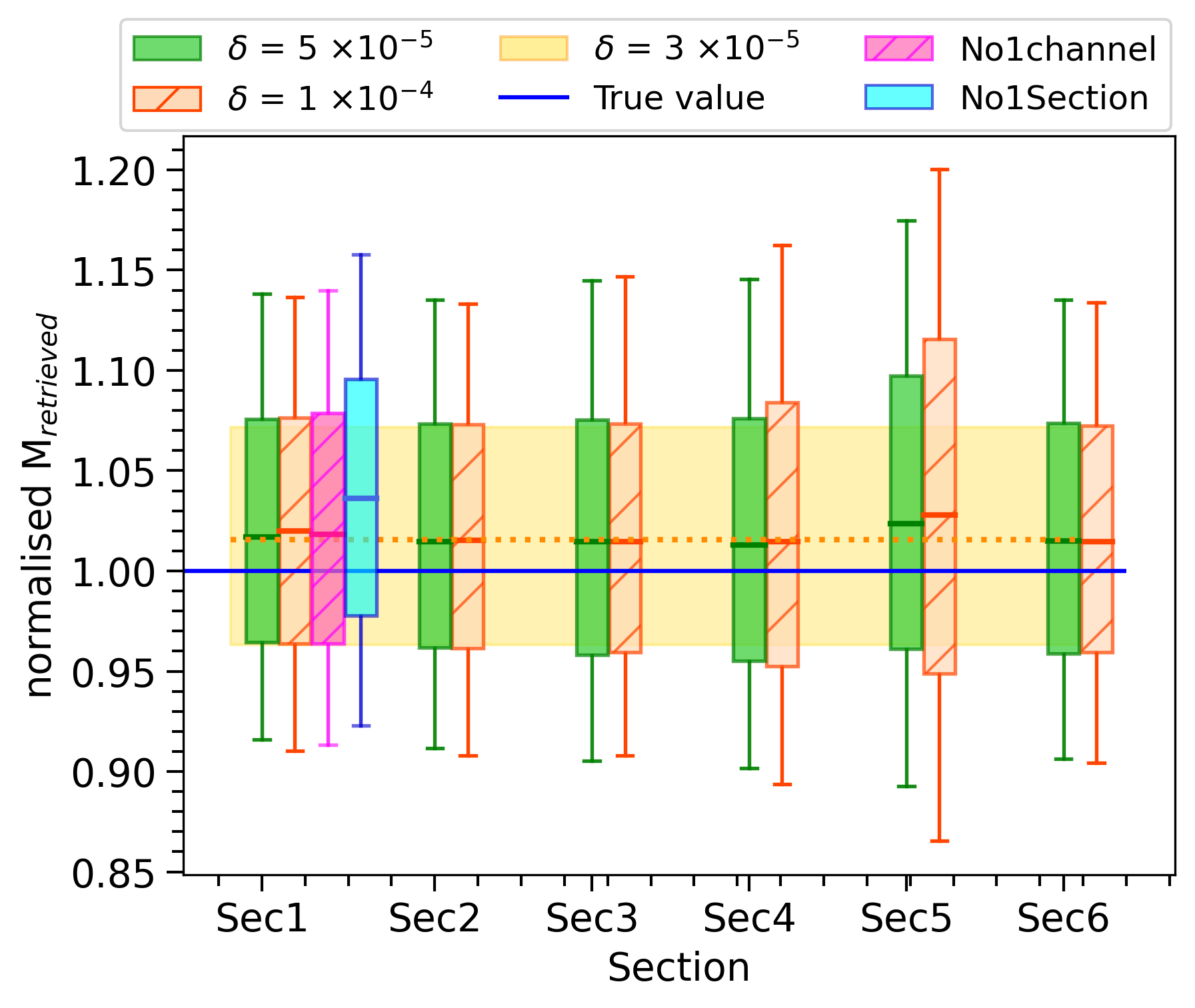}} 
        \\
        \subfloat[H$_2$O mixing ratio \label{fig:H2O_SN}]{\includegraphics[scale=0.41]{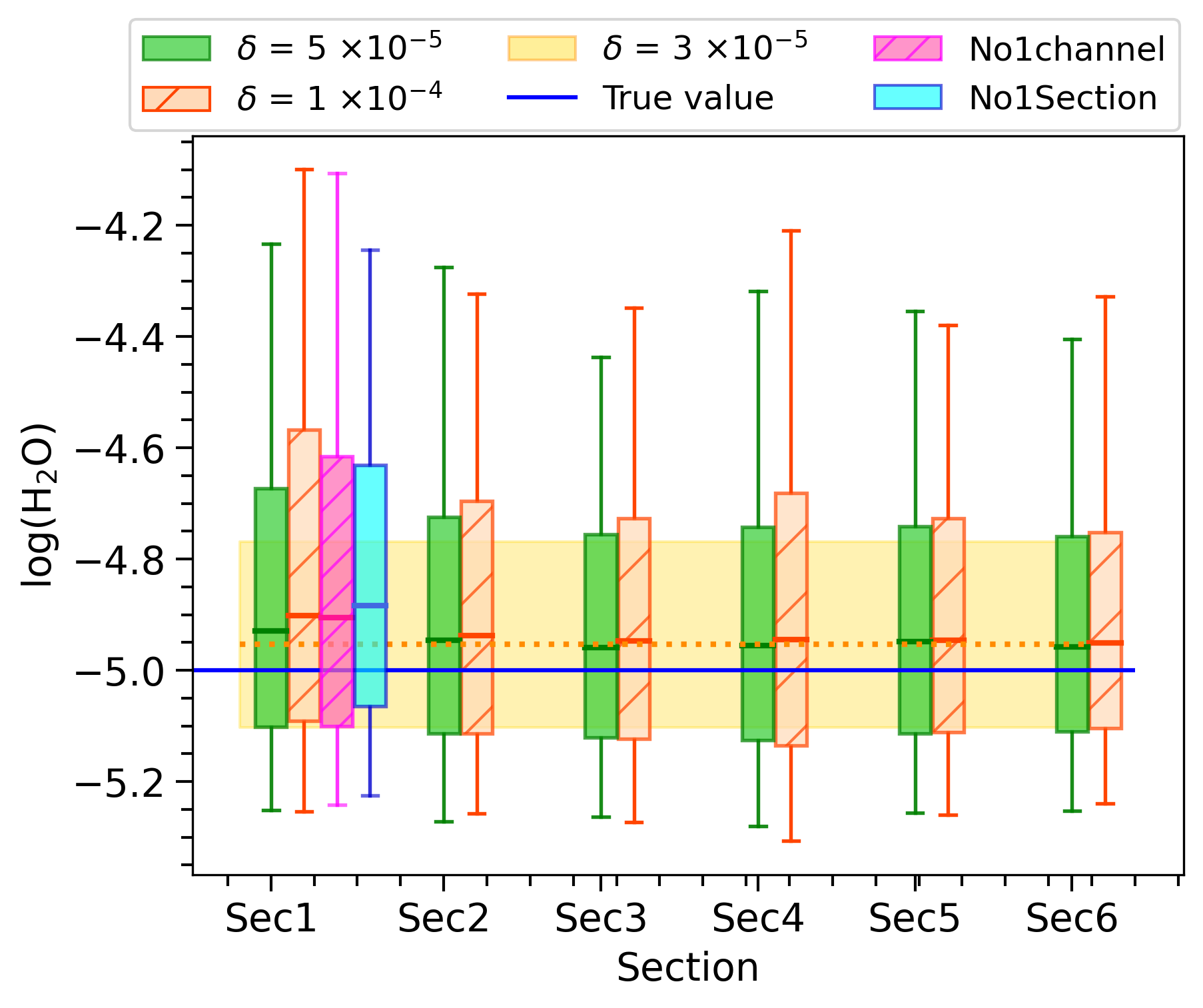}}
        \subfloat[CH$_4$ mixing ratio\label{fig:CH4_SN}]{\includegraphics[scale=0.41]{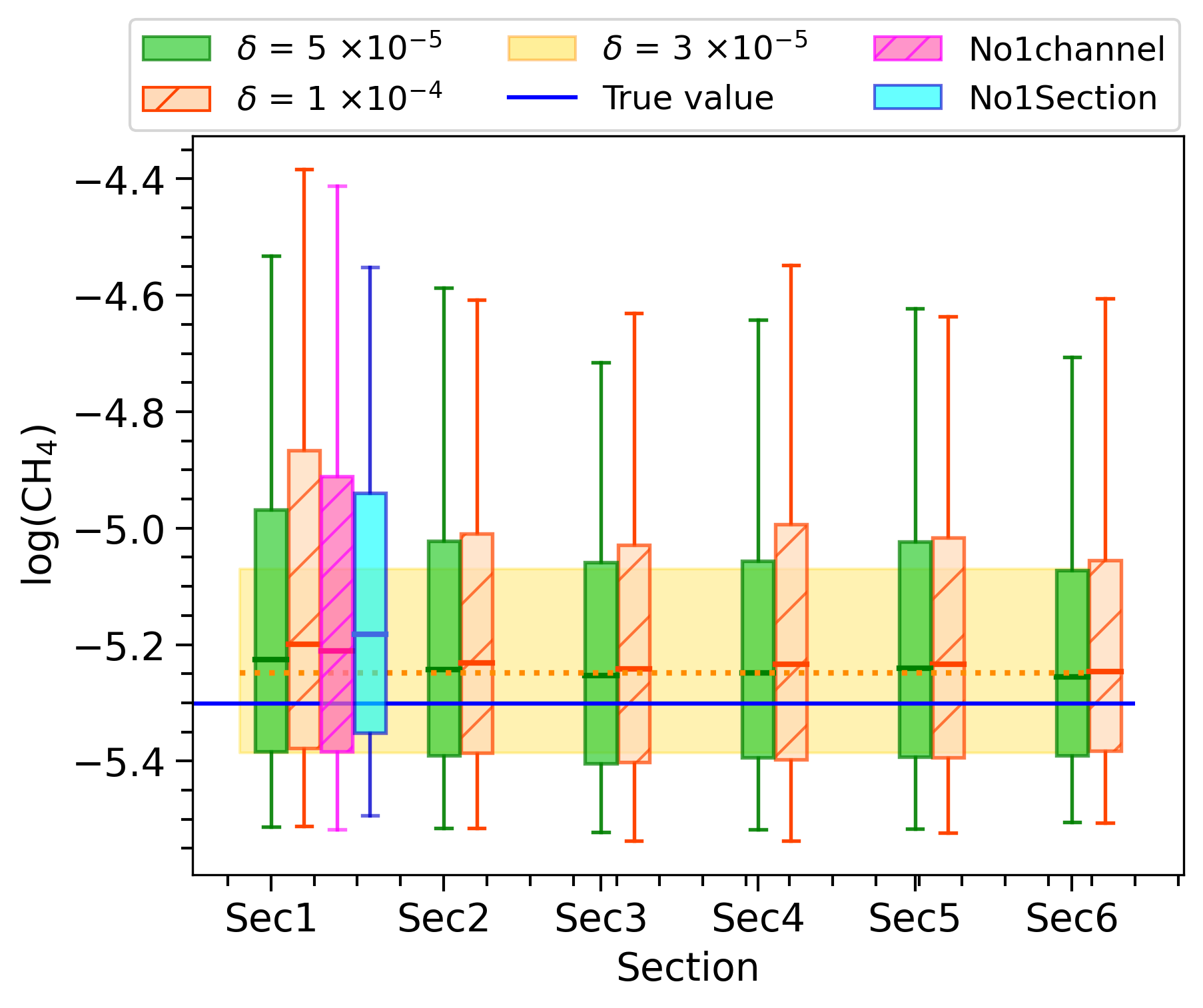}}
        \subfloat[CO mixing ratio\label{fig:CO_SN}]{\includegraphics[scale=0.41]{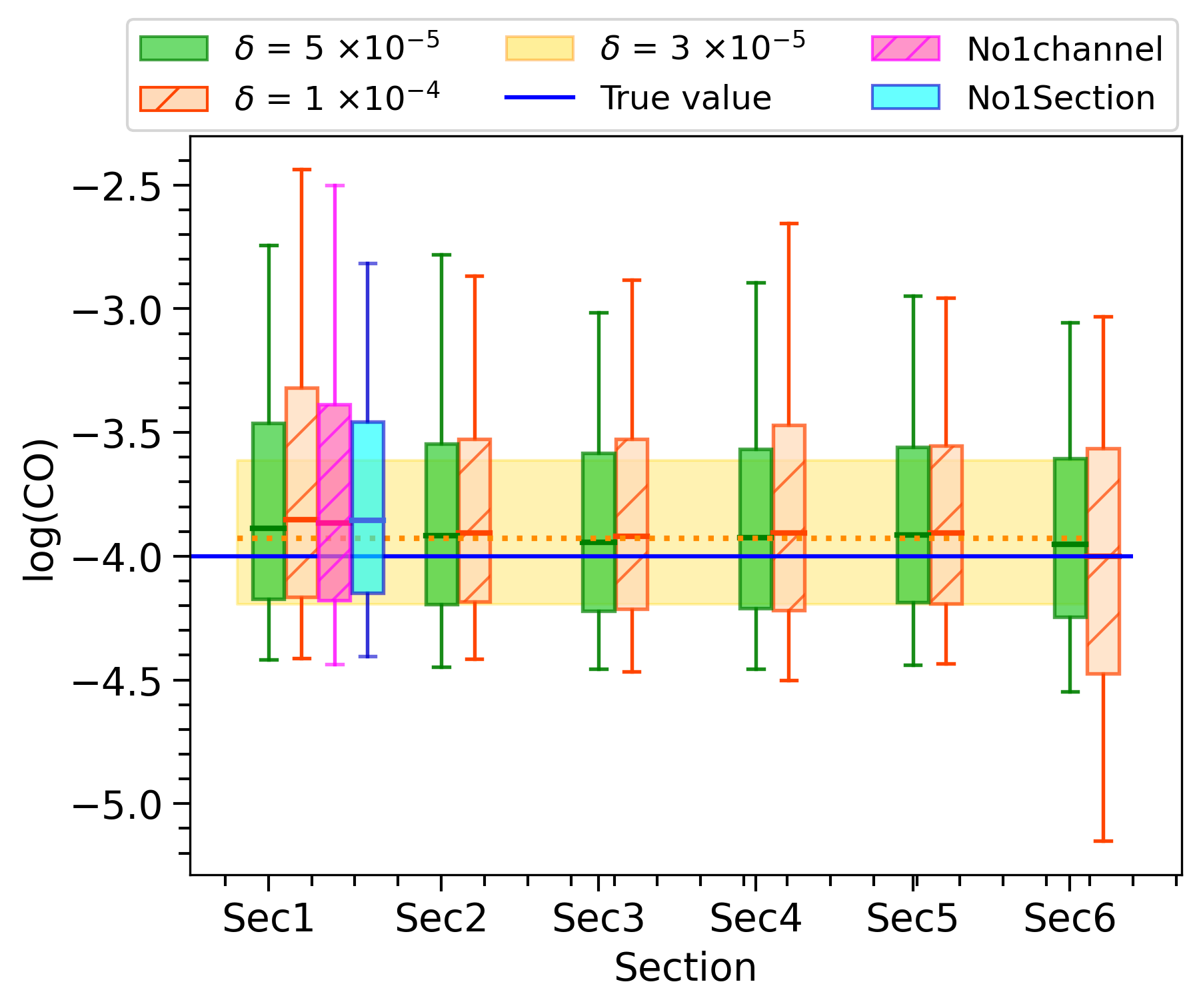}}
    }
    \caption{Test of the impact of the S/N in each of the selected range of the spectrum performed on the primordial atmosphere of the hot Jupiter around an 8th magnitude G star. In green, we show the retrieval performed considering an error of about 5 $\times$ 10$^{-5}$. In orange, the retrieval obtained with an error of about 1 $\times$ 10$^{-4}$.  Yellow band highlights the values retrieved in the original case ($\delta \simeq$ 3 $\times$ 10$^{-5}$). Magenta and cyan boxes represent the distributions of the values obtained by performing the retrieval without the first point or without the entire section 1 of Fig. \ref{fig:split}, respectively. Blue line highlights the true value.}
    \label{fig:SNsection}
 \end{figure*}
 
 \begin{figure}
    \centering
    \includegraphics[scale=0.5]{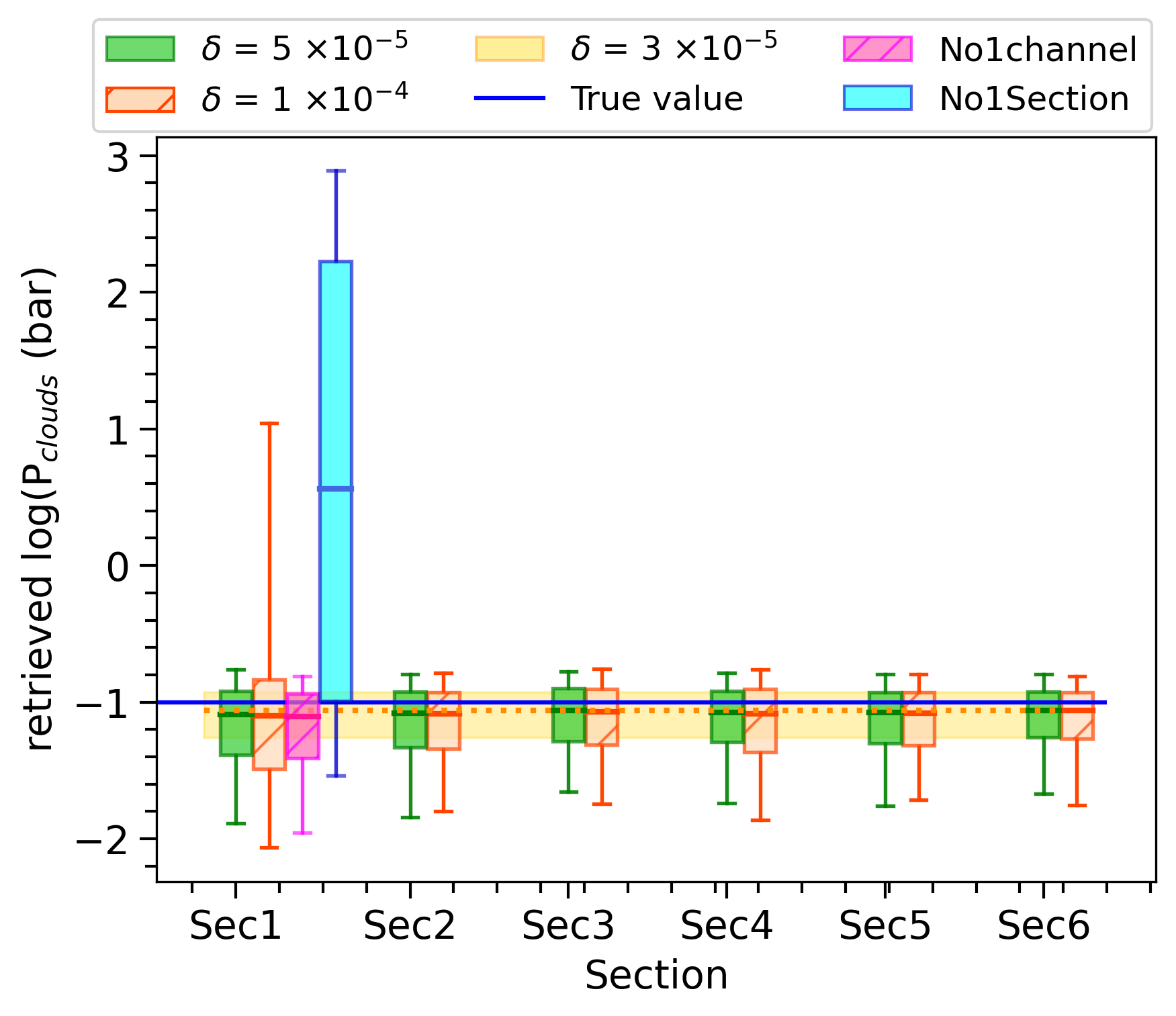}
    \caption{Impact of the S/N in each of the selected range of the spectrum on the retrieved clouds pressure. Colour scale and description of the figure are the same as in Fig. \ref{fig:SNsection}}.
    \label{fig:SNclouds}
 \end{figure}
 
For each retrieval, we changed the error bars of the points within one of the six ranges of the spectrum from $\delta$ = 3 $\times$ 10$^{-5}$ (the yellow band in Fig. \ref{fig:SNsection}) to  $\delta$ = 5 $\times$ 10$^{-5}$ (the green boxes) and $\delta$ = 10$^{-4}$ (the orange boxes).
Also, to better understand the contribution of the spectrum at low wavelengths, we decided to performed other two cases: in the first case we totally excluded the first point (the magenta box in Fig. \ref{fig:SNsection}) and in the second case we excluded all the points of the first section (the cyan box).

Figure \ref{fig:SNclouds} suggests that we are not able to correctly retrieve the clouds pressure when we entirely exclude the points of the first section. This result confirms that the wavelength range between 0.5 and 2 $\mu$ contains information of the features of the clouds, as also suggested by \citet{Yip2021AJ....161....4Y,Yip2021AJ....162..195Y}. This result highlights the importance of the continuous wavelengths coverage of the blue end of the spectrum that allow us to fit for more complicated cloud models and probe the presence of species such as H$_2$O and CH$_4$.

For all the other parameters, except the temperature, we can see an increase in the uncertainties with a decrease in the S/N in the first section. 
Since, in the first section, the main contribution to the spectrum is due to the clouds component, this result suggests that all the parameters of the retrieval, except the temperature, are impacted by the clouds pressure knowledge. However, whilst the cloud is harder to constrain, in cases of low-altitude clouds, we are still able to constrain the atmospheric parameters, which is encouraging if there are cases where we cannot use the shortwave region.
\subsection{Secondary atmosphere}\label{Sec:Secondary}
\begin{figure*}[h!]
 \centering
        {\subfloat[Radius\label{fig:Radius_Secondary}]{\includegraphics[scale=0.42]{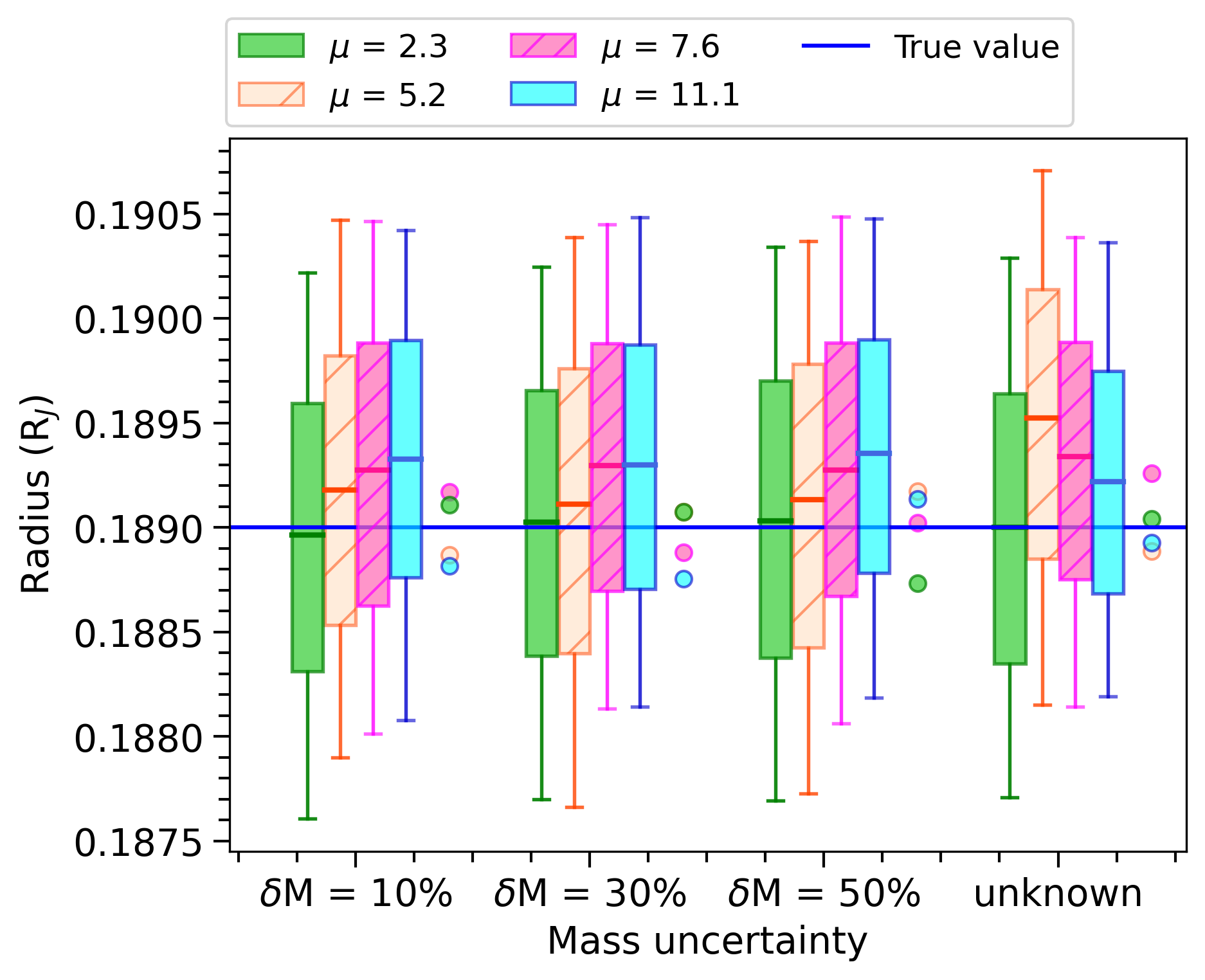}}
        \subfloat[Temperature\label{fig:Temp_Secondary}]{\includegraphics[scale=0.42]{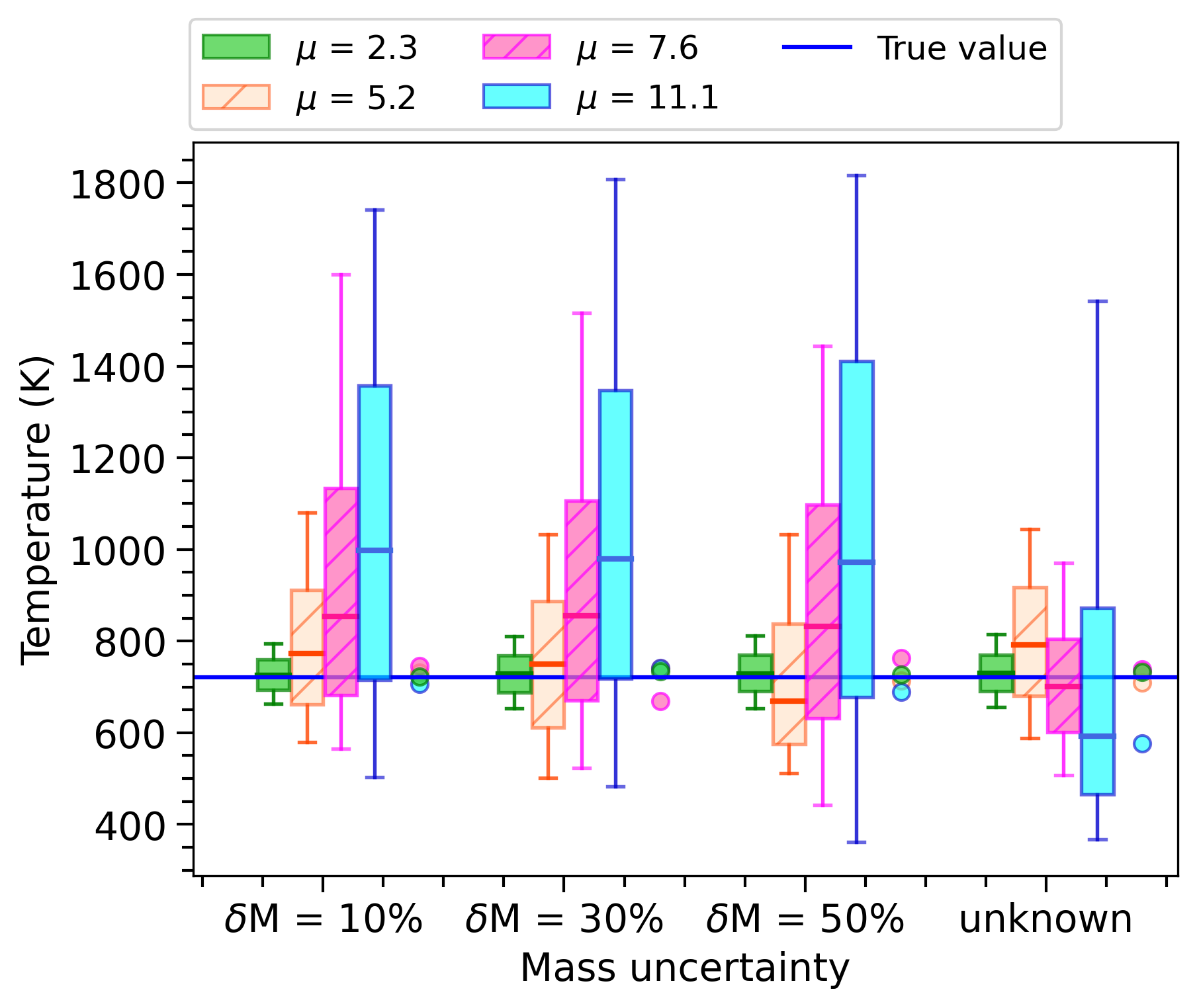}}
        \subfloat[Mass\label{fig:Mass_Secondary}]{\includegraphics[scale=0.42]{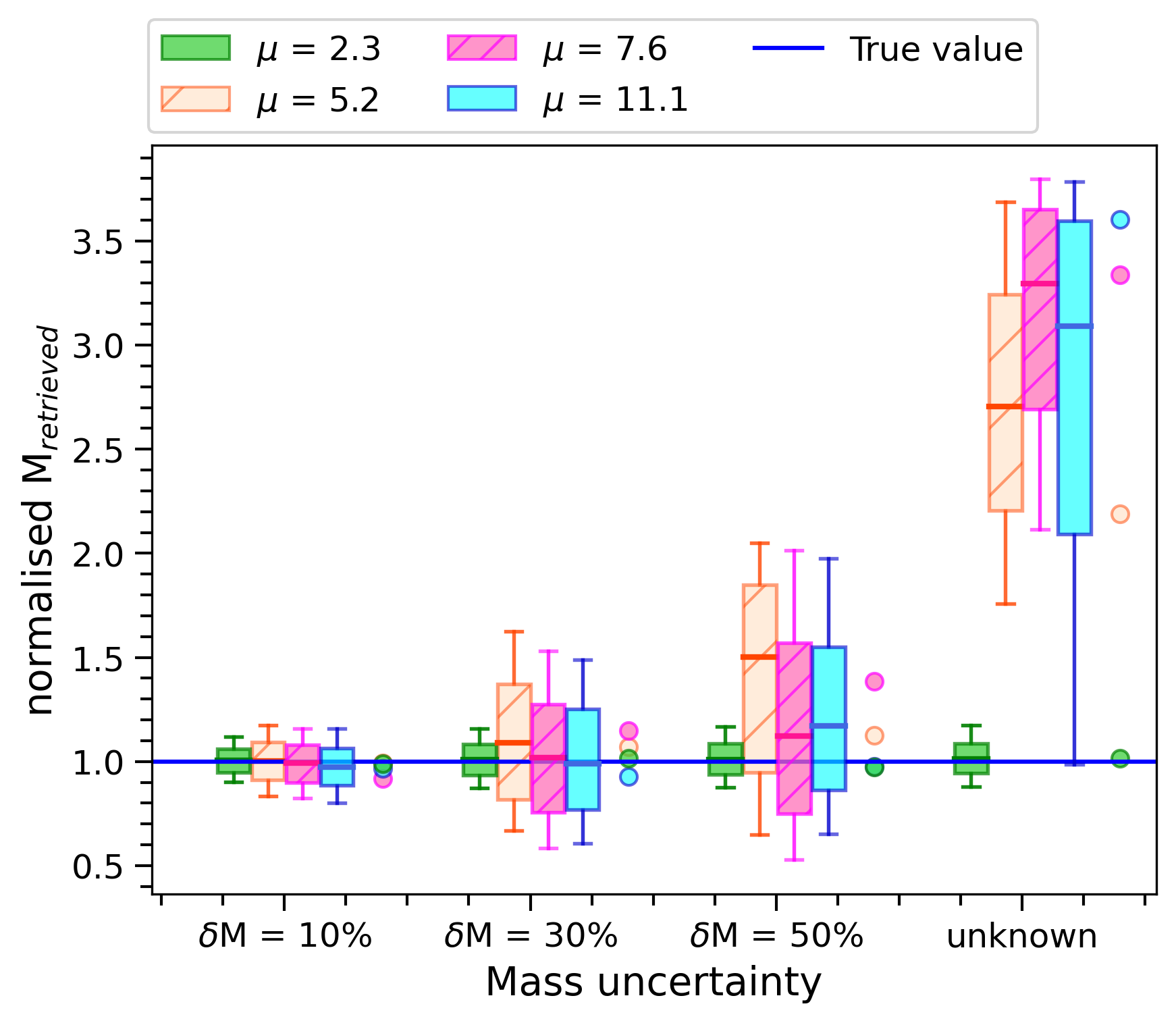}} \\
        \subfloat[H$_2$O mixing ratio \label{fig:H2O_Seycondar}]{\includegraphics[scale=0.42]{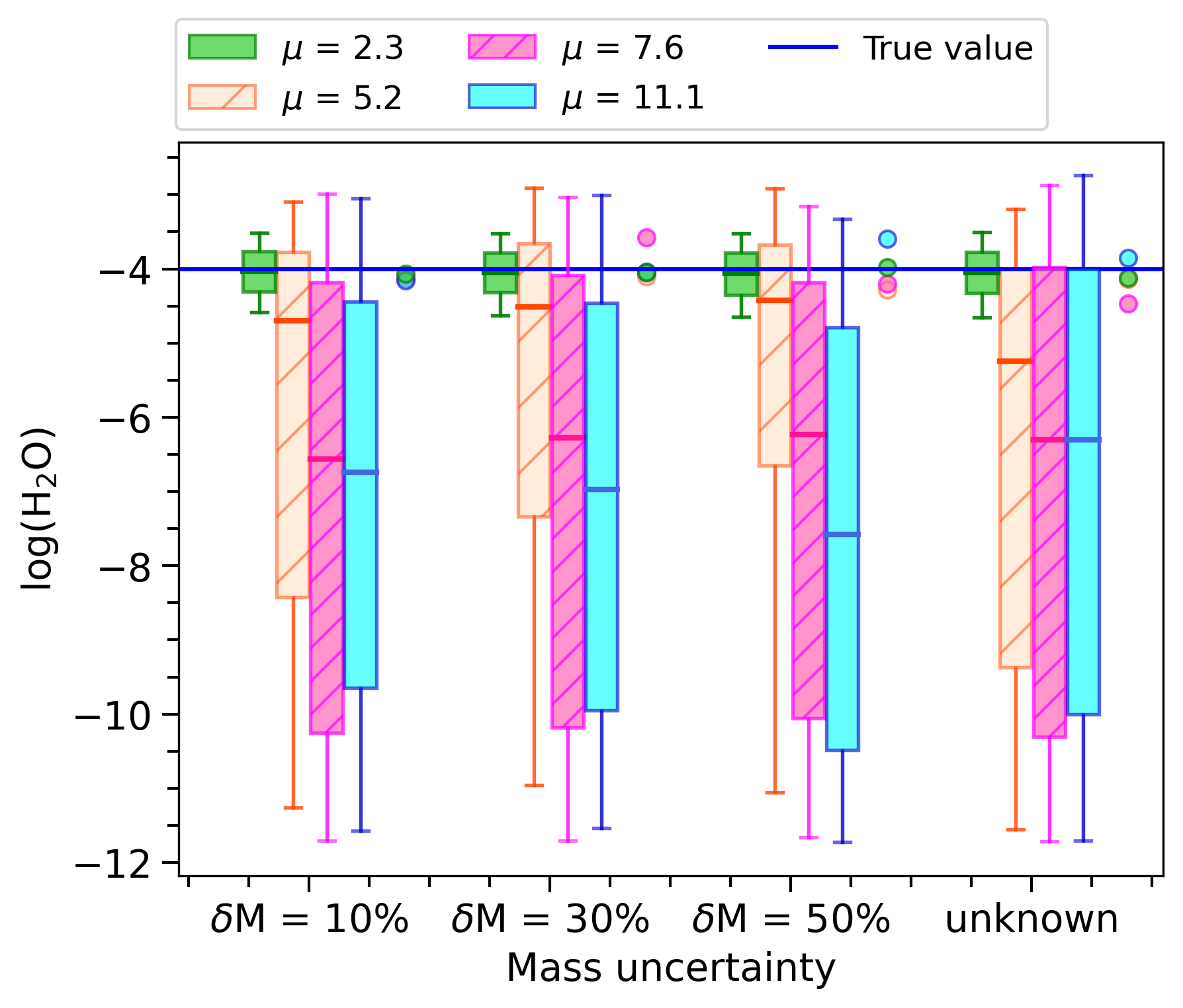}}
        \subfloat[CH$_4$ mixing ratio\label{fig:CH4_Secondary}]{\includegraphics[scale=0.42]{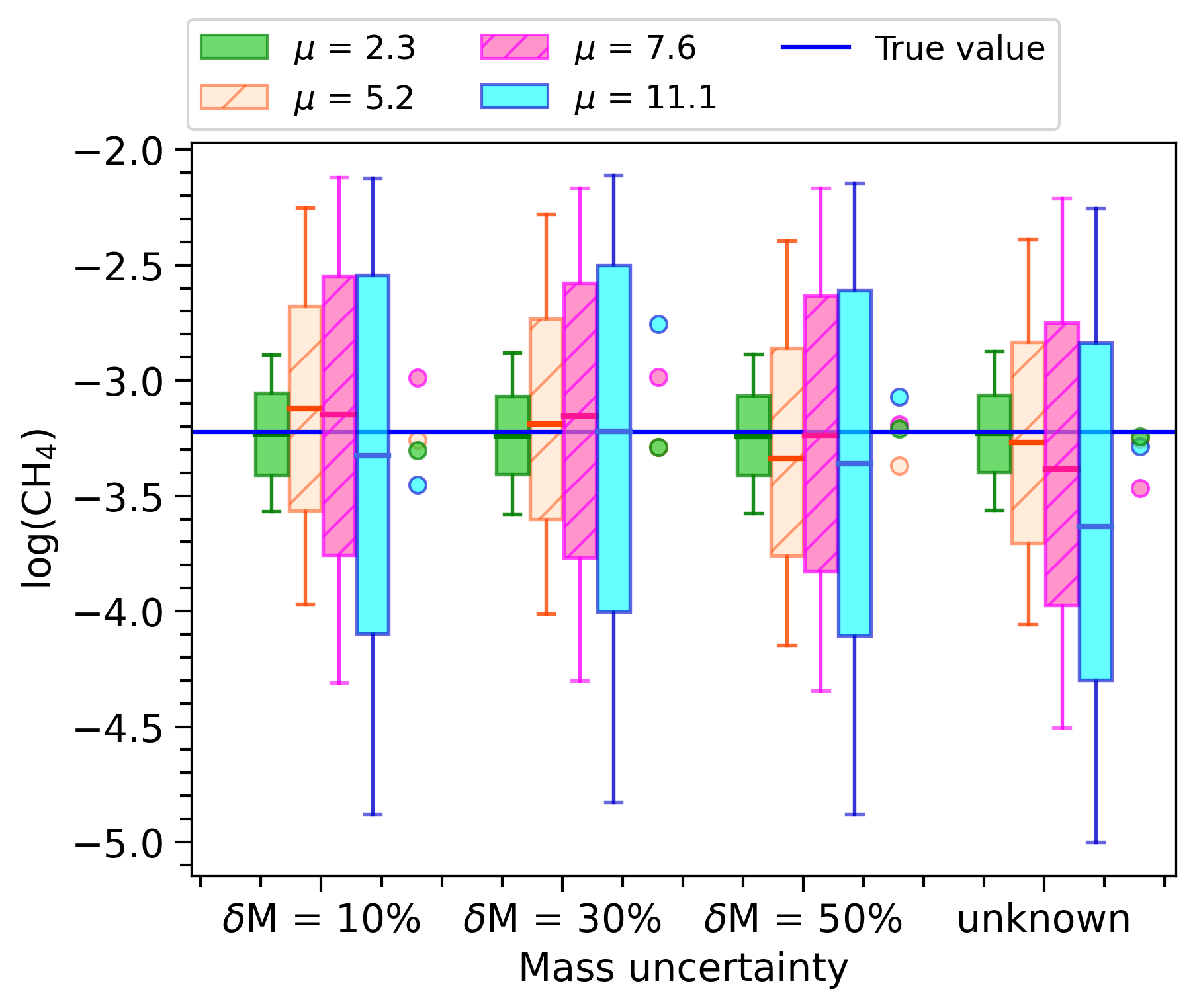}}
        \subfloat[N$_2$/He \label{fig:N2He_Secondary}]{\includegraphics[scale=0.42]{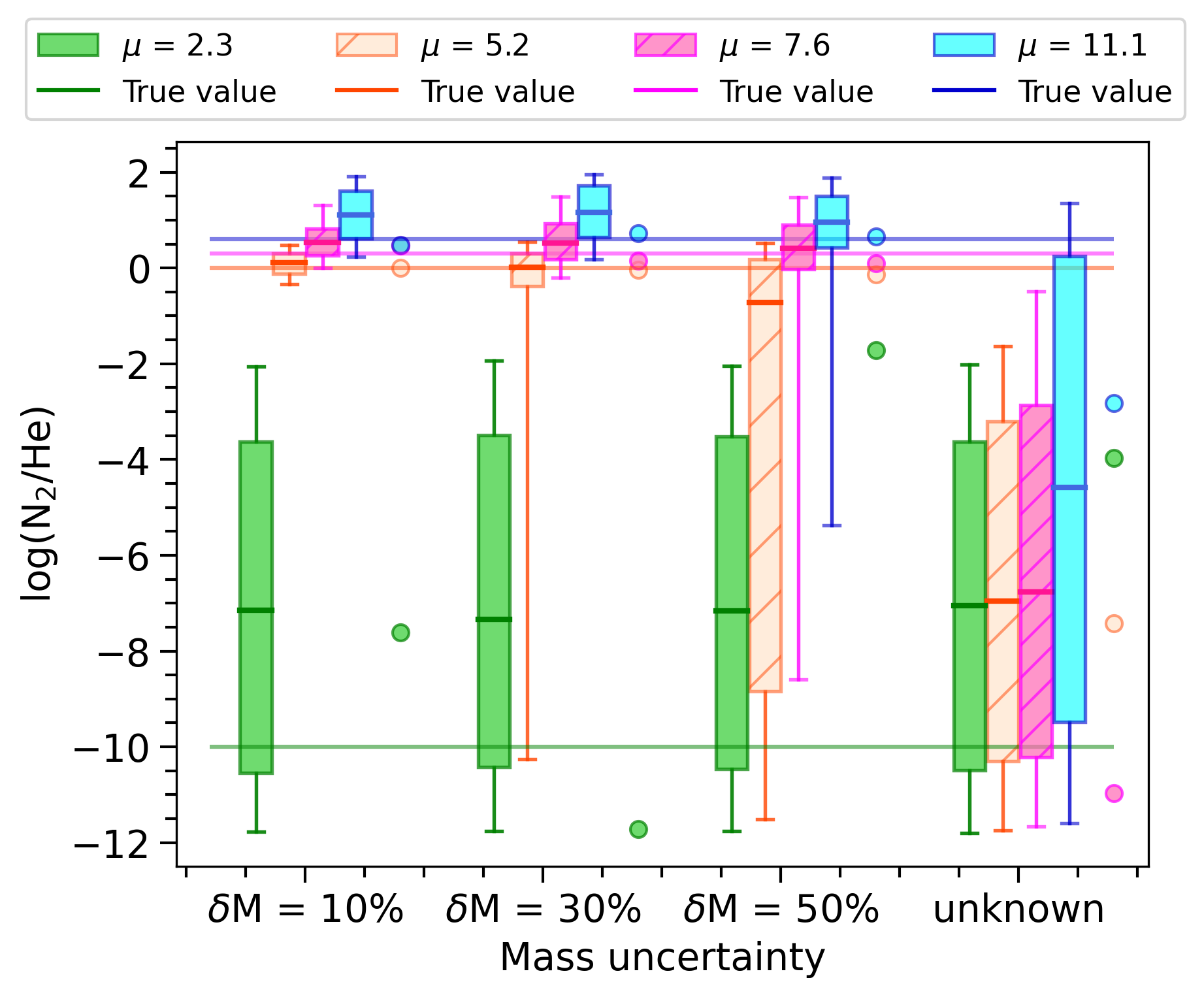}}
    }
    \caption{Impact of the mass uncertainties on the retrieval for different scenarios of heavy N$_2$-dominated secondary atmospheres represented by increasing values of $\mu$ (2.3 in green, 5.2 in orange, 7.6 in magenta, and 11.1 in cyan). Blue line highlights the true value. The points alongside the boxes highlight the MAP (maximum-a-posteriori) parameters obtained for each analysed case. The size of the box and the error bar represent the points within 1$\sigma$ and 2$\sigma$ of the median of the distribution (highlighted with solid lines), respectively.}
    \label{fig:Secondary}
 \end{figure*}

\begin{figure}[h!]
    \centering
    \includegraphics[scale=0.55]{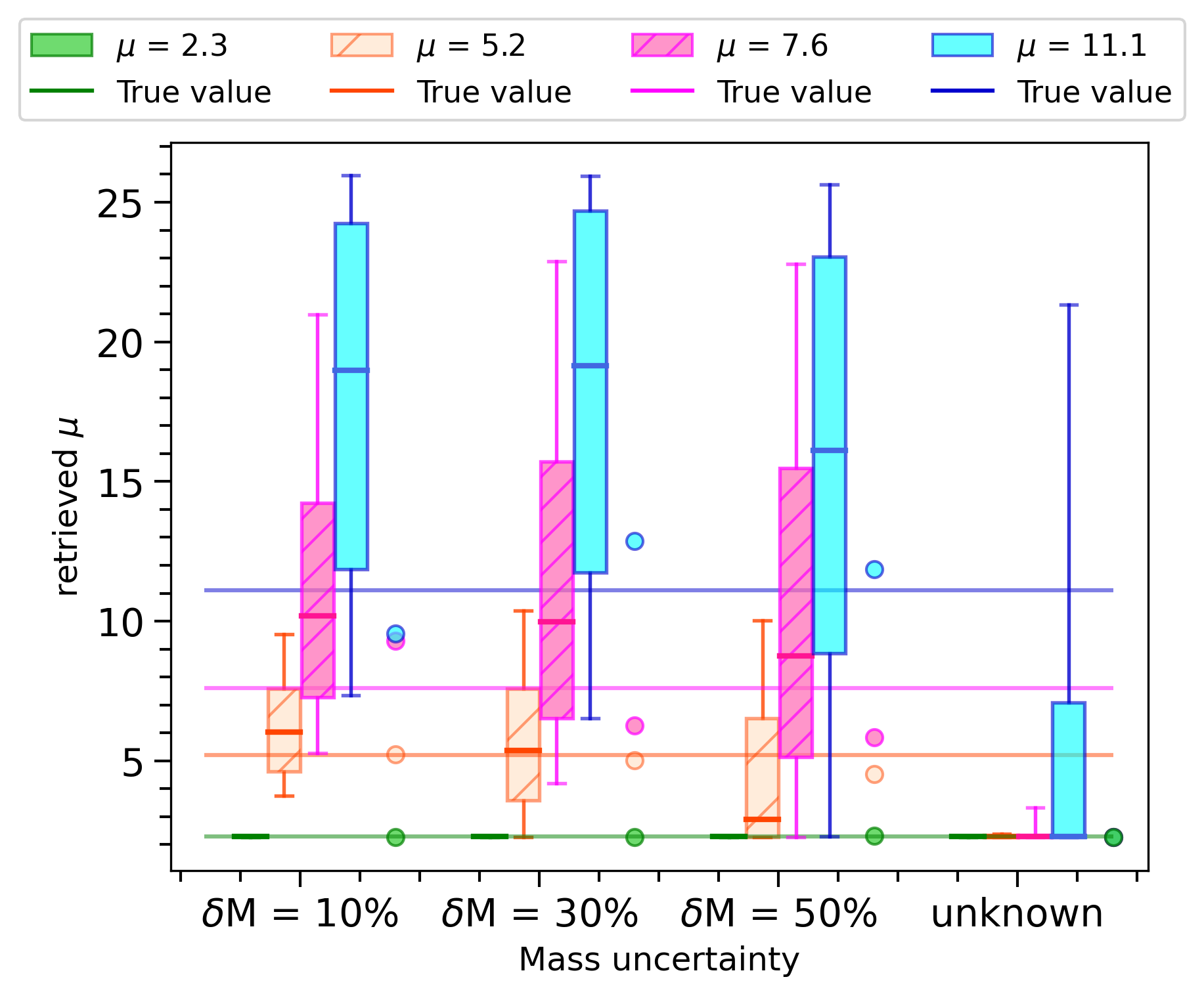}
    \caption{Impact of the mass uncertainties on the retrieved mean molecular weight for different scenarios of heavy N$_2$-dominated secondary atmospheres represented by increasing values of $\mu$ (2.3 in green, 5.2 in orange, 7.6 in magenta, and 11.1 in cyan). Colour and description of the figure are the same as in Fig. \ref{fig:Secondary}}.
    \label{fig:Secondarymu}%
 \end{figure}
 
The atmospheric retrieval of Earths and super-Earths is challenging because the mean molecular weight, $\mu$, is completely unconstrained (with the assumption of $\mu \sim$ 2.3 no longer valid). Furthermore, diatomic background gases, such as H$_2$ and N$_2$ referred to as spectrally inactive gases, do not exhibit strong vibrational absorptions bands, so they have not directly observable features in the spectrum. Additionally, lower-mass planets tend to not have precise mass measurements.  

To investigate how the mass uncertainties could impact in the retrieval of low-mass planets, we considered a secondary atmosphere consisting of elements heavier than H/He.
The super-Earth simulated here is based on HD 97658b. The parameters used in our model are reported in Table \ref{tab:Secondary}.

We considered a N$_2$-dominated atmosphere and we used the inactive gas N$_2$ to increase the mean molecular weight of the atmosphere and simulate a heavy atmospheres around a rocky planet. 
We also included H$_2$O and CH$_4$ as trace gases fixing their absolute abundances at 10$^{-4}$ and 6 $\times$ 10$^{-4}$, respectively. The rest of the atmosphere is filled with a combination of H$_2$ and He. We considered four different scenarios with different values for the mean molecular weight ($\mu$ = 2.3, N$_2$/He = 10$^{-10}$; $\mu$ = 5.2, N$_2$/He = 1; $\mu$ = 7.6, N$_2$/He = 2;  $\mu$ = 11.1, and N$_2$/He = 4) to explore different compositions of the atmosphere. The highest considered mean molecular weight was selected to have  atmospheric features detectable by an instrument such as Ariel, but other worst-case scenarios for this planet could exist. For instance, a pure Venus-like CO$_2$ atmosphere would not be detectable without impacting the science objectives of the Ariel mission. Also, in order to test the impact of the mass uncertainties, we performed the retrieval considering a mass uncertainty of about 10\%, 30\%, and 50\%, along with a case in where the mass is totally unknown (by using very large boundary for the mass parameter). 
In Fig. \ref{fig:Secondary}, we show the impact of the mass uncertainties on the atmospheric retrievals of different scenarios where we considered heavy secondary atmospheres represented by increasing values of $\mu$. In Appendix \ref{app:tabelle}, we summarise the results and in Appendix \ref{app:corner}, we report some examples of corner plots obtained from these analyses. We note that a mass estimation with an uncertainty equal or lesser than 50\% could help us to better constrain the mean molecular weight for all the tested cases with different mean molecular weight. 

From this plot we do not see significant differences in the retrieved atmospheric parameters obtained when performing the retrieval in cases where we know the mass with different uncertainties; this is not surprising, because for the high mean molecular weight atmosphere the scale height is relatively small, so changes in the gravity will not produce such large differences in the spectrum. However, as expected, the retrieved mass shows a correlation with the mass uncertainty.

In Fig. \ref{fig:Secondary}, we highlight the maximum a posteriori (MAP) parameters with circle points.
In cases with higher values for $\mu,$ some discrepancies appear in the temperature and H$_2$O retrieved values with respect to the true values, and in some cases, we obtained retrieved values that are not within 1$\sigma$ of the true values. However, we note that in these cases, even when we have a larger distribution, the MAP values obtained from the retrieval are totally consistent with the true values.

In addition, we note some discrepancies in the retrieved MAP of N$_2$/He when $\mu$ = 2.3. This result suggests that we are not able to constrain this ratio. In these cases, we can only define a possible range of values and some performed tests have demonstrated that this result does not depend on the choice of prior limits.

Furthermore, from Fig. \ref{fig:Secondarymu}, we can see a slight trend between the mean molecular weight and the mass uncertainties. 
In particular, we note that for a mass uncertainty lower than 50\%, we are able to retrieve the mean molecular weight with a higher accuracy (mostly if we consider the MAP values) with respect to the unknown mass cases and, in particular, for the heavier atmospheres, along with a slight increase precision when we perform the retrieval with a mass uncertainty of 10\%. This is probably due to the higher accuracy and precision in the retrieval of the N$_2$/He when we consider a mass uncertainties of 10\%.

All these results suggest that we should be able to correctly retrieve the atmospheric parameters of a secondary atmosphere with a clear sky, even when we know the mass with an uncertainty of 50\%, despite the fact that we have considered our worst-case scenario to assess the degeneracy between the mass and the mean molecular weight. Our analysis, also suggests that this degeneracy is intrinsic to secondary atmospheres -- and not directly connected with the mass uncertainty. Despite this, and precisely by virtue of this degeneracy, a more accurate estimate of the mass obtained from an independent determination could help to break the degeneracy, thus increasing the accuracy in the determination of the abundances of the fill gases.

We also tested the atmospheric retrieval of an analogue scenario, but considering a 10.5th magnitude stars, however. The results obtained from this test are reported in Table \ref{table:Secondarymv10}. In this case, we obtained similar results with respect to the previous case in which we considered an 8th magnitude star. However, due to the lower S/N, we obtained a larger uncertainties for all the parameters, including those for the cases with lower mean molecular weight.

\subsection{Cloudy secondary atmosphere}\label{Sec:ClodySecondary}
\begin{figure}[h!]
    \centering
    \includegraphics[scale=0.55]{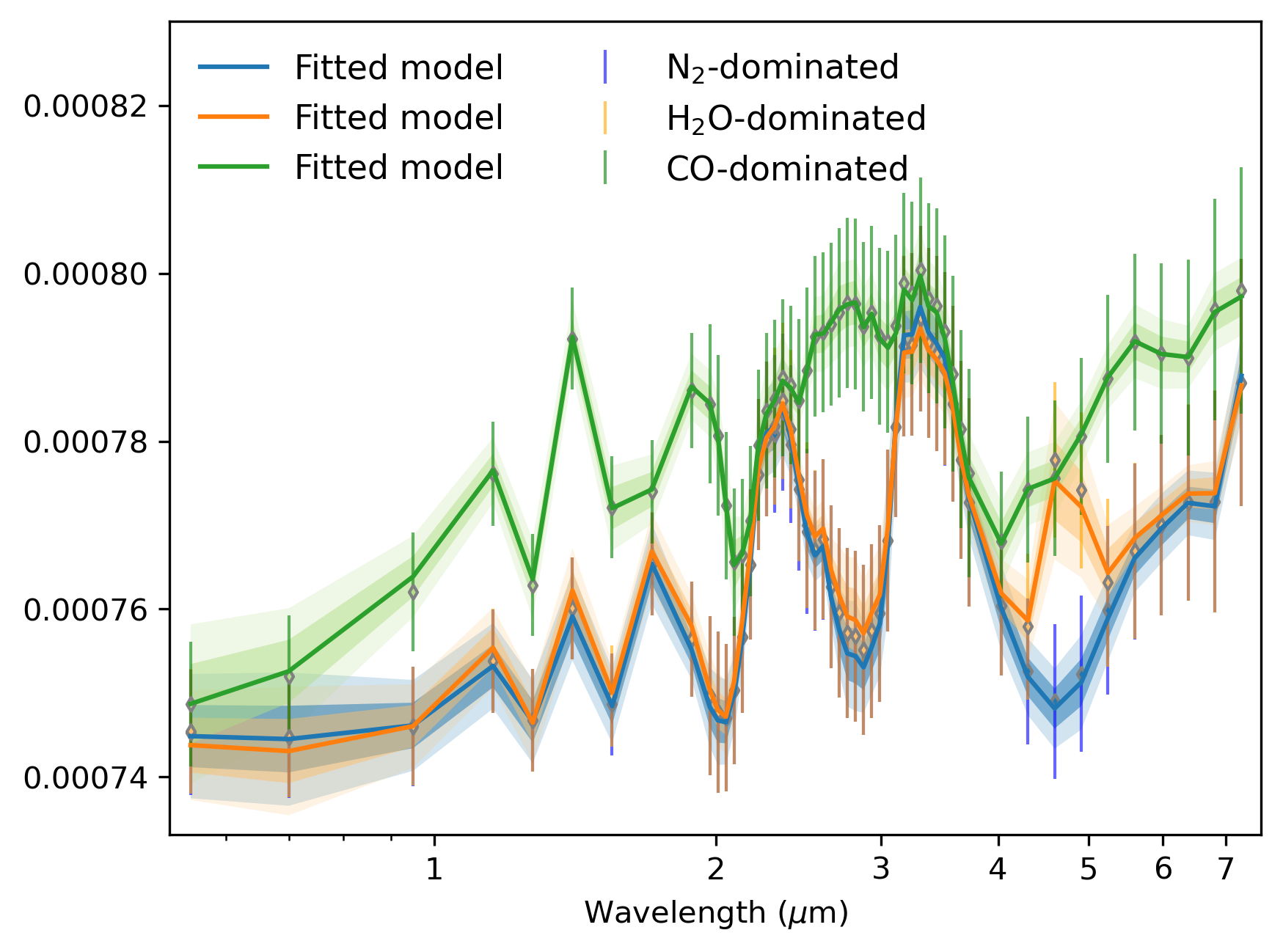}
    \caption{Comparison between the observed spectrum and the fitted model obtained for a N$_2$- (blue), CO- (green), and H$_2$O-dominated (orange) atmospheres, in the case of $\mu$ = 5.2 and with a P$_{clouds}$ = 5 $\times$ 10$^{-2}$ bar.}
    \label{fig:Comparison}%
 \end{figure}

Finally, we investigated the case of cloudy secondary atmospheres. We note that small planets might not have a H2-dominated atmosphere and the dominant gas is often unknown. 
We decided to investigate three different scenarios: in the first, we considered a nitrogen-dominated atmosphere representative of a rocky planet to investigate the retrieval results and compared it with the case without clouds (see Sec. \ref{Sec:Secondary}). Also, in order to provide evidence of the difference in the atmospheric retrieval when an active gas dominates in the transmission spectrum, we analysed the second and third scenarios -- where we considered a H$_2$O-dominated and a CO-dominated atmosphere, respectively. 
Atmospheres dominated by species such as H$_2$O or CO would have traceable molecular features directly observable in the spectrum, as we can see from Fig. \ref{fig:Comparison}, where we compare the observed spectrum and the fitted model obtained for a N$_2$-, H$_2$O- and CO-dominated atmosphere in the case of $\mu$ = 5.2 and with a P$_{clouds}$ = 5 $\times$ 10$^{-2}$ bar. These cases represent a more favourable scenario for the inverse models with respect to the N$_2$-dominated ones.
 
In Appendix \ref{app:tabelle}, we report the results obtained from the analysis of the cloudy secondary atmosphere in the three different scenarios and in all the configuration of mean molecular weight for the clouds pressure 10$^{-1}$ and 10$^{-3}$ bar.

\paragraph{\textbf{N$_2$-dominated Atmosphere}}
\begin{figure*}[h!]
 \centering
        {\subfloat[Radius\label{fig:Radius_N2dom5.2Secondary}]{\includegraphics[scale=0.42]{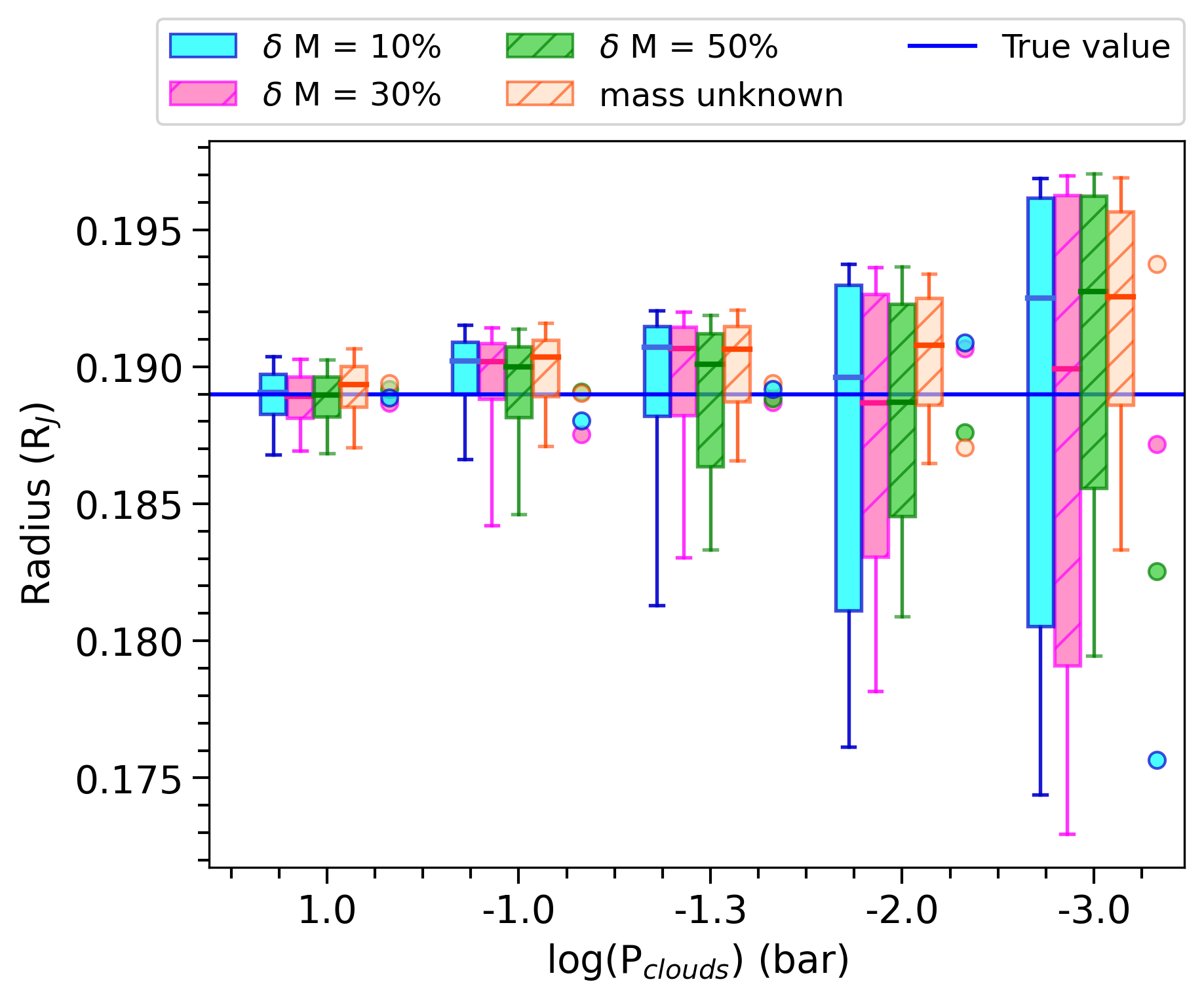}}
        \subfloat[Temperature\label{fig:Temp_N2dom5.2Secondary}]{\includegraphics[scale=0.42]{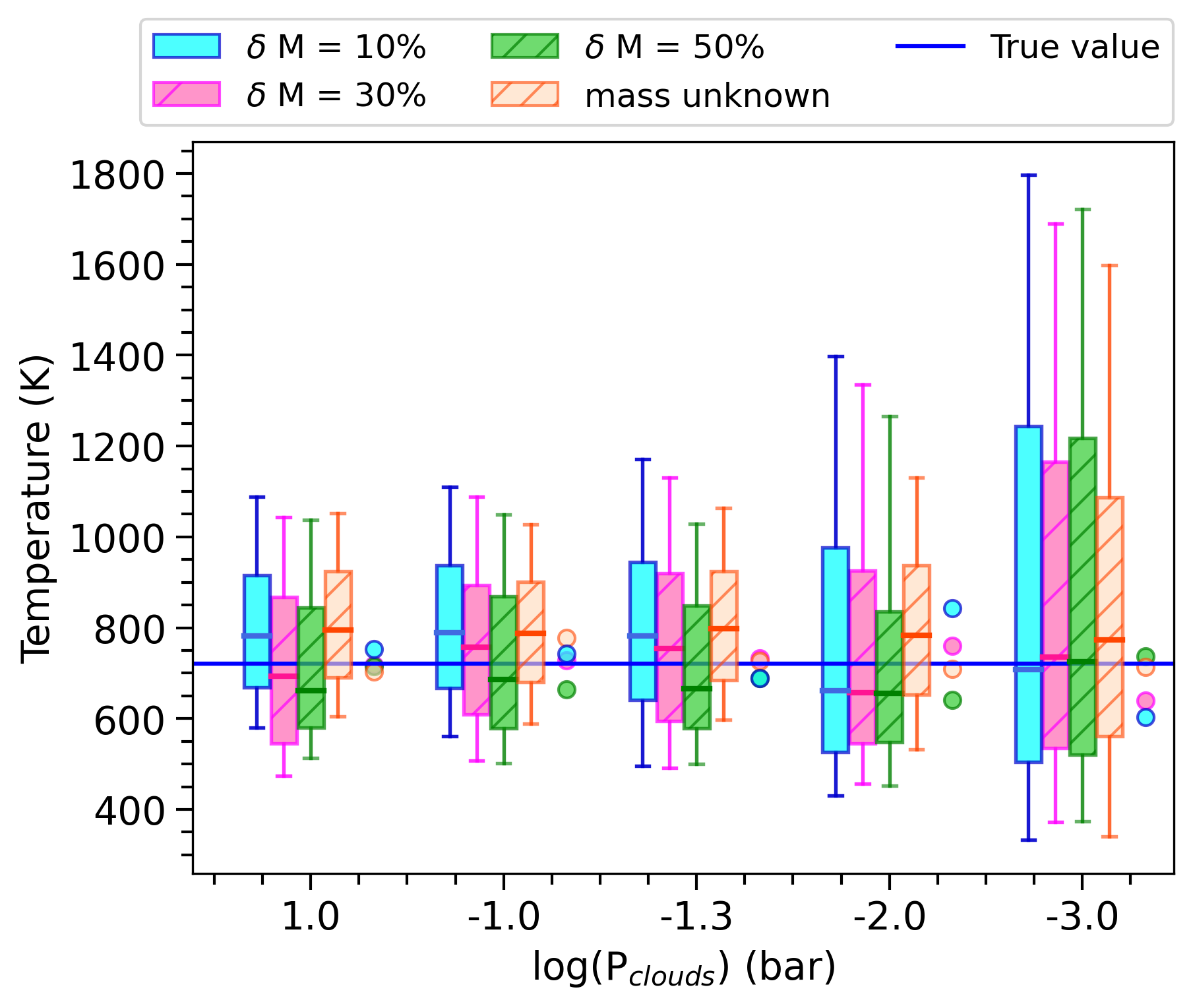}}
        \subfloat[Mass\label{fig:Mass_N2dom5.2Secondary}]{\includegraphics[scale=0.42]{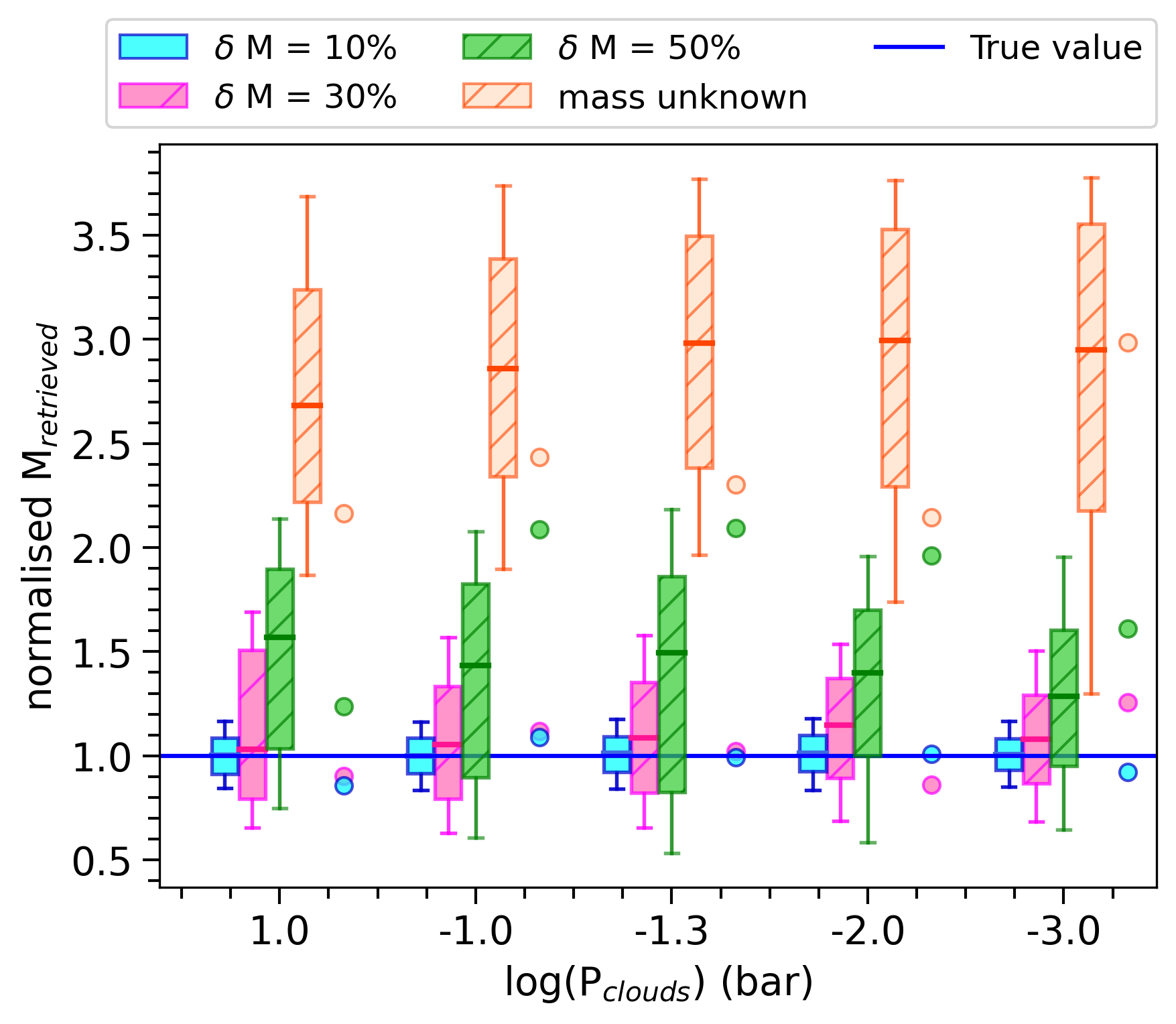}} 
        \\
        \subfloat[Clouds\label{fig:Clouds_N2dom5.2Secondary}]{\includegraphics[scale=0.42]{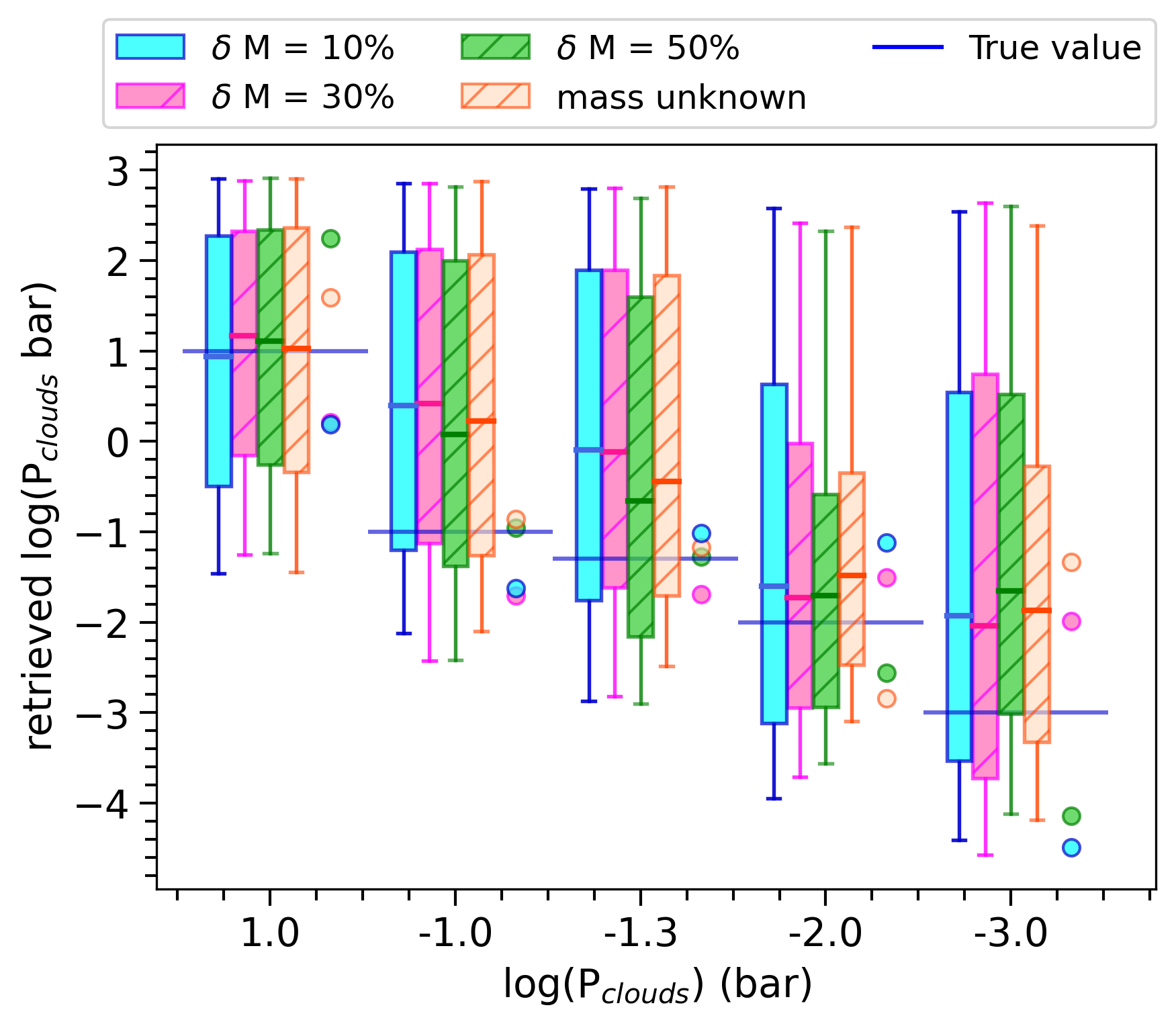}}
        \subfloat[$\mu$\label{fig:mu_N2dom5.2Secondary}]{\includegraphics[scale=0.42]{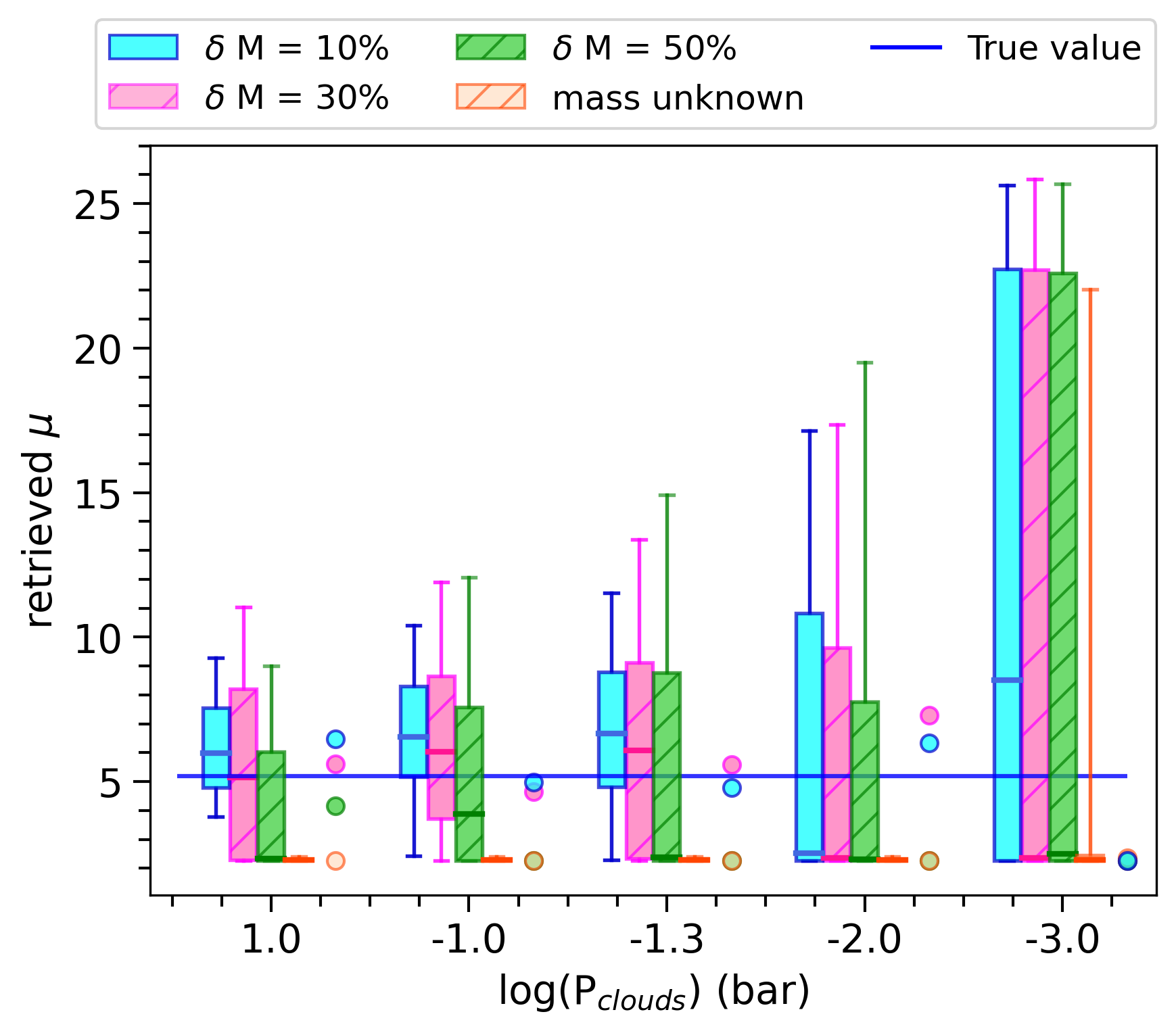}} 
        \subfloat[H$_2$O mixing ratio \label{fig:H2O_N2dom5.2Seycondar}]{\includegraphics[scale=0.42]{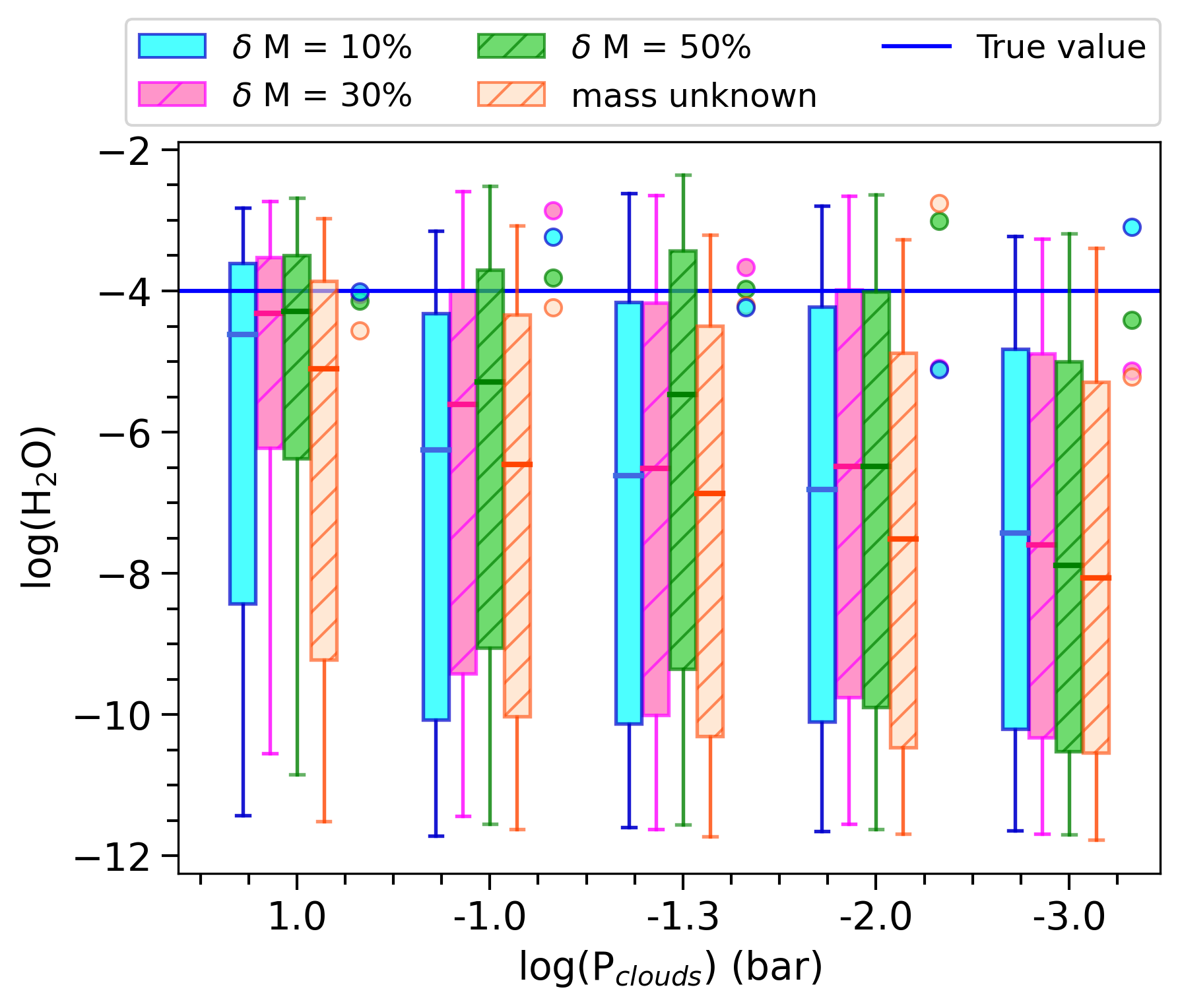}}
        \\
        \subfloat[CH$_4$ mixing ratio\label{fig:CH4_N2dom5.2Secondary}]{\includegraphics[scale=0.42]{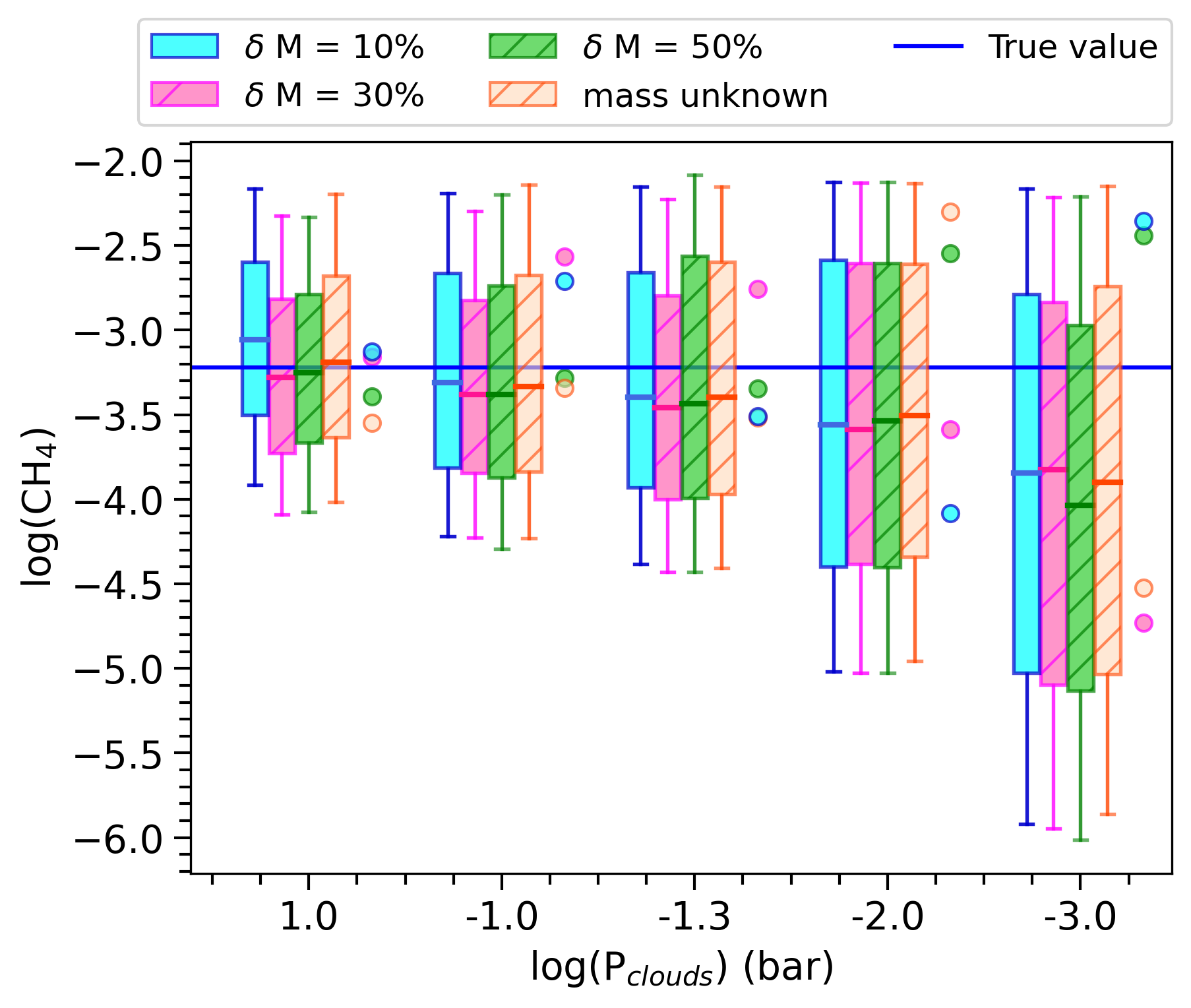}}
        \subfloat[N$_2$/He \label{fig:N2He_N2dom5.2Secondary}]{\includegraphics[scale=0.42]{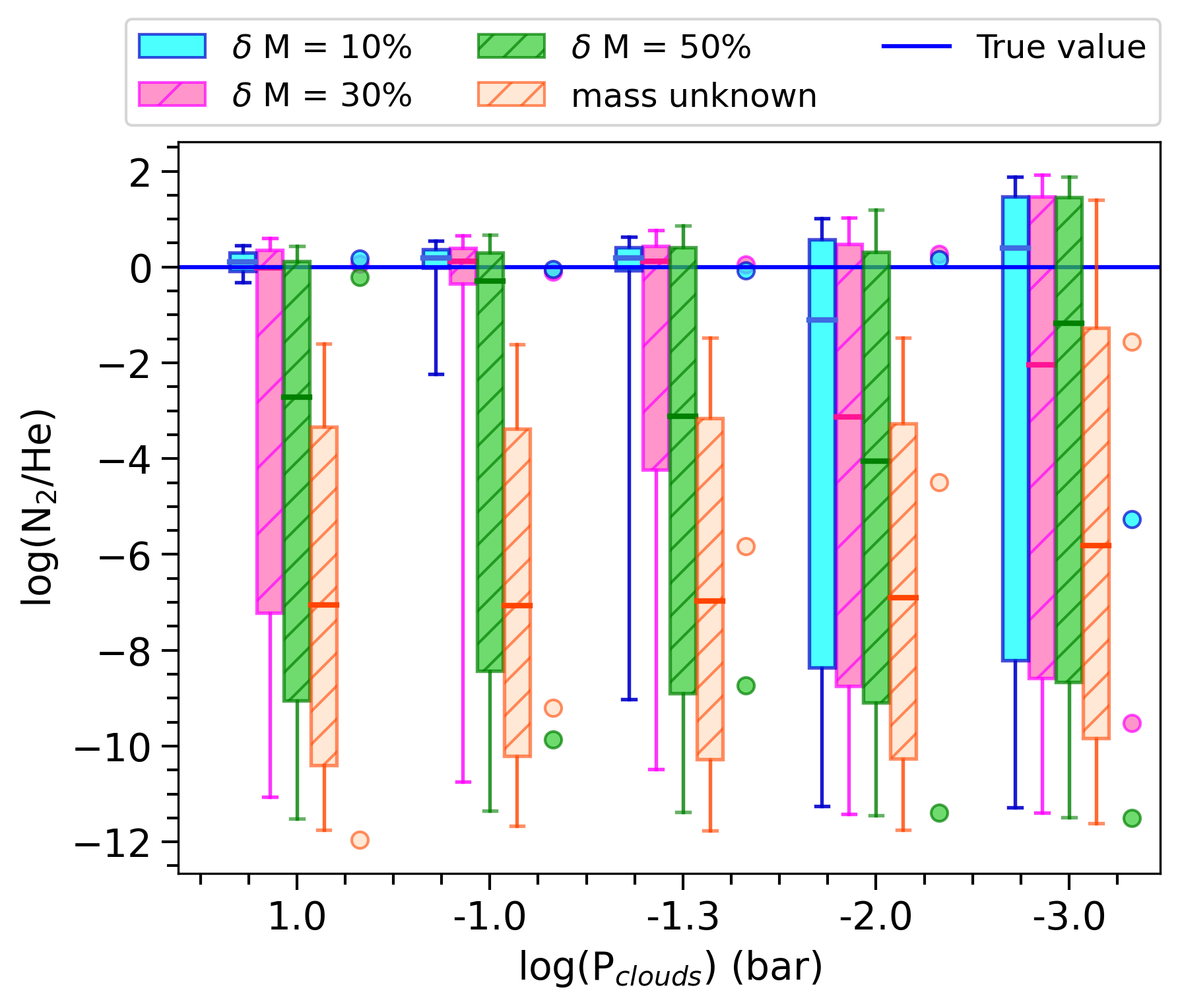}}
    }
    \caption{Impact of the mass uncertainties on the retrieval for different scenarios of cloudy secondary N$_2$-dominated atmospheres in the case of $\mu$=5.2. The different coloured boxes represent the different mass uncertainties. Blue line highlights the true value. The points alongside the boxes highlight the MAP  parameters obtained for each analysed case. The size of the box and the error bar represent the points within 1$\sigma$ and 2$\sigma$ of the median of the distribution (highlighted with
solid lines), respectively.}
    \label{fig:N25.2_Secondary}
 \end{figure*}
 

In Fig. \ref{fig:N25.2_Secondary}, we show the case of $\mu$ = 5.2, where we compare the results obtained for different cloud pressure measurements. In this scenario, the atmosphere is lighter and presents a better signal. 
From Fig. \ref{fig:N2He_N2dom5.2Secondary} we can see that with a mass uncertainty equal or lesser than 30\%, we significantly increase the accuracy and the precision on the retrieval of N$_2$/He --  particular in cases with higher cloud pressure; however, if we consider the MAP values, we increase the accuracy also in the worst scenario with lower cloud pressure. 
These results are reflected in the determination of the mean molecular weight. Indeed, from Fig. \ref{fig:mu_N2dom5.2Secondary}, we may note that with a mass uncertainty equal or lesser than 30\%, we are able to retrieve the mean molecular weight and (as we would expect) the width of the values distributions increase (and, consequently, the uncertainties associated to the median values as well) while decreasing the cloud pressure. It seems that the mass uncertainty does not impact the retrievals of the CH$_4$ mixing ratio. 
The H$_2$O mixing ratio (see Fig. \ref{fig:H2O_N2dom5.2Seycondar}) shows some discrepancies between the retrieved values and the true values, although the MAP values are compatible with the true values. However, these results do not show a correlation with the mass uncertainty, since they could be connected with the discrepancies shown in the clouds pressure retrieval (see Fig. \ref{fig:Clouds_N2dom5.2Secondary}).

\begin{figure*}[h!]
 \centering
        {\subfloat[Radius\label{fig:Radius_N2dom11Secondary}]{\includegraphics[scale=0.42]{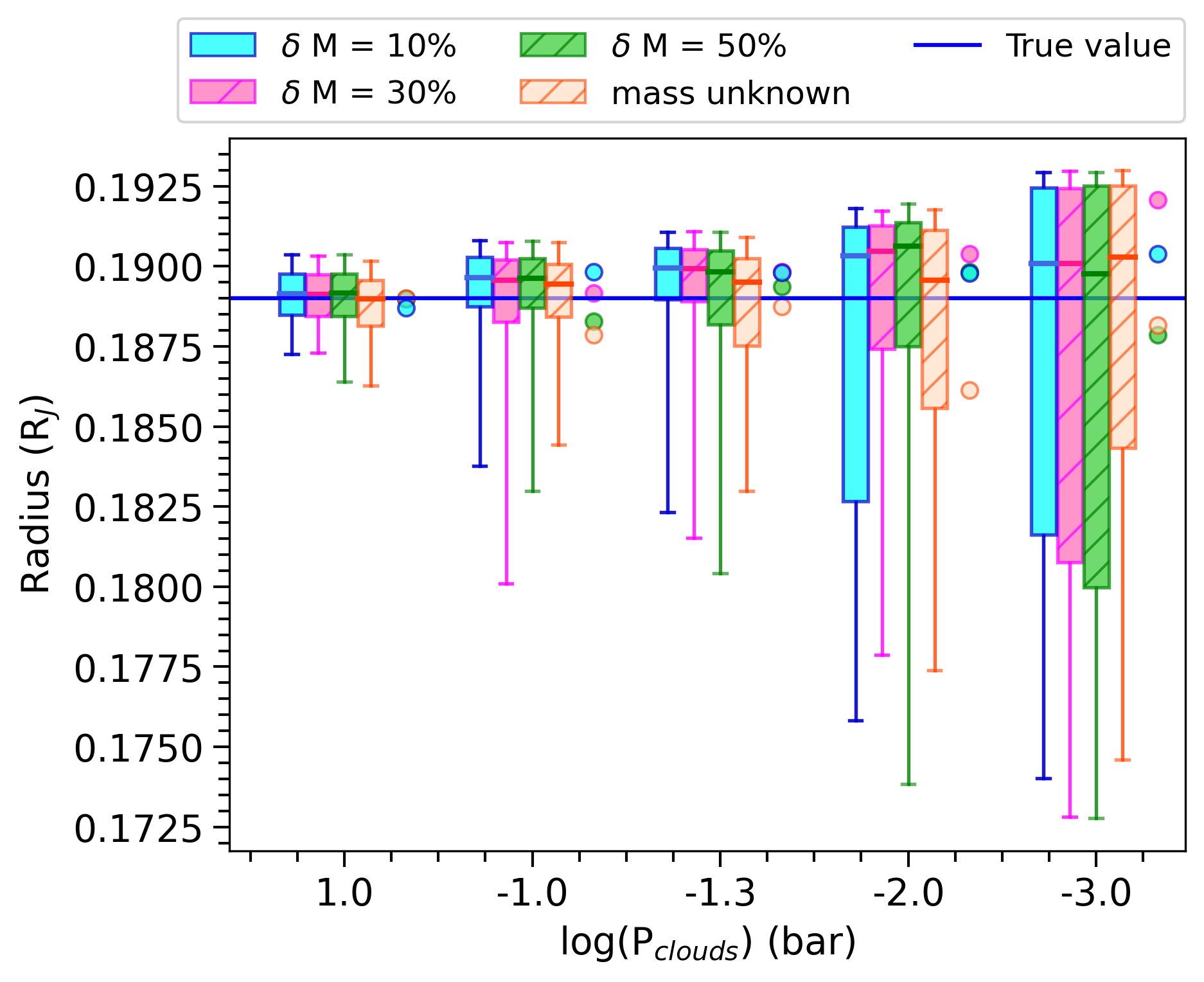}}
        \subfloat[Temperature\label{fig:Temp_N2dom11Secondary}]{\includegraphics[scale=0.42]{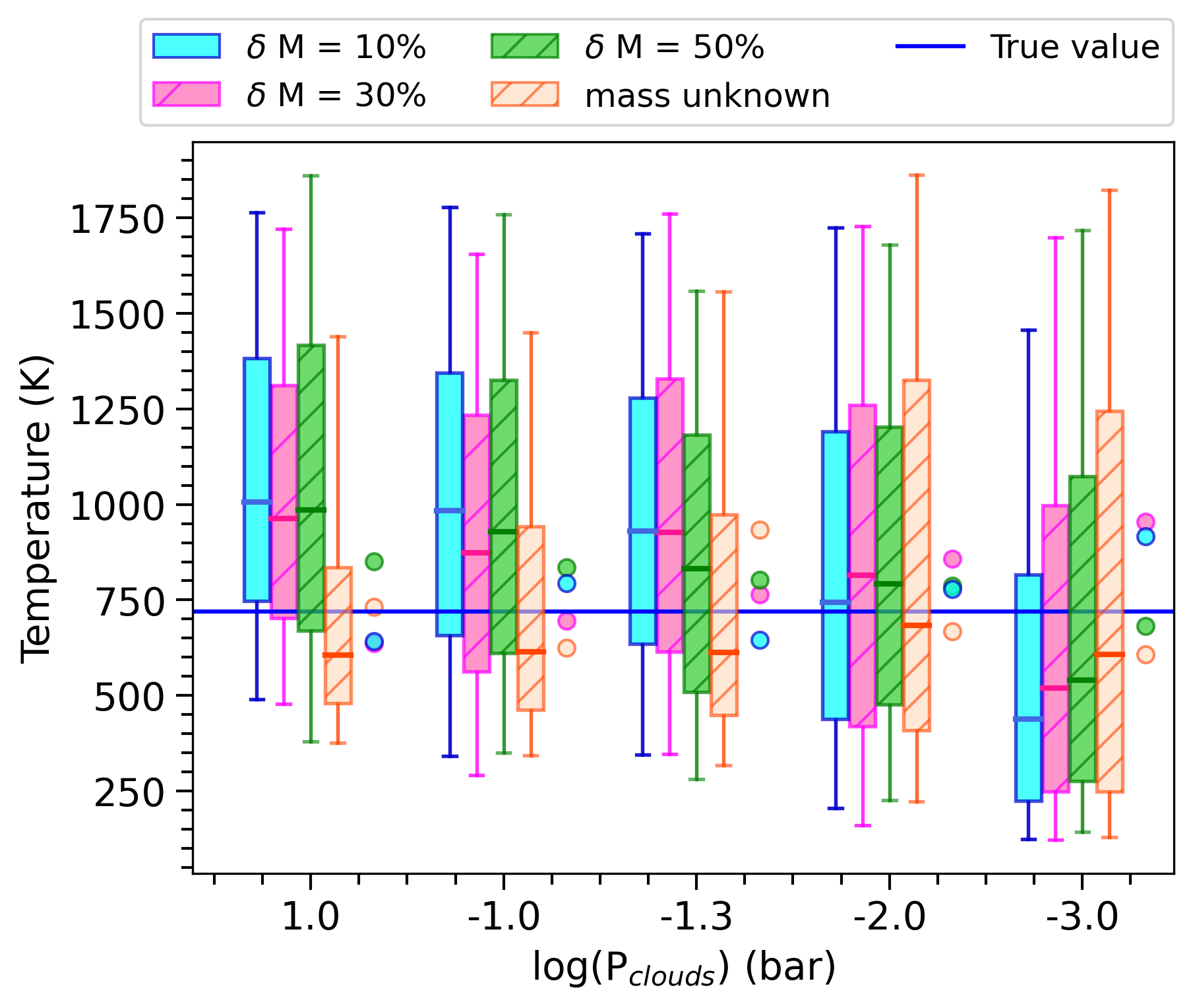}}
        \subfloat[Mass\label{fig:Mass_N2dom11Secondary}]{\includegraphics[scale=0.42]{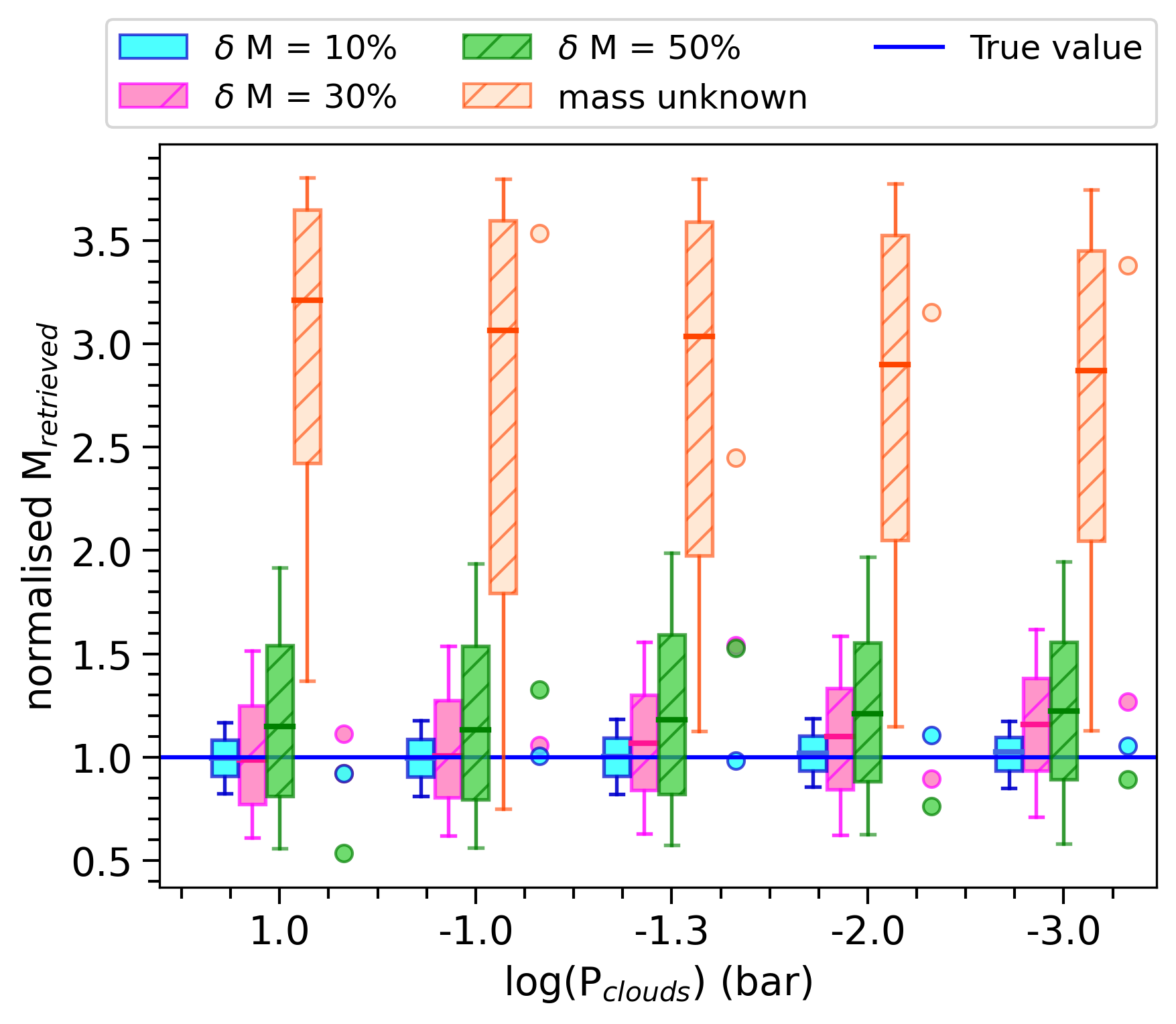}}
        \\
        \subfloat[Clouds\label{fig:Clouds_N2dom11Secondary}]{\includegraphics[scale=0.42]{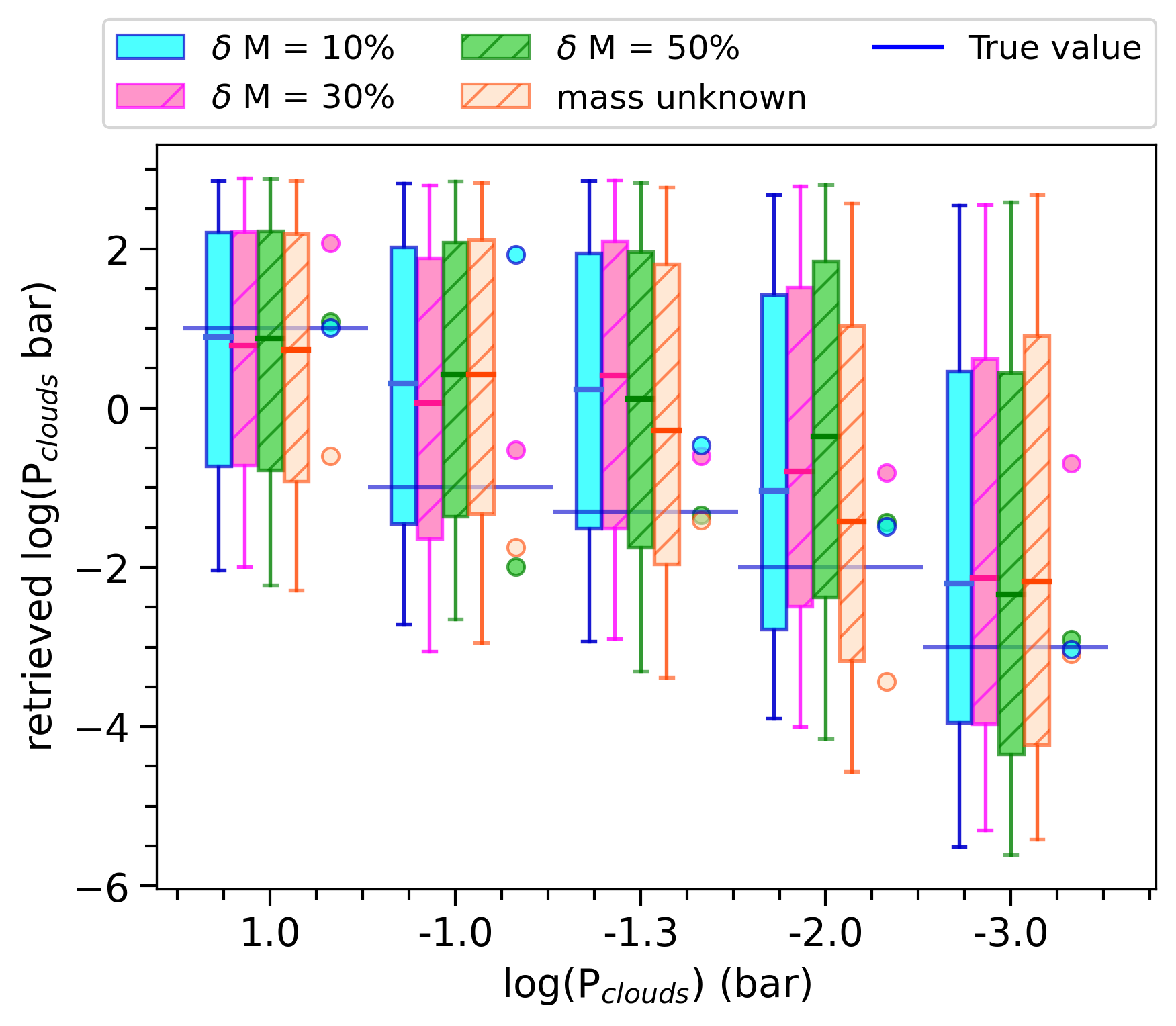}}
        \subfloat[$\mu$\label{fig:mu_N2dom11Secondary}]{\includegraphics[scale=0.42]{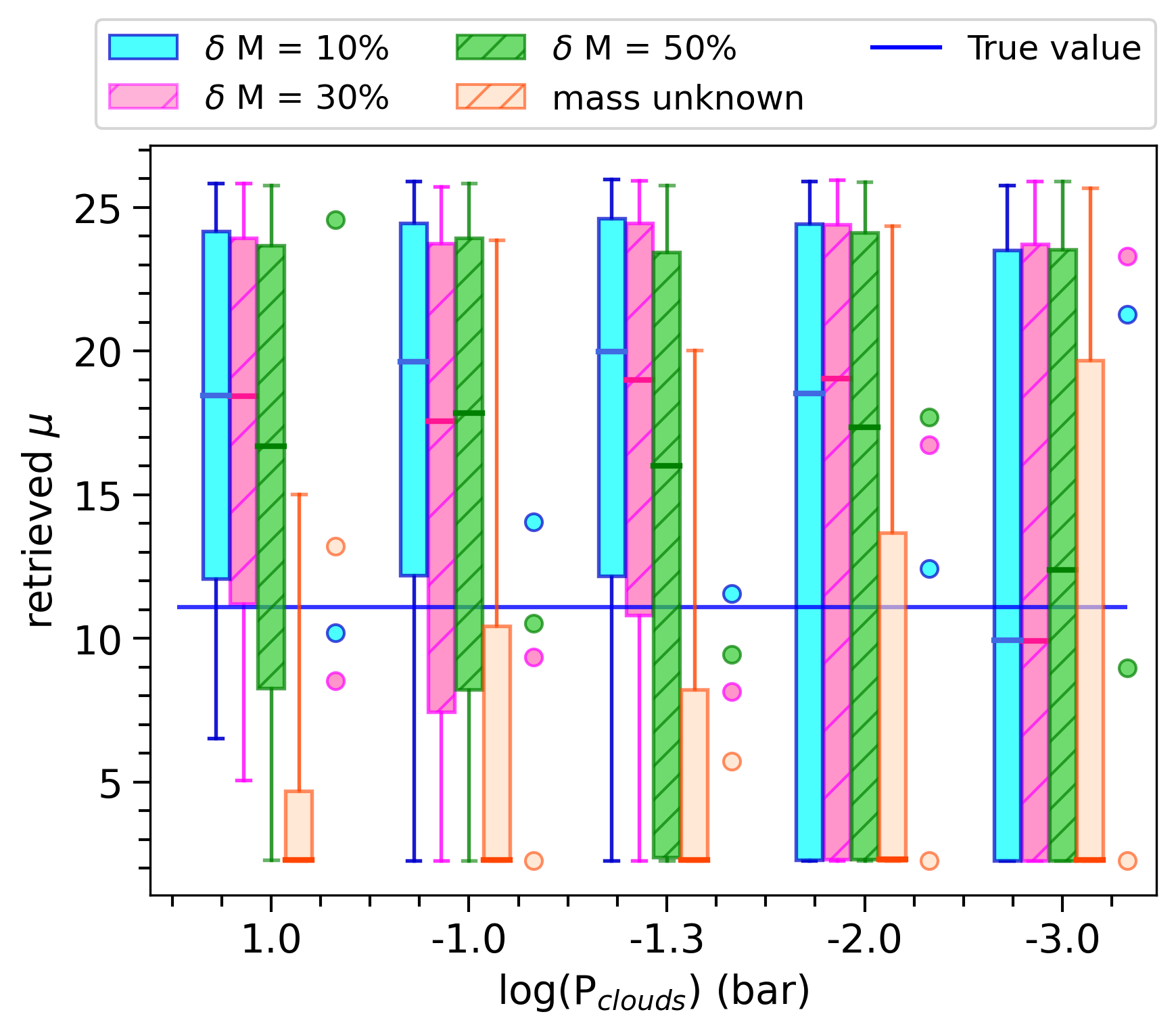}} 
        \subfloat[H$_2$O mixing ratio \label{fig:H2O_N2dom11Seycondar}]{\includegraphics[scale=0.42]{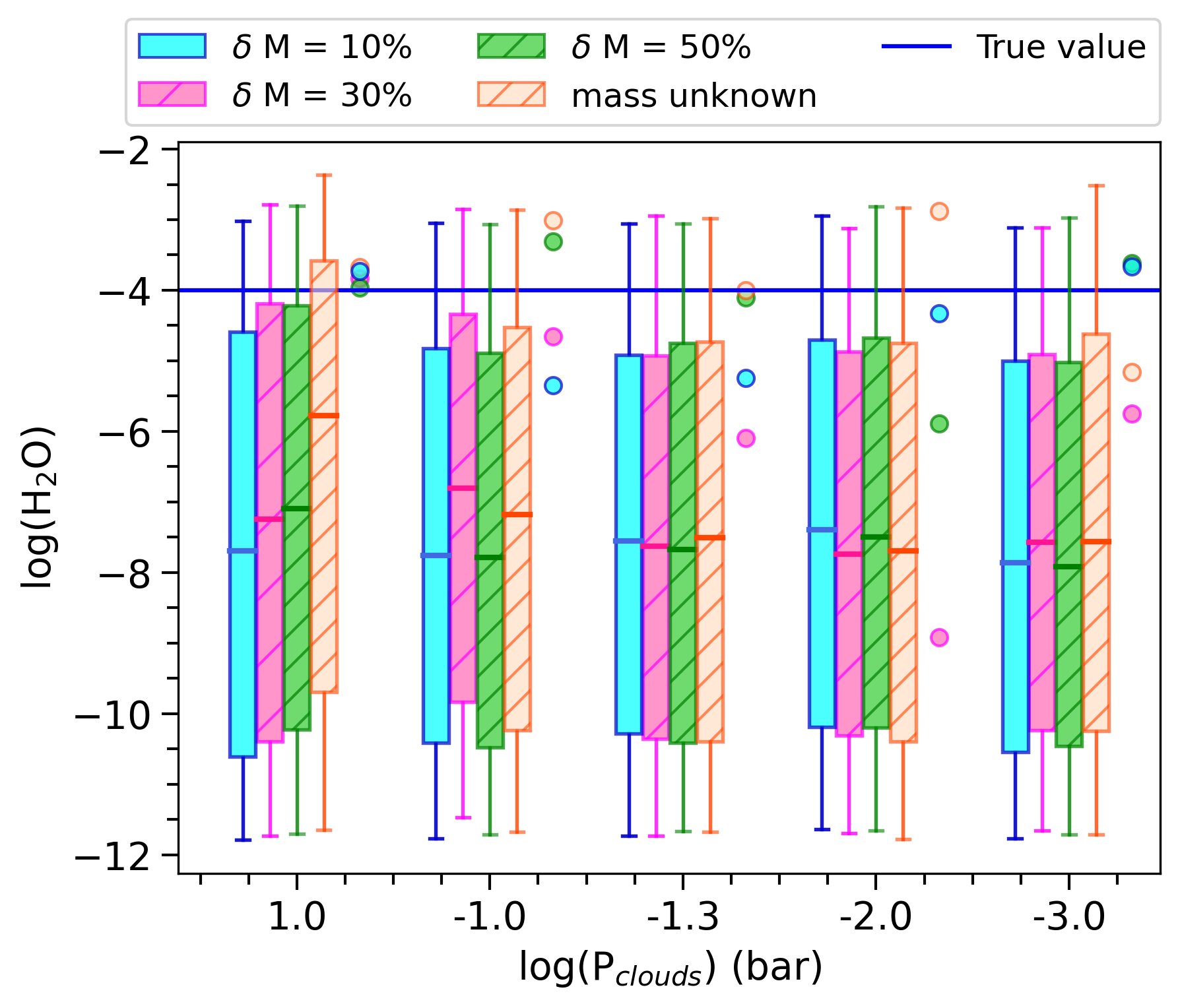}}
        \\
        \subfloat[CH$_4$ mixing ratio\label{fig:CH4_N2dom11Secondary}]{\includegraphics[scale=0.42]{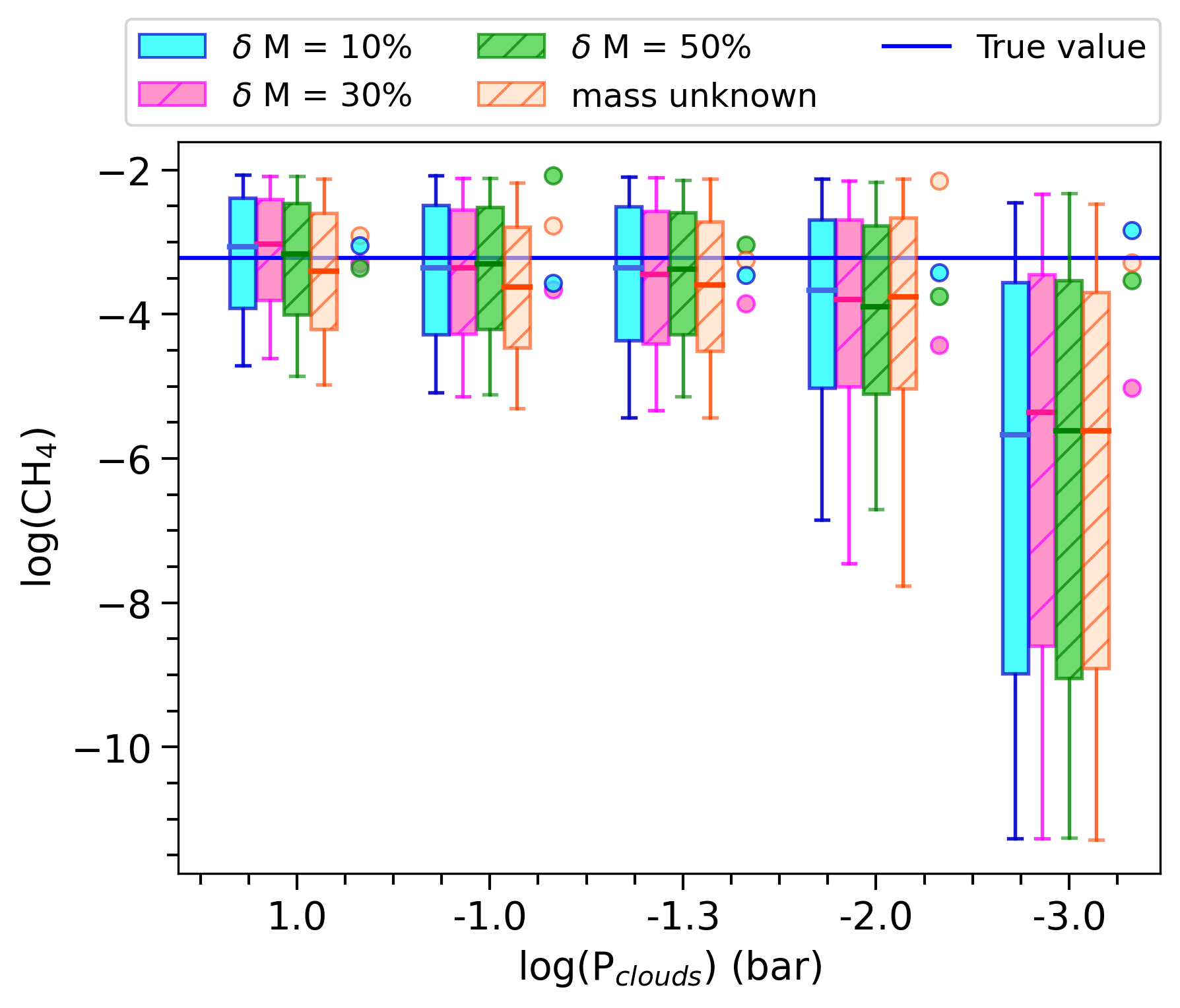}}
        \subfloat[N$_2$/He \label{fig:N2He_N2dom11Secondary}]{\includegraphics[scale=0.42]{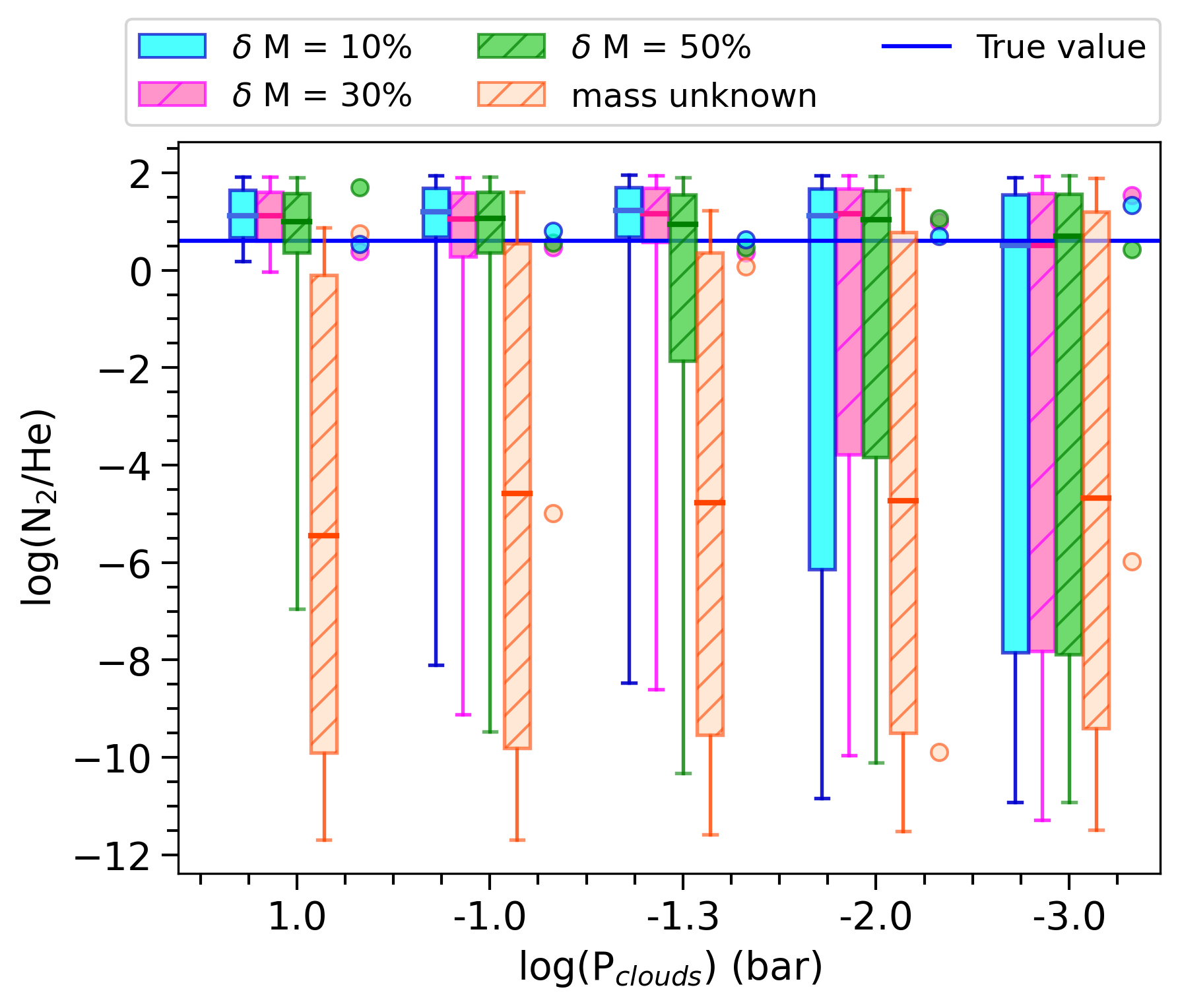}}
    }
    \caption{Results obtained from the retrieval of N$_2$-dominated atmosphere in the case of $\mu$=11.1. Differently coloured boxes represent the different mass uncertainties. Blue line highlights the true value. The points alongside the boxes highlights the MAP parameters obtained for each analysed case. The size of the box and the error bar represent the points within 1$\sigma$ and 2$\sigma$ of the median of the distribution (highlighted with
solid lines), respectively.}
    \label{fig:N211_Secondary}
 \end{figure*}

In Fig. \ref{fig:N211_Secondary},  we consider the heaviest scenario ($\mu$ = 11.1). In this case, we are not able to constrain the mean molecular weight. From Fig. \ref{fig:mu_N2dom11Secondary}, we can see that the retrieved $\mu$ tends to be larger than the true value, but these results are not correlated with the mass uncertainties. Additionally, (and as expected) the retrieved $\mu$ present larger uncertainties when the cloud pressure decreases.

In Fig. \ref{fig:Clouds_N2dom11Secondary}, it is suggested that the mass uncertainties do not impact the retrieved cloud pressure. Indeed, we do not see significant discrepancies in the retrieved distribution with respect to the mass uncertainty. However, we do note a better compatibility between the true values and the MAP values when we consider mass uncertainties of 10\%. 

With regard to the atmospheric parameters, the CH$_4$ mixing ratio is also adequately retrieved when the cloud pressure get closer to 10$^{-3}$ bar; whereas we are not able to accurately retrieve the H$_2$O mixing ratio, particularly for cloud pressure lower than 10$^{-2}$ bar. Here, additional observations are needed to increase the S/N and to constrain the mean molecular weight.

\paragraph{\textbf{H$_2$O and CO-dominated Atmosphere}}
\begin{figure*}[h!]
 \centering
        {\subfloat[Radius\label{fig:Radius_H2Odom5.2Secondary}]{\includegraphics[scale=0.42]{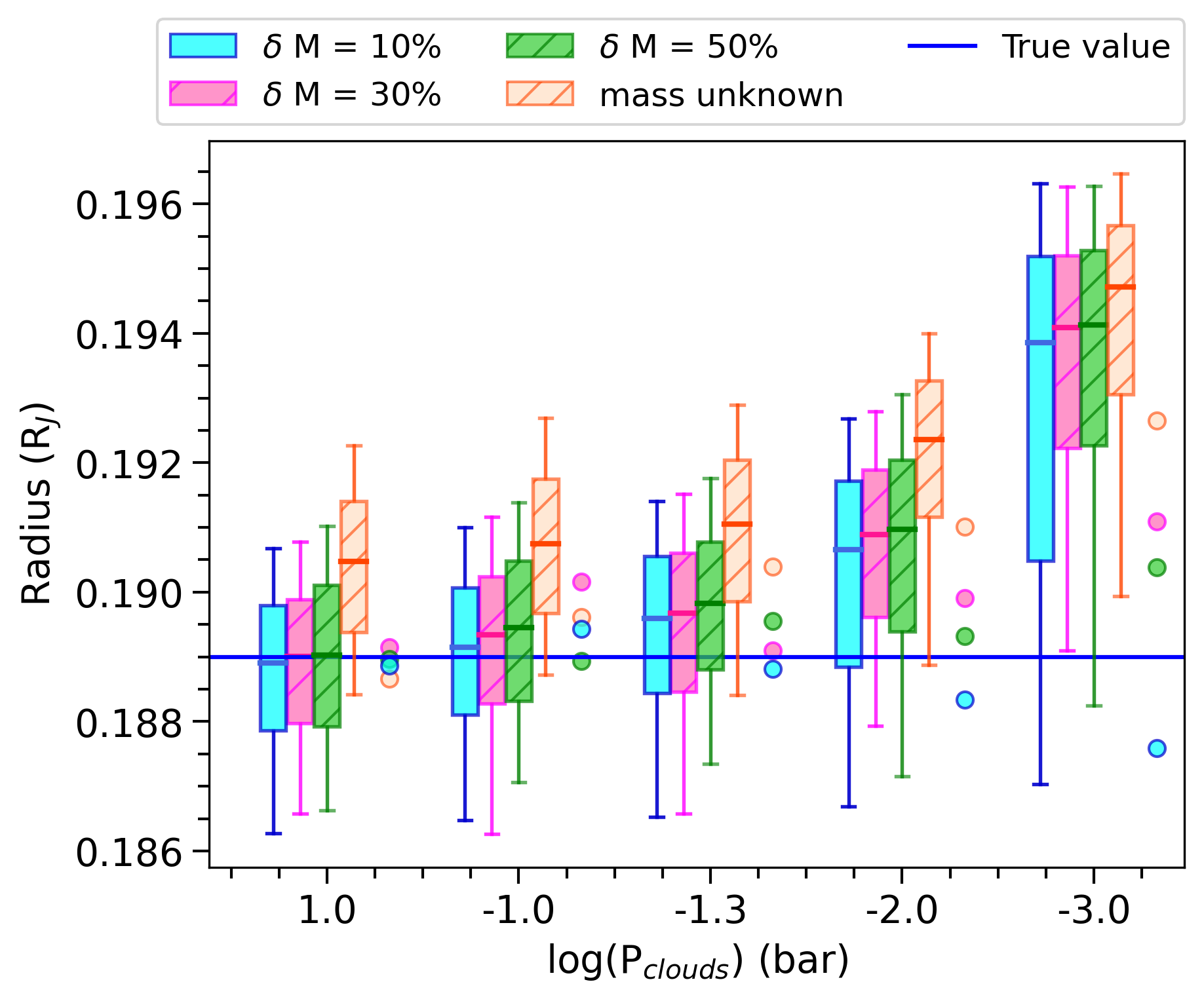}}
        \subfloat[Temperature\label{fig:Temp_H2Odom5.2Secondary}]{\includegraphics[scale=0.42]{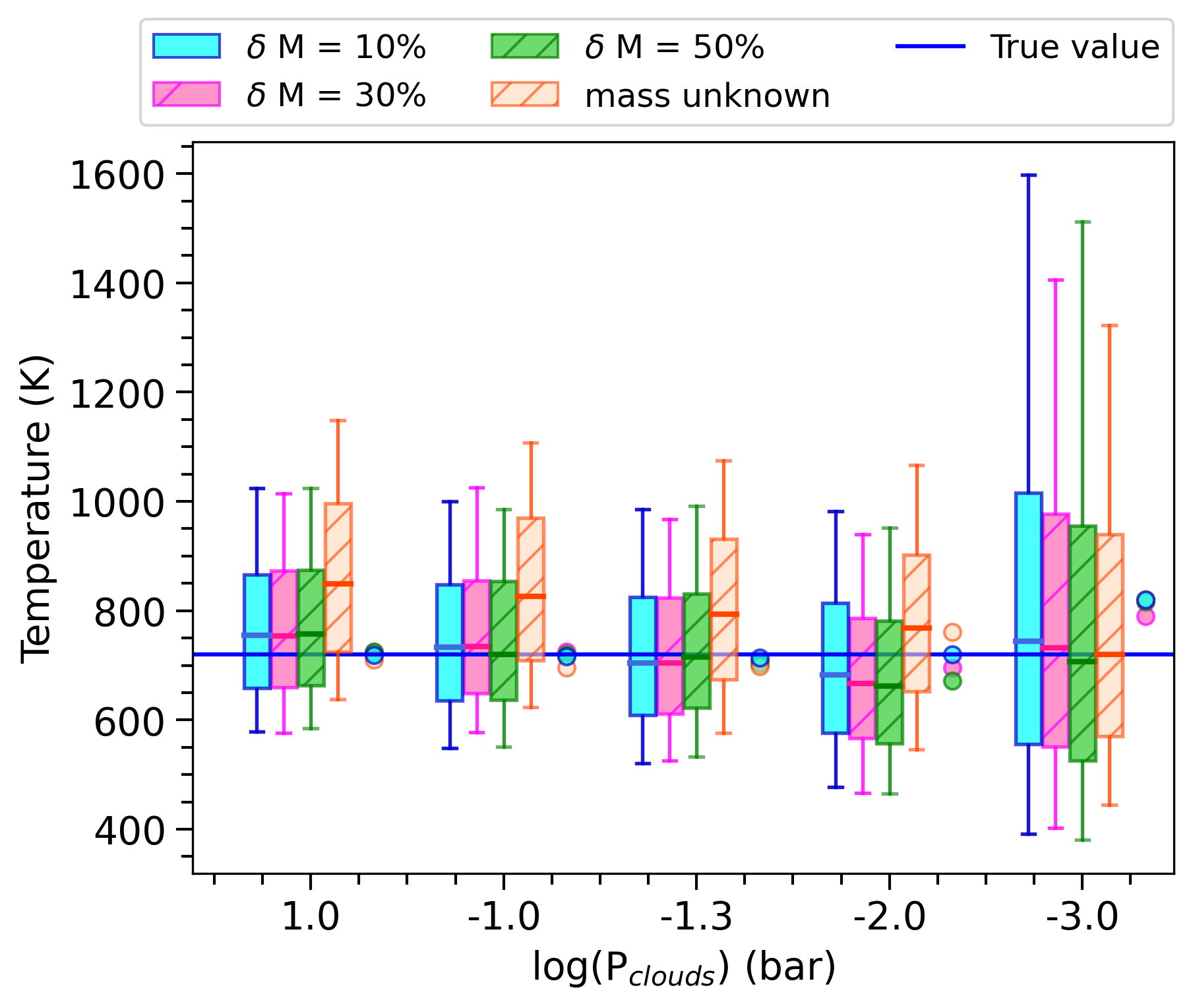}}
        \subfloat[Mass\label{fig:Mass_H2Odom5.2Secondary}]{\includegraphics[scale=0.42]{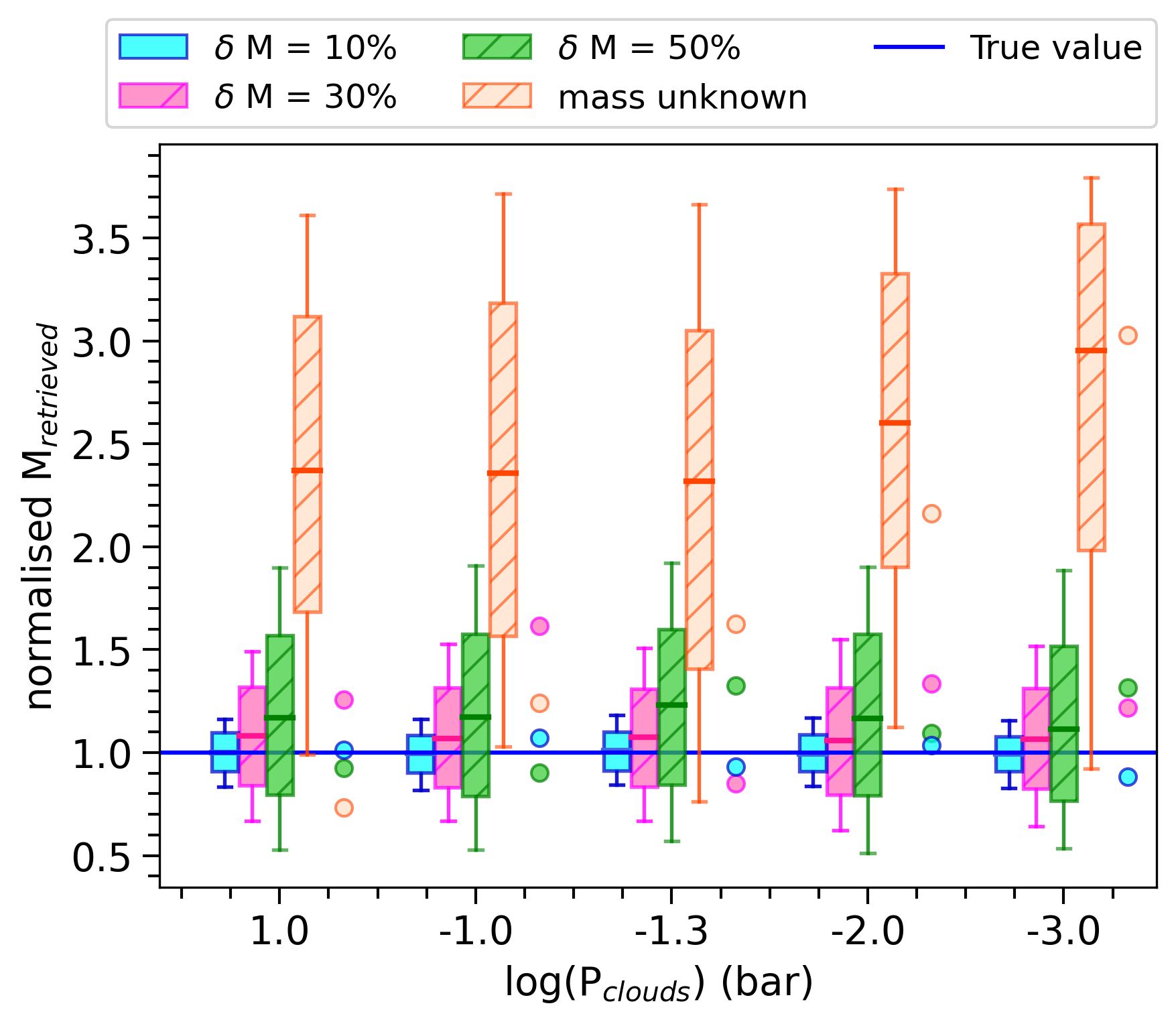}} 
        \\
        \subfloat[Clouds\label{fig:Clouds_H2Odom5.2Secondary}]{\includegraphics[scale=0.42]{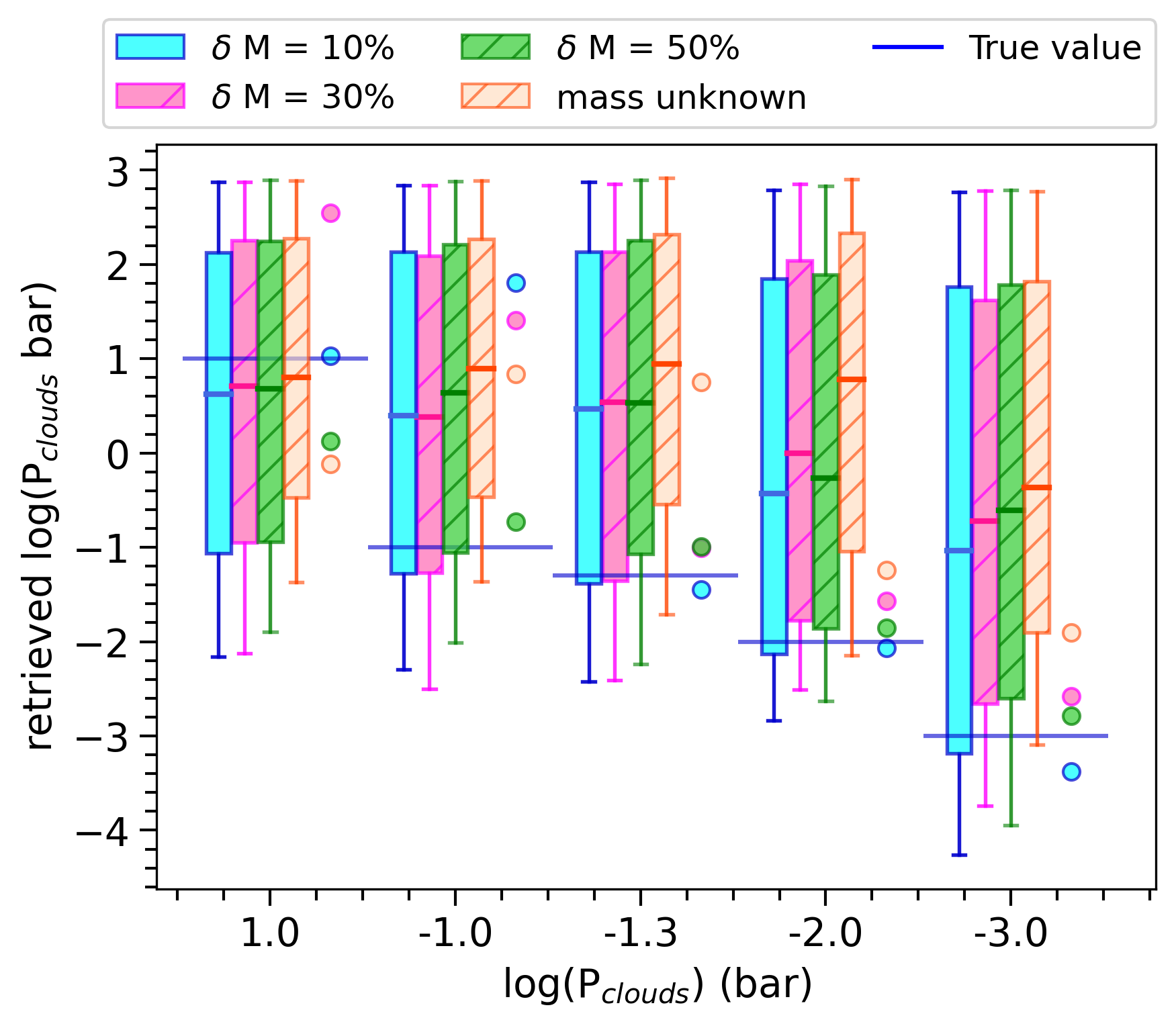}}
        \subfloat[$\mu$\label{fig:mu_H2Odom5.2Secondary}]{\includegraphics[scale=0.42]{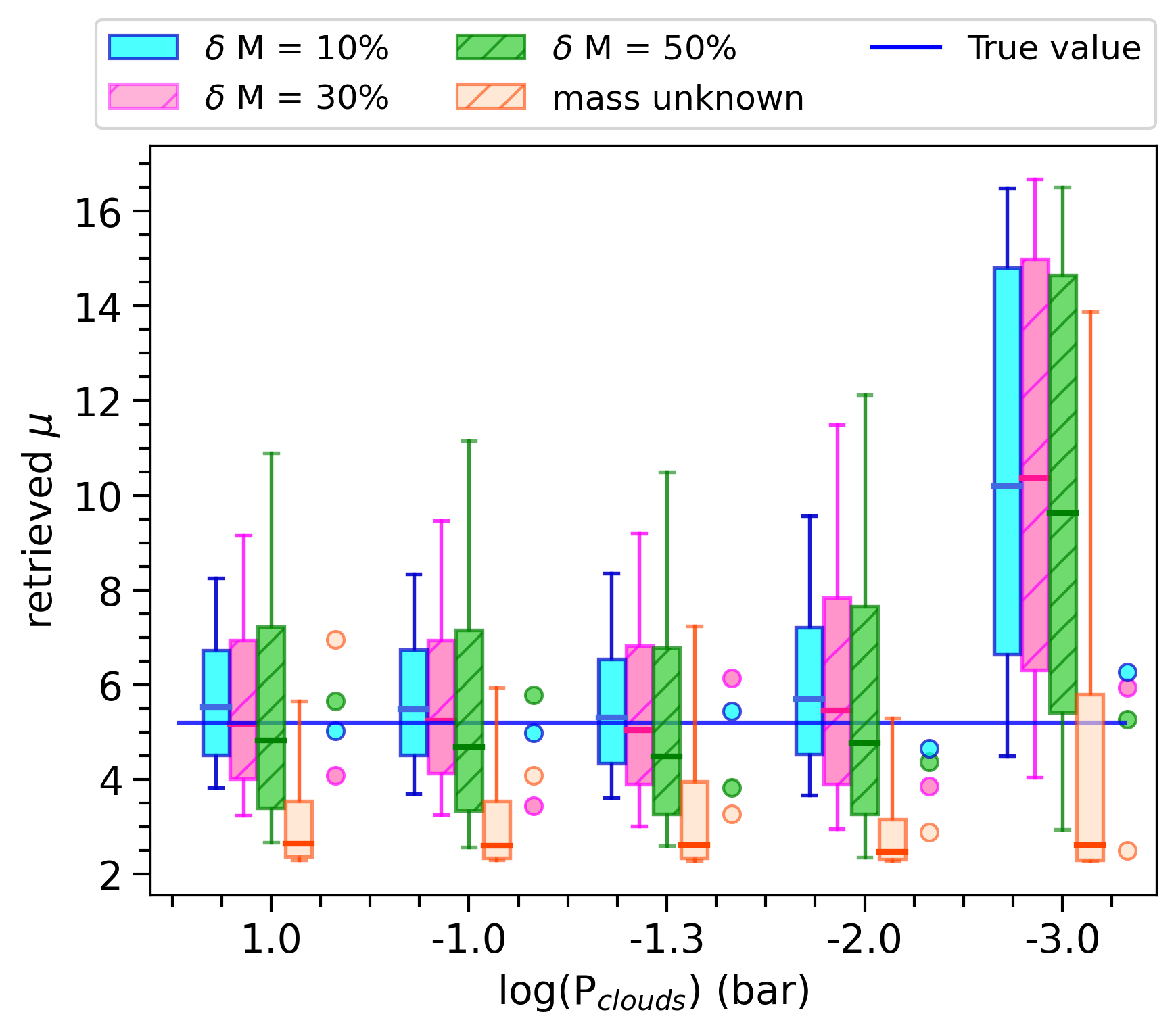}} 
        \subfloat[CH$_4$ mixing ratio\label{fig:CH4_H2Odom5.2Secondary}]{\includegraphics[scale=0.42]{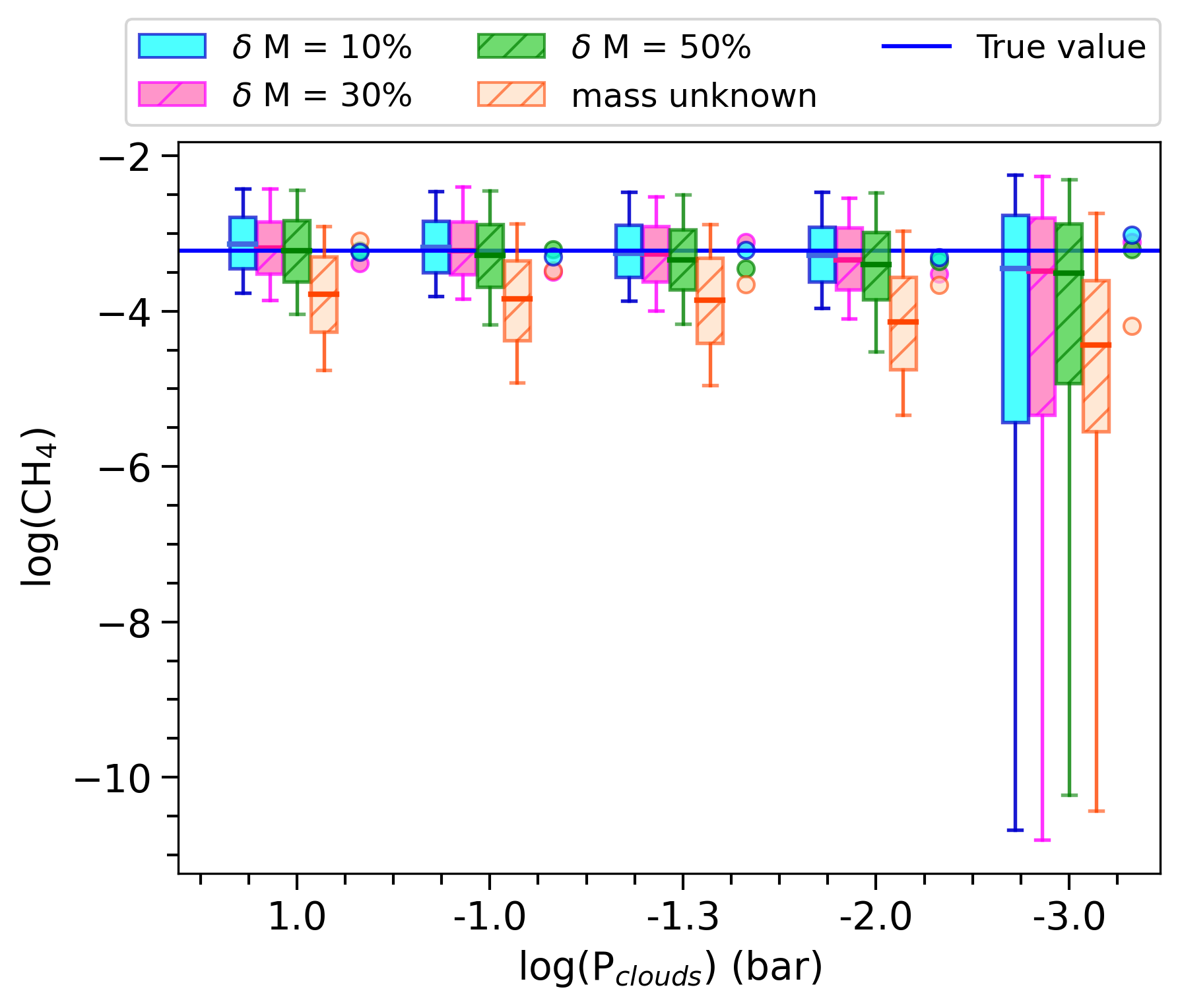}}
        \\
        \subfloat[H$_2$O/He \label{fig:H2OHe_H2Odom5.2Secondary}]{\includegraphics[scale=0.42]{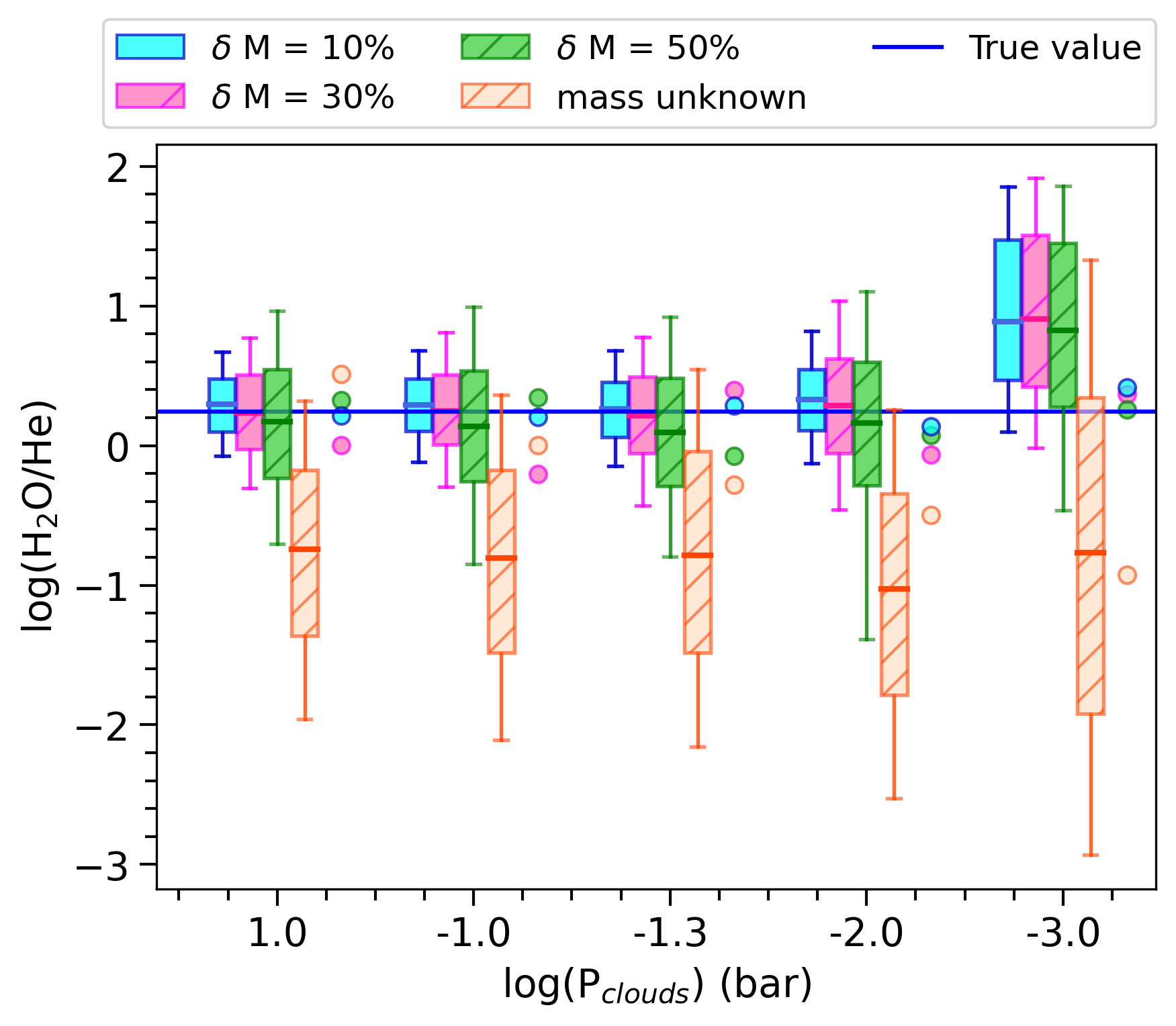}}
    }
    \caption{Retrieval results obtained from different scenarios of cloudy secondary H$_2$O-dominated atmosphere in the case of $\mu$=5.2. Differently coloured boxes represent the different mass uncertainties. Blue line highlights the true value. The points alongside the boxes highlight the MAP  parameters obtained for each analysed case. The size of the box and the error bar represent the points within 1$\sigma$ and 2$\sigma$ of the median of the distribution (highlighted with
solid lines), respectively.}
    \label{fig:H2O5.2_Secondary}
 \end{figure*}
 
In Fig. \ref{fig:H2O5.2_Secondary}, we show the results obtained from the retrieval of H$_2$O-dominated secondary atmosphere ($\mu$ = 5.2). 
In this case, the mass uncertainties do not significantly impact the retrieval.
Here, we are able to constrain the H$_2$O/He  with a slightly increased accuracy for lower mass uncertainties. 
Also, the cloud pressure and the mean molecular weight are adequately retrieved, even in cases with lower cloud pressure as well. In the worst  scenarios considered, when the clouds pressure get closer to 10$^{-3}$ bar, the retrieved $\mu$ is within 2$\sigma$ of the true value, while the MAP value its closer to the true value.
 
Additionally, we performed the retrieval for our worst-case scenario, $\mu$ = 11.1 (see Fig. \ref{fig:H2O11_Secondary} in Appendix \ref{app:figure}). From this test, we can confirm that for this target we need more observations in order to achieve an adequate S/N. The lower S/N values prevent us from correctly retrieve the mean molecular weight, which for all cases is higher than the true value. The accuracy in the retrieved $\mu$, namely, within 2$\sigma$ of the true value, does not depends on the the mass uncertainties. The uncertainties of the CH$_4$ increase by several orders of magnitude with respect to the case $\mu$ = 5.2. This is because, in the case of $\mu$ = 11.1, the water features tend to dominate the methane features present in the redder region of the spectrum, leading to greater uncertainty in the retrieval of CH$_4$. 
 
An analogous behaviour is seen for the CO-dominated atmosphere (see Fig. \ref{fig:CO5.2_Secondary} and Fig. \ref{fig:CO11_Secondary} in Appendix \ref{app:figure}). In particular, in this scenario we note a slight trend with the mass uncertainties in the retrieved CH$_4$ mixing ratio (see Fig. \ref{fig:CH4_COdom5.2Secondary}). 
This increased accuracy in the retrieved CH$_4$ mixing ratio could be linked to the presence of a prominent CO feature in the redder part of the spectrum that allows us to better describe and constrain the CH$_4$ component.

\begin{figure*}[h!]
 \centering
        {\subfloat[Radius\label{fig:Radius_COdom5.2Secondary}]{\includegraphics[scale=0.42]{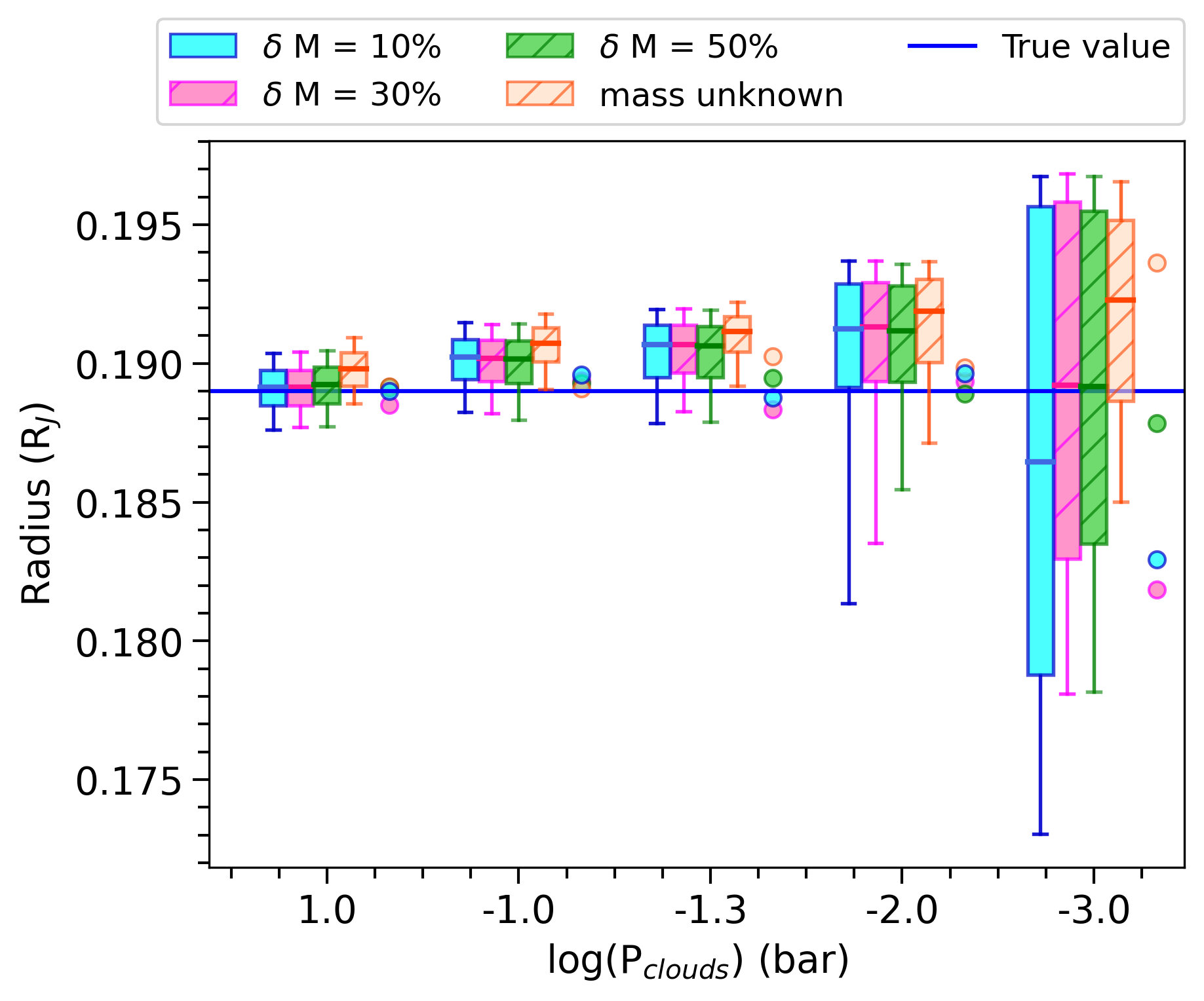}}
        \subfloat[Temperature\label{fig:Temp_COdom5.2Secondary}]{\includegraphics[scale=0.42]{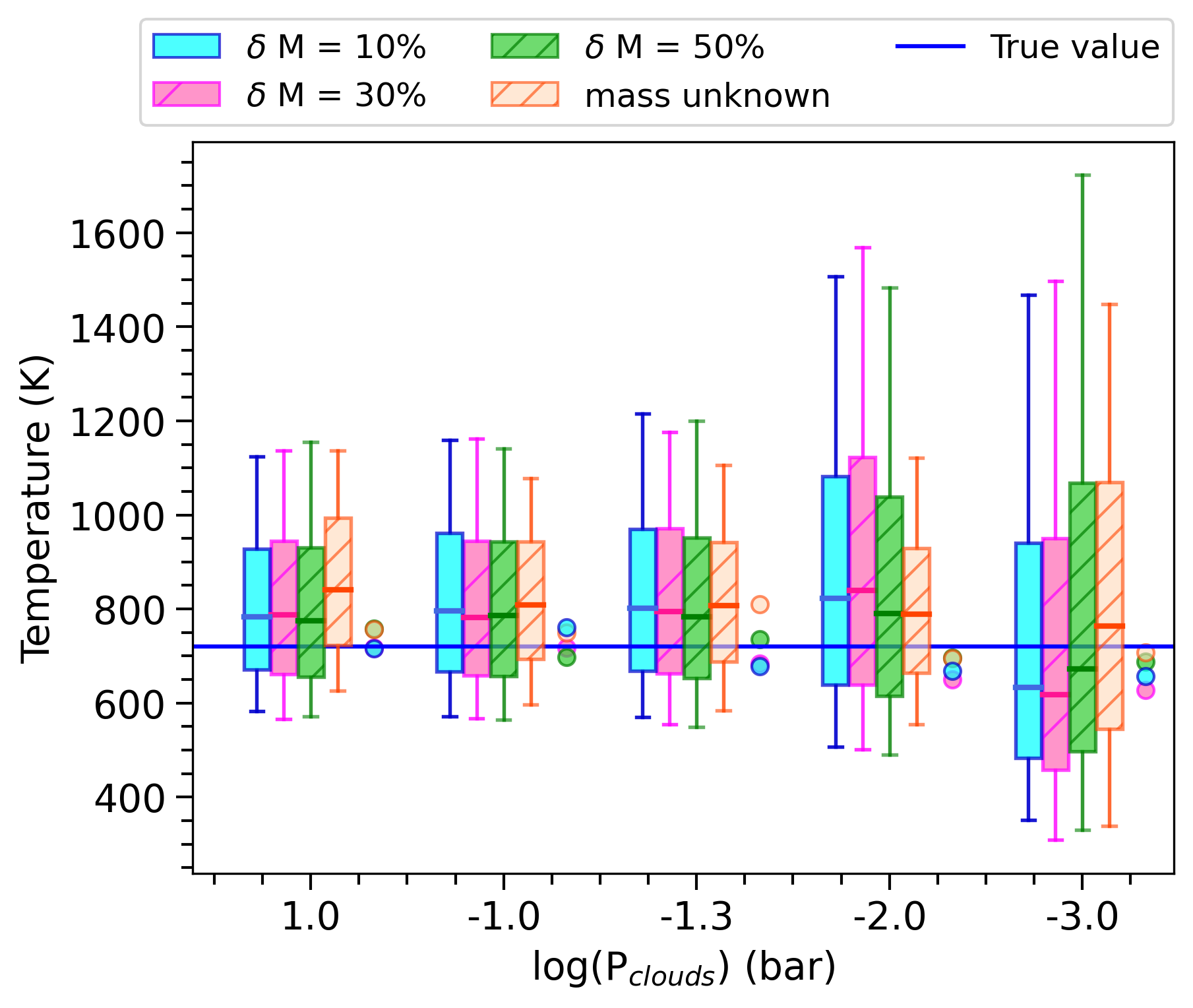}}
        \subfloat[Mass\label{fig:Mass_COdom5.2Secondary}]{\includegraphics[scale=0.42]{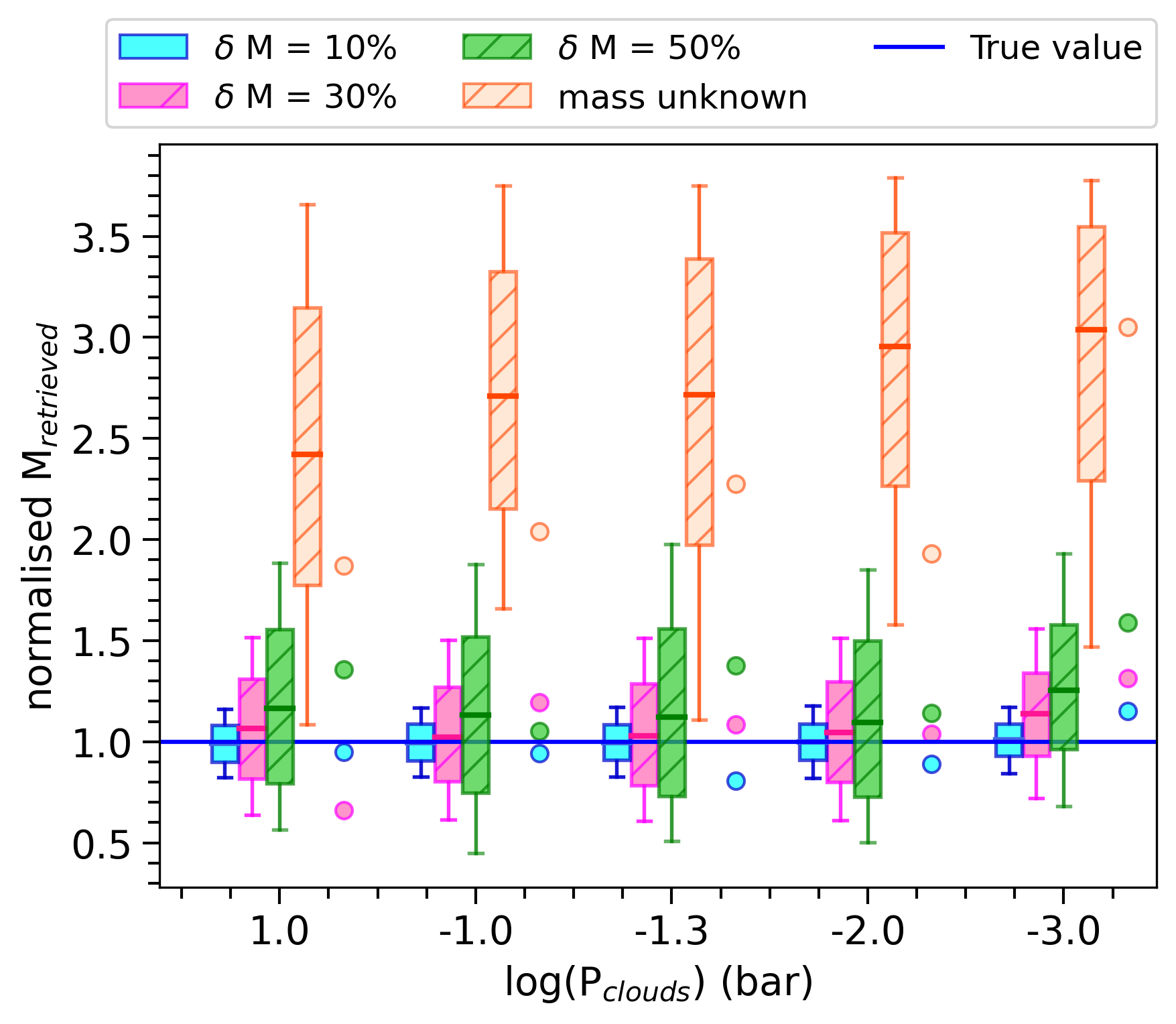}} 
        \\
        \subfloat[Clouds\label{fig:Clouds_COdom5.2Secondary}]{\includegraphics[scale=0.42]{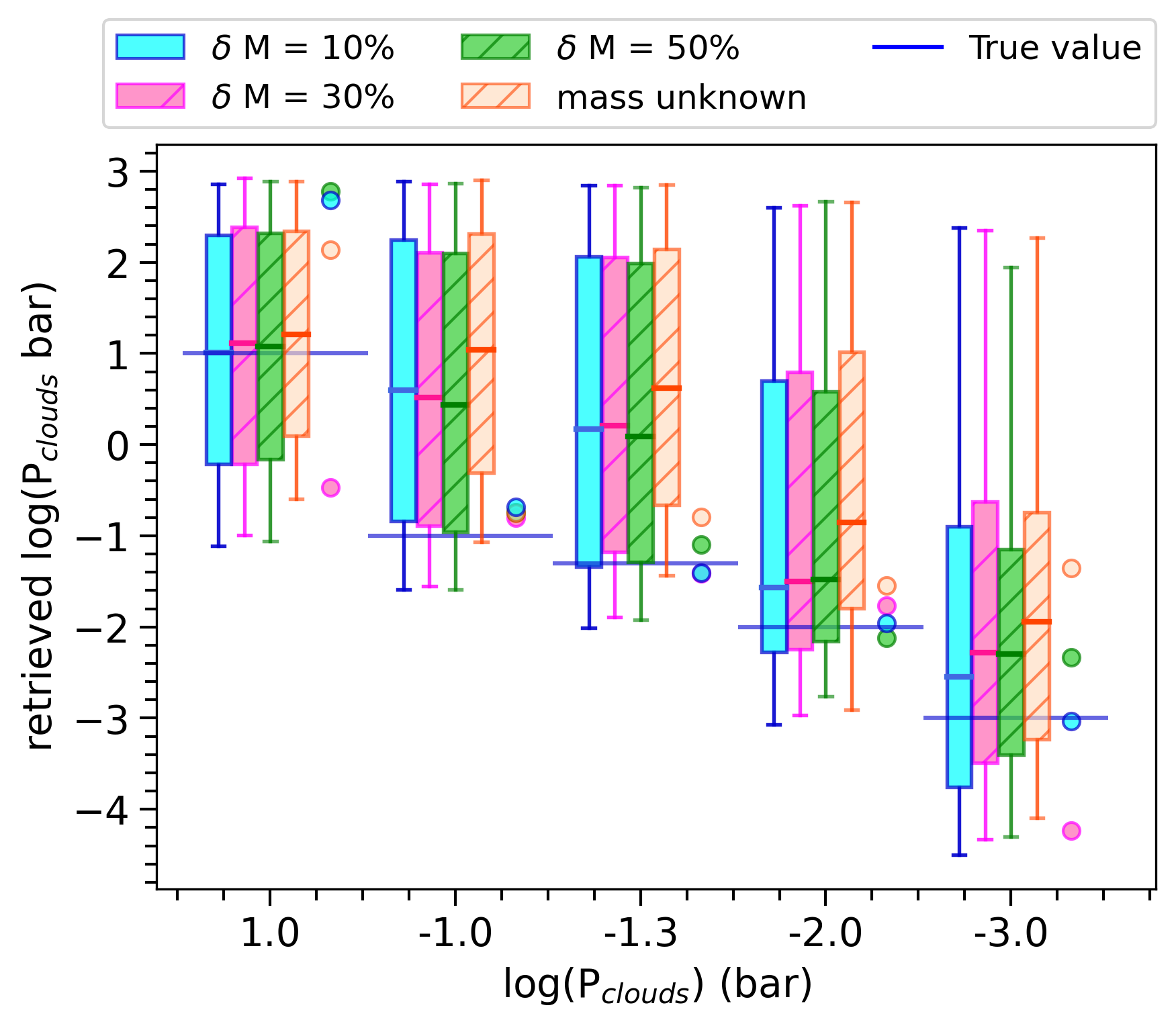}}
        \subfloat[$\mu$\label{fig:mu_COdom5.2Secondary}]{\includegraphics[scale=0.42]{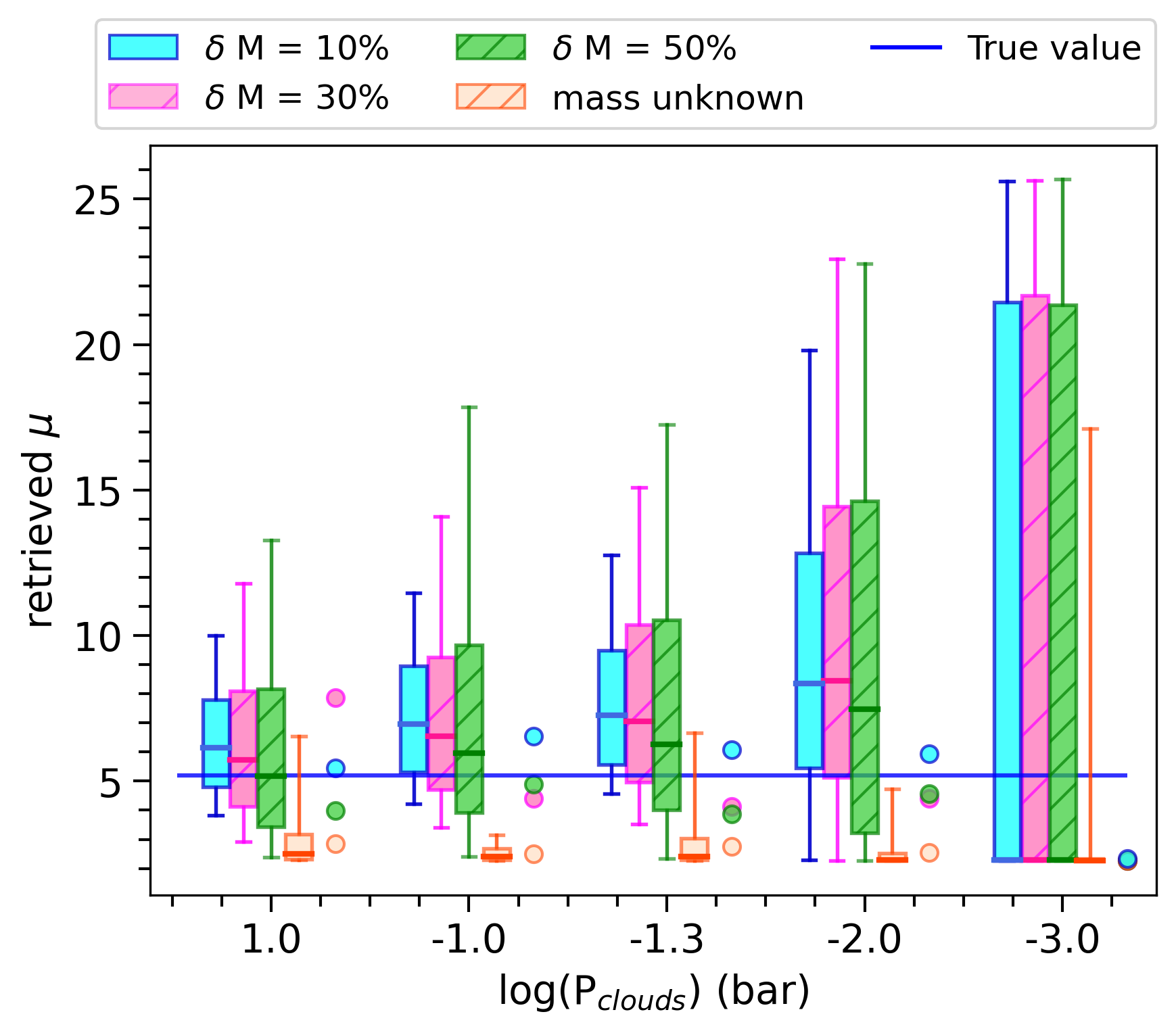}} 
        \subfloat[H$_2$O mixing ratio\label{fig:H2O_COdom5.2Secondary}]{\includegraphics[scale=0.42]{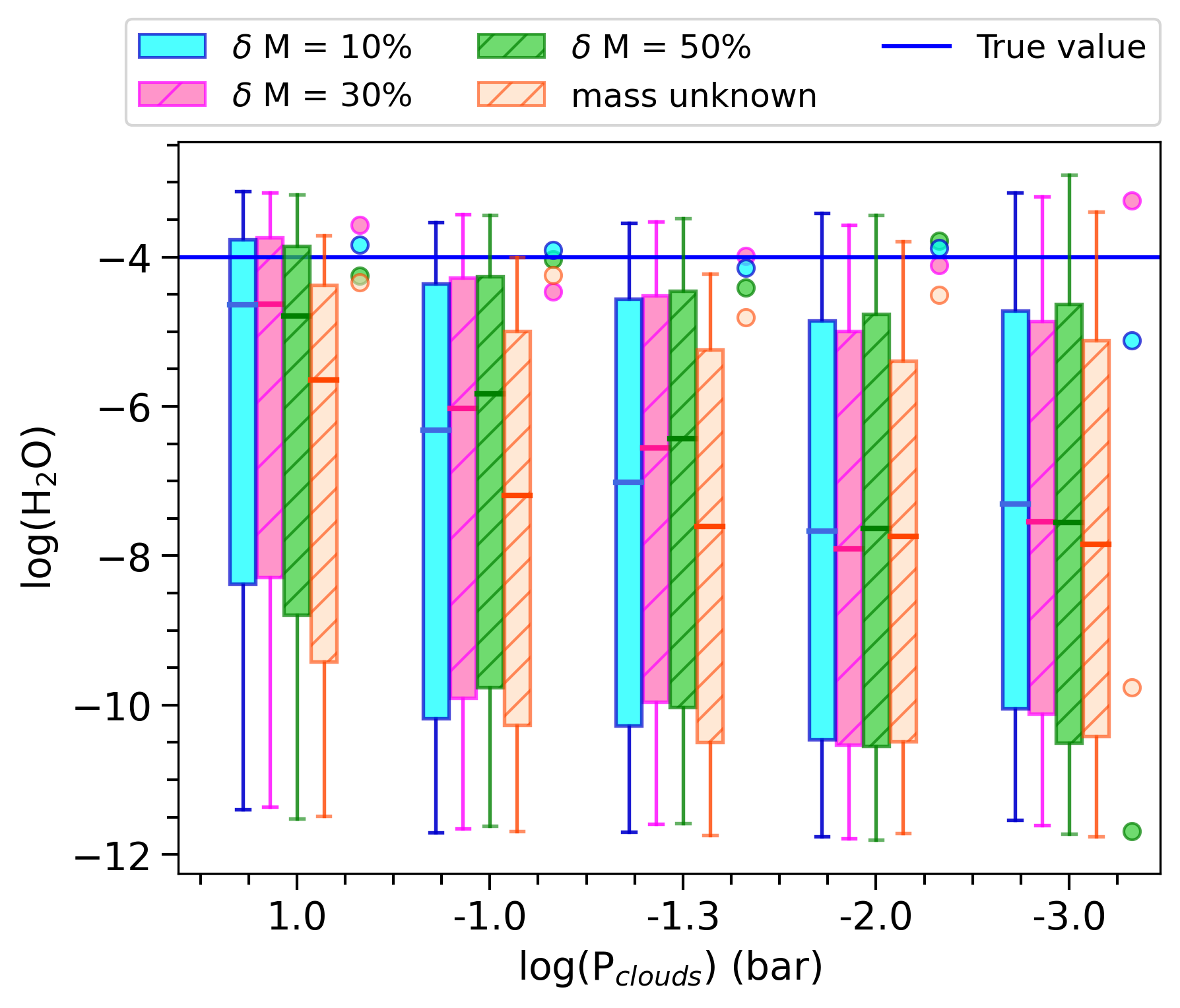}}
        \\
        \subfloat[CH$_4$ mixing ratio\label{fig:CH4_COdom5.2Secondary}]{\includegraphics[scale=0.42]{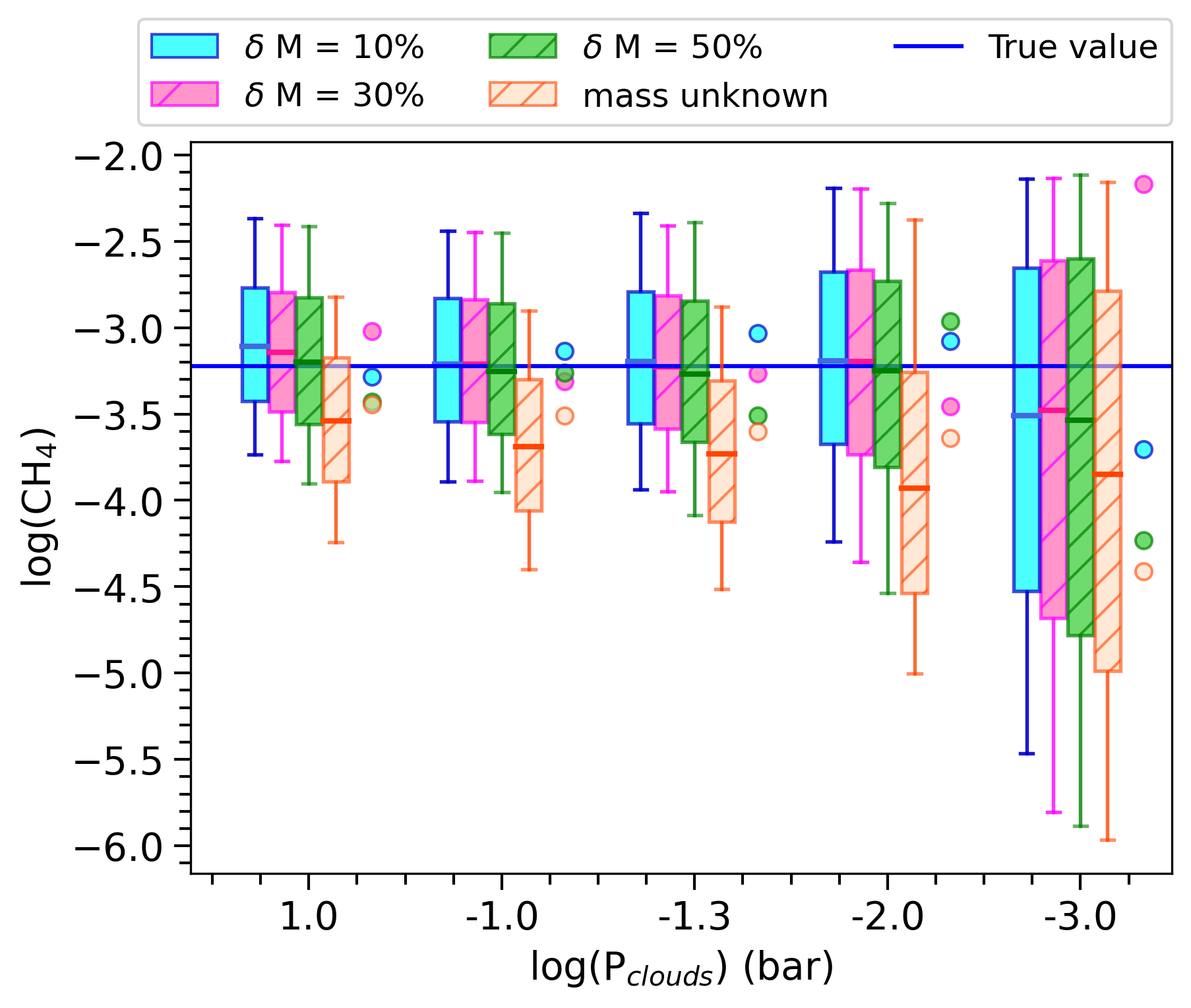}}
        \subfloat[CO/He \label{fig:COHe_COdom5.2Secondary}]{\includegraphics[scale=0.42]{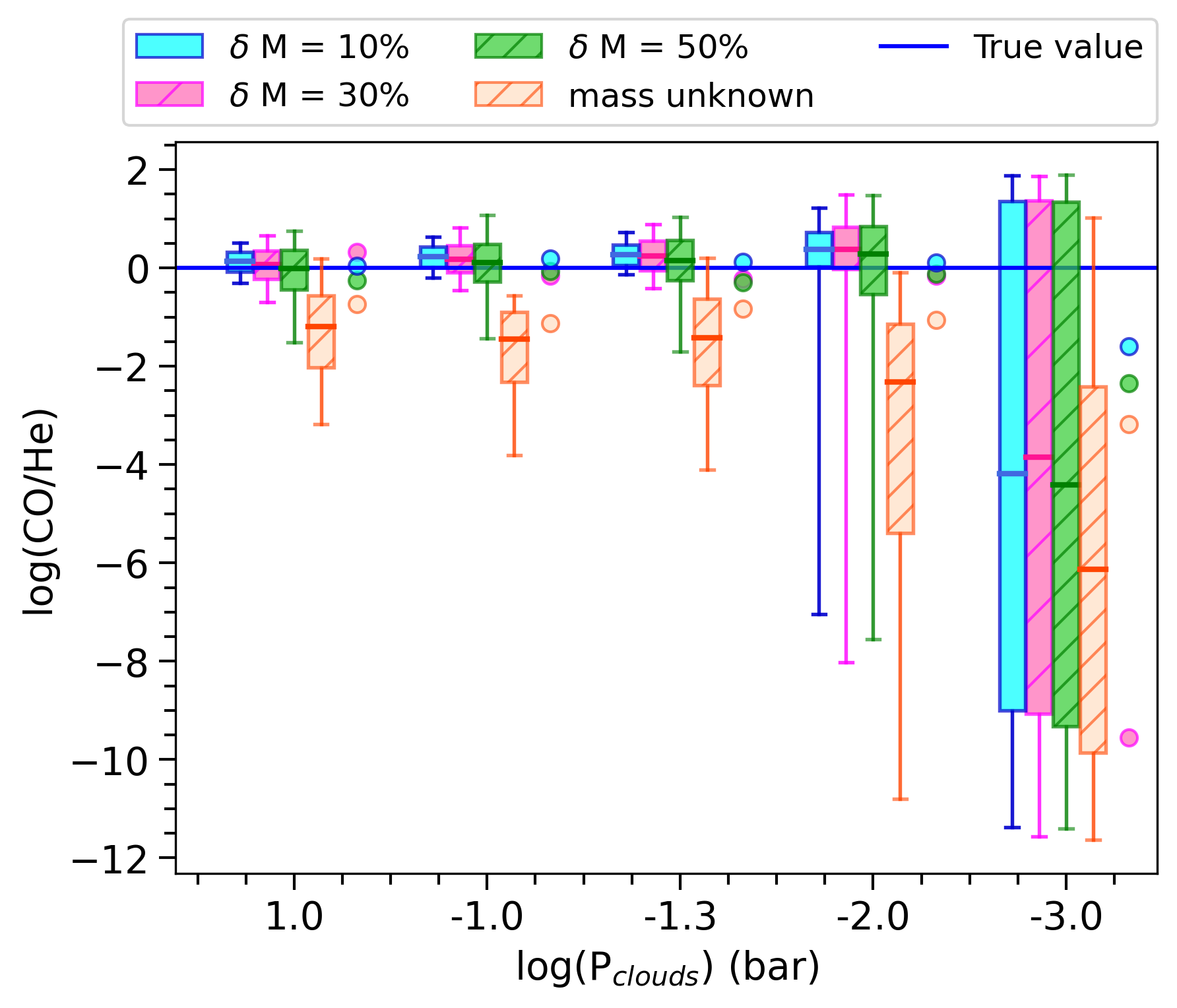}}
    }
    \caption{Retrieval results obtained from different scenarios of cloudy secondary CO-dominated atmosphere in the case of $\mu$=5.2. The different coloured boxes represent the different mass uncertainties. Blue line highlights the true value. Points alongside the boxes highlight the MAP  parameters obtained for each analysed case. The size of the box and the error bar represent the points within 1$\sigma$ and 2$\sigma$ of the median of the distribution (highlighted with
solid lines), respectively.}
    \label{fig:CO5.2_Secondary}
 \end{figure*}
 
These results confirm that with a more favourable scenario, represented by an atmosphere with a main gas producing spectral signature, we could be able to better constrain the atmospheric parameters and the mean molecular weight. However, these results do not appear to be strongly correlated with the mass uncertainty, although in some cases, a better estimate of the mass can help obtain more accurate retrievals.

\section{Conclusions}\label{sec:conclusions}
In this paper, we detail our process of performing several tests to investigate the impact of planetary mass uncertainties in atmospheric retrieval and to identify the cases where mass measurements and their appropriate precision are needed in the presence of clear or cloudy and primary or secondary atmospheres in the context of the ESA Ariel Mission. 

We considered different scenarios to determine the level of planet mass precision required for robust atmospheric characterisation. We selected three representative targets from the Ariel MRS. For the primordial atmosphere, we considered a hot Jupiter and a hot Neptunian. In addition, we also tested the importance of the S/N on the retrievals and the role of the spectral bands. We also investigated the retrieval of a secondary atmosphere of a super-Earth, also in presence of clouds. For each planet, we conducted the retrievals with varying levels of precision for the mass measurements. Our conclusions are as follows:
\begin{enumerate}
    \item In the hot Jupiter case, we were able to accurate retrieve the atmospheric composition of the atmosphere, with an accuracy that does not depends on the mass uncertainty. In the worst-case  scenario analysed here, when the clouds pressure get closer to $10^{-3}$ bar, there is a small discrepancy in the retrieval of the radius that disappears when we performed the retrieval considering a mass uncertainty of about 40\% or lower. For all the other parameters, the uncertainties increase for high-altitude clouds, which can be partially mitigated by increasing the S/N. 
    
    \item We could use the atmospheric analysis to estimate the mass of hot Jupiters with greater precision; for example, we can increase the precision level of the mass estimation from 40\% to 10-25\% depending on  the presence of clouds.
    
    \item For faint stars, the uncertainties on all the fitted parameters increase, confirming the relevance of S/N, independently from our knowledge of the mass.
    
    \item Analogous considerations can be made about the hot Neptunian case. We note increased uncertainties in presence of high altitude clouds and, in particular, a worse estimation of the CO mixing ratio and of the temperature when the cloud pressure gets closer to $10^{-3}$ bar. However, these results are independent from the planetary mass uncertainties. 
    
    \item Studying\ how the S/N values at different wavelength ranges impact the retrieval  highlights the importance of the blue end of the Ariel spectrum, without which we could not be able to retrieve the cloud pressure, bringing on less accurate determinations of other relevant parameters. 
    
    \item In the N$_2$-dominated secondary atmosphere case, when we do not consider the presence of clouds, a minimum knowledge of the mass (of about 50\%) allows us to significantly increase the accuracy and the precision of the retrieval, which only slightly improves further if we consider a better estimation of the mass. 
    
    \item For a cloudy N$_2$-dominated secondary atmosphere, we need an estimation of the mass with an uncertainty of about 50\% to correctly retrieve the mean molecular weight. The uncertainties on all the parameters increase for a cloud pressure lower than $10^{-2}$ bar. A better estimation of the mass could moderately help in the determination of the atmospheric parameters, in particular with regard to increasing the accuracy of the maximum probability values.  
    
    \item The test performed for a H$_2$O- and CO-dominated atmosphere highlights that in the presence of a main gas producing spectral signatures, we should be able to better constrain the atmospheric parameters and the mean molecular weight. Additionally, in this case, a minimum uncertainty of 50\% on the mass is sufficient to measure the atmospheric parameters. 
\end{enumerate}

Our analysis indicates that even in the worst-case scenarios investigated in this work, it is sufficient to have a 50\% mass precision level   to obtain an accurate atmospheric characterisation.
This implies that next-generation transmission spectra contain the information content necessary to independently constrain planetary mass (see also \citet{deWit2013Sci...342.1473D}) and, thus, even an a priori uncertainty as large as 50\% on the mass does not affect retrieval. On the other hand, going into an atmospheric characterisation without any knowledge of a planetary mass could compromise our ability to retrieve the atmospheric composition in cloudy primary atmospheres and in secondary atmospheres.

These results can be used in the preparation and target prioritisation of  RV surveys supporting atmospheric characterisation studies.
In the preparation of the Ariel mission, this work can help in defining the strategy of a RV monitoring for those targets included in the MRS that still lack mass measurements.

\begin{acknowledgements}
      We would like to thank Joanna Barstow for a thorough referee report, which greatly improved the quality and the clarity of the paper. The authors
      acknowledge the partial support of the ARIEL ASI-INAF agreement no. 2021-5-HH.0.
\end{acknowledgements}
    \balance{
    \bibliographystyle{aa}
    \bibliography{Main_Mass.bib}}

   \onecolumn
   \begin{appendix}
   \section{N$_2$-dominated clear sky secondary atmosphere  $\delta$M=50\%}\label{app:corner}
   \begin{figure*}[h!]
   \centering
   \includegraphics[scale=0.42]{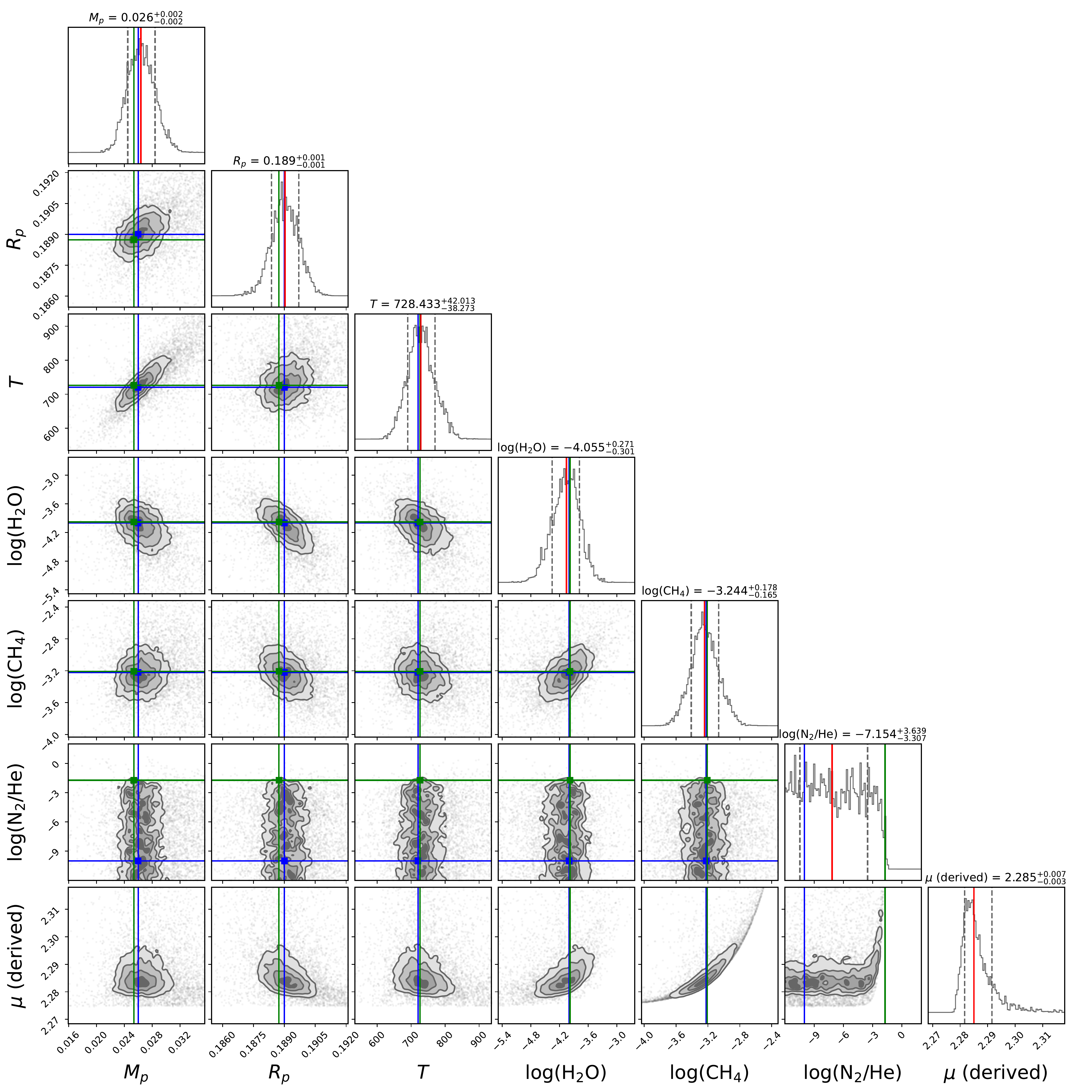}
   \caption{Retrieval results obtained for N$_2$-dominated clear sky secondary atmosphere in the case of $\mu$=2.3 and $\delta$M = 50\%. The blue, green, and red vertical solid lines highlight the true, MAP, and median values, respectively, while the vertical dashed-lines represent the values at 1$\sigma$ from the median.}
    \label{fig:cornermu2.3}
 \end{figure*}
 
 \begin{figure*}[h!]
   \centering
   \includegraphics[scale=0.42]{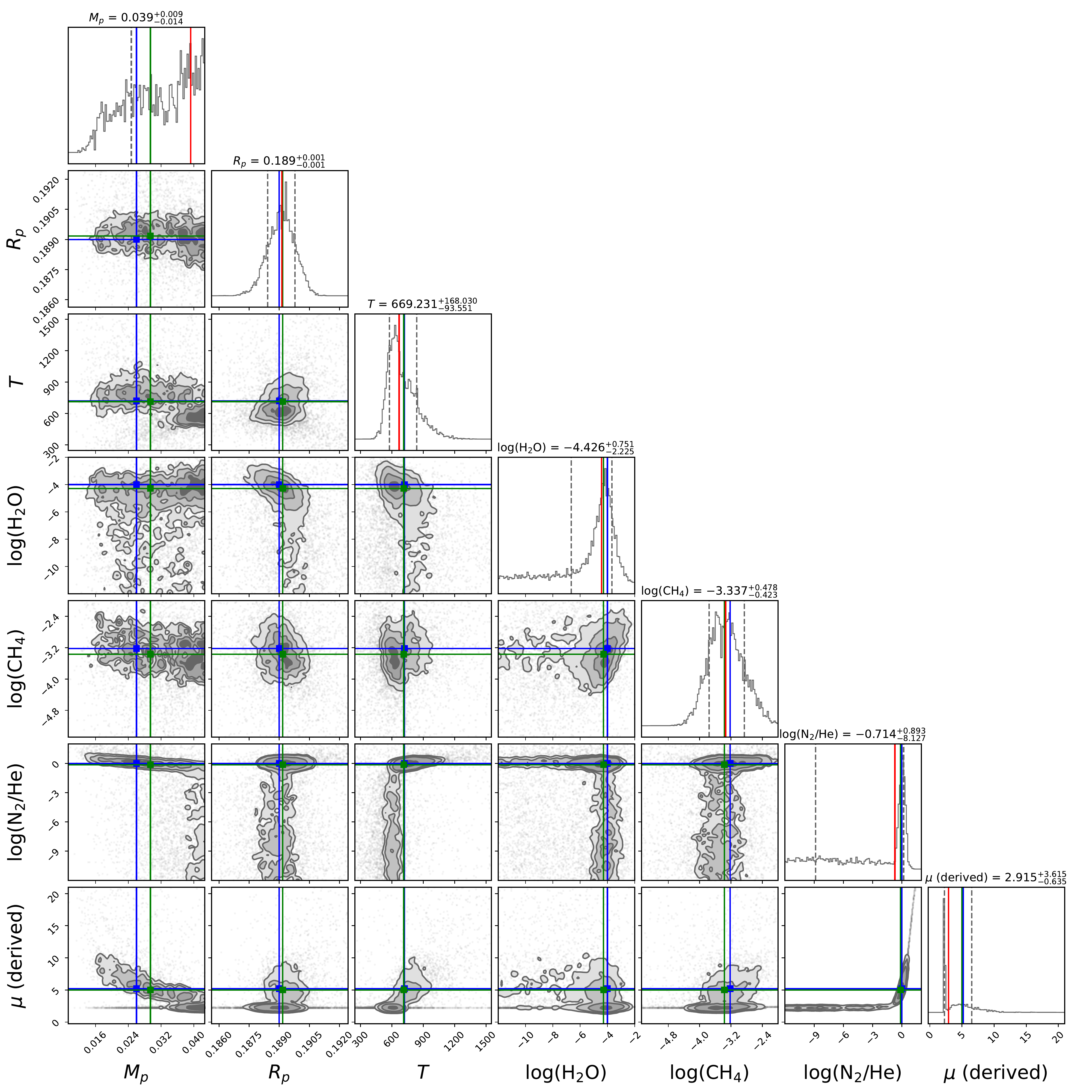}
   \caption{Retrieval results obtained for N$_2$-dominated clear sky secondary atmosphere in the case of $\mu$=5.2 and $\delta$M = 50\%. The blue, green, and red vertical solid lines highlight the true, MAP, and median values, respectively, while the vertical dashed-lines represent the values at 1$\sigma$ from the median.}
    \label{fig:cornermu5.2}
 \end{figure*}
 
 \begin{figure*}[h!]
   \centering
   \includegraphics[scale=0.42]{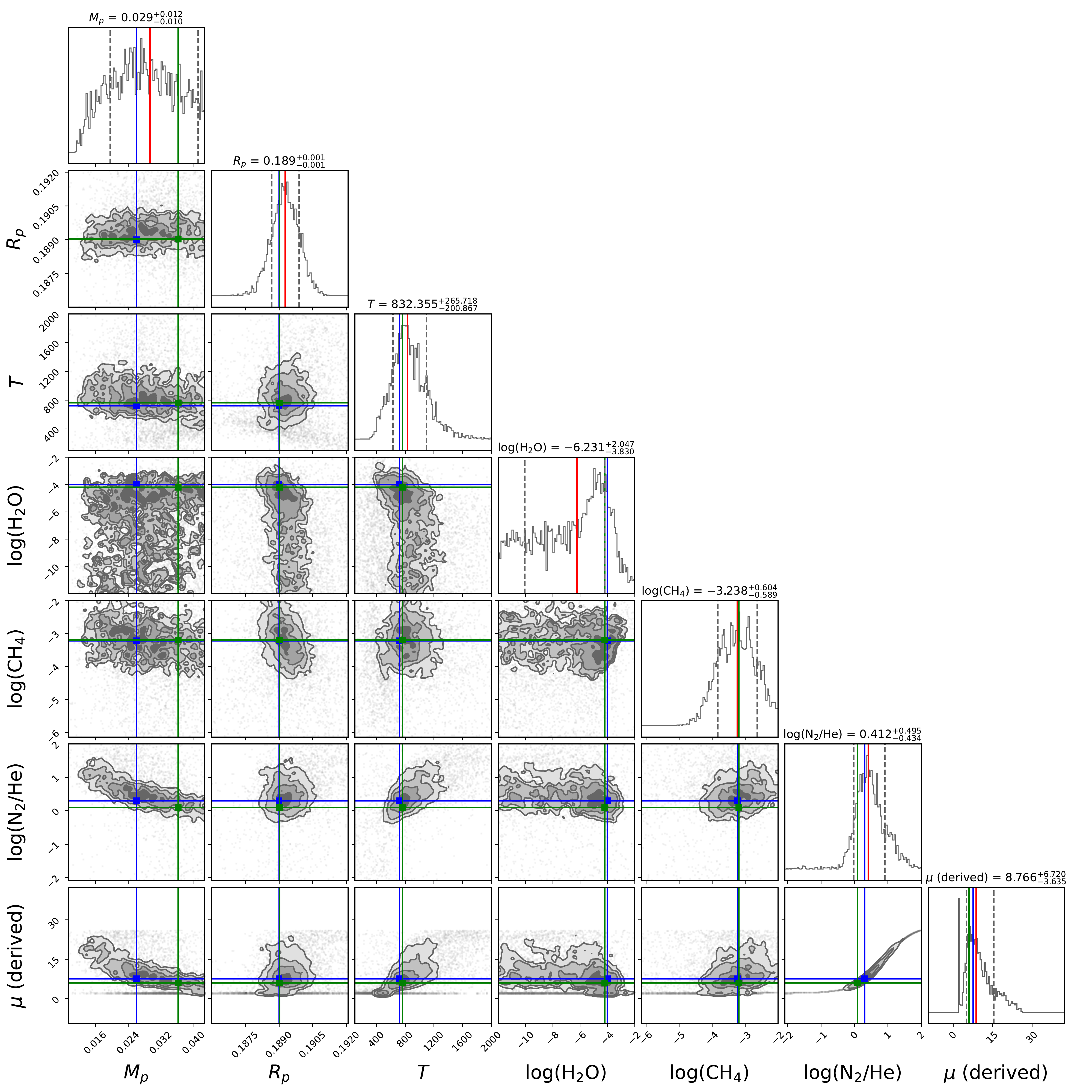}
   \caption{Retrieval results obtained for N$_2$-dominated clear sky secondary atmosphere in the case of $\mu$=7.6 and $\delta$M = 50\%. The blue, green, and red vertical solid lines highlight the true, MAP, and median values, respectively, while the vertical dashed-lines represent the values at 1$\sigma$ from the median.}
    \label{fig:cornermu7.6}
 \end{figure*}
 
 \begin{figure*}[h!]
   \centering
   \includegraphics[scale=0.42]{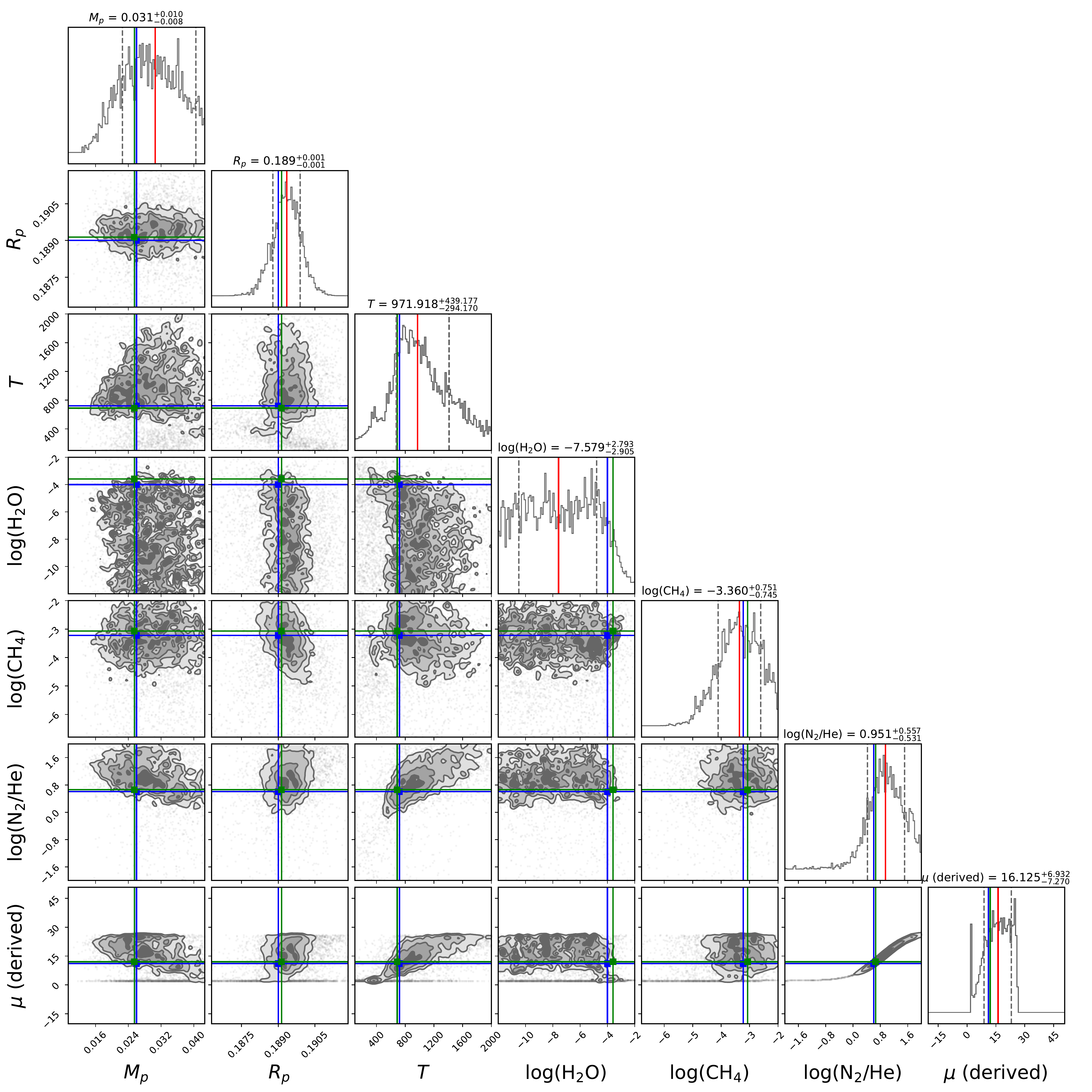}
   \caption{Retrieval results obtained for N$_2$-dominated clear sky secondary atmosphere in the case of $\mu$=11.1 and $\delta$M = 50\%. The blue, green, and red vertical solid lines highlight the true, MAP, and median values, respectively, while the vertical dashed-lines represent the values at 1$\sigma$ from the median.}
    \label{fig:cornermu11.1}
 \end{figure*}
 
 \clearpage
   
   \section{H$_2$O- and CO-dominated atmospheres for $\mu$ = 11.1 }\label{app:figure}
   
   \begin{figure*}[h!]
 \centering
        {\subfloat[Radius\label{fig:Radius_H2Odom1Secondary}]{\includegraphics[scale=0.42]{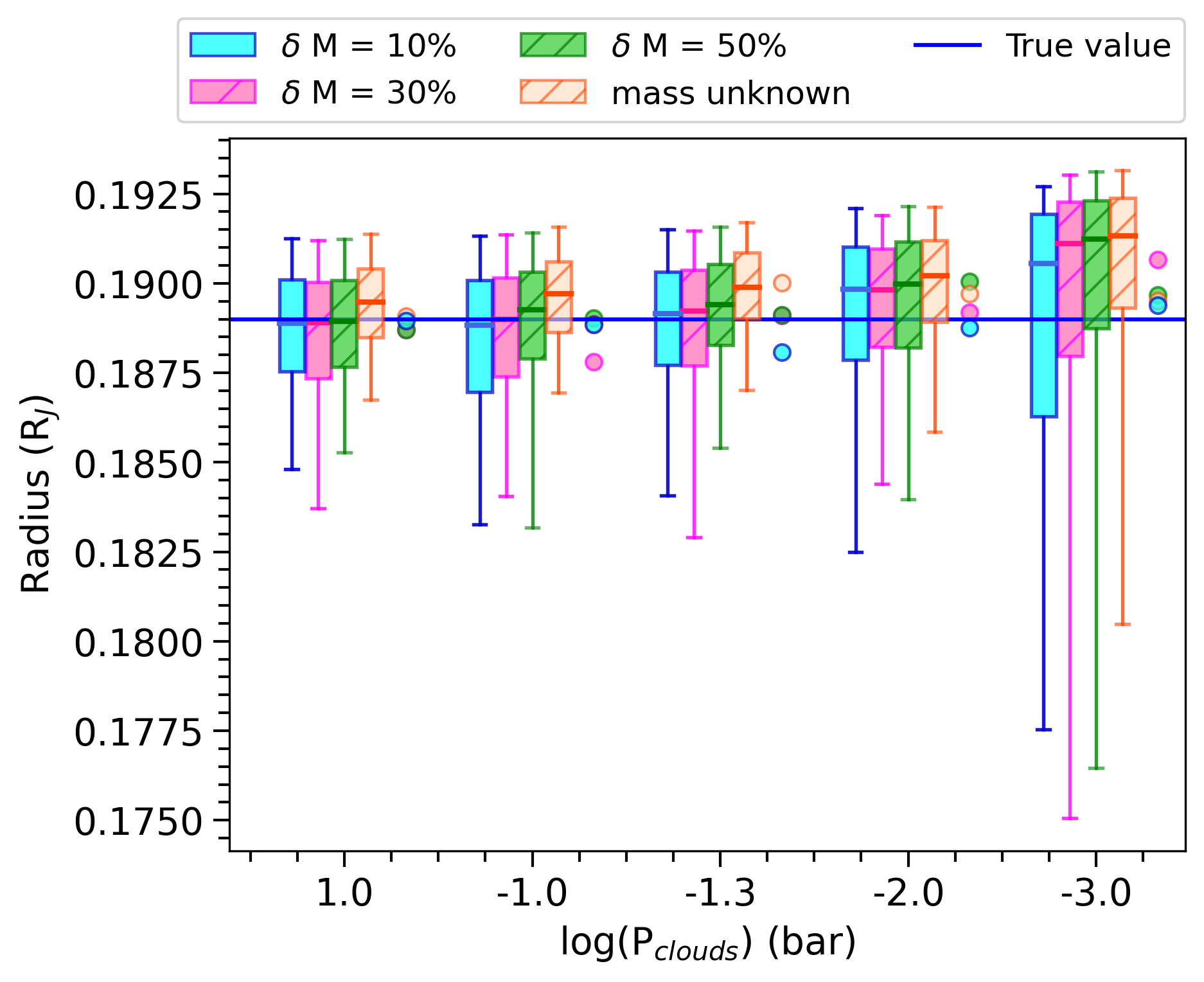}}
        \subfloat[Temperature\label{fig:Temp_H2Odom1Secondary}]{\includegraphics[scale=0.42]{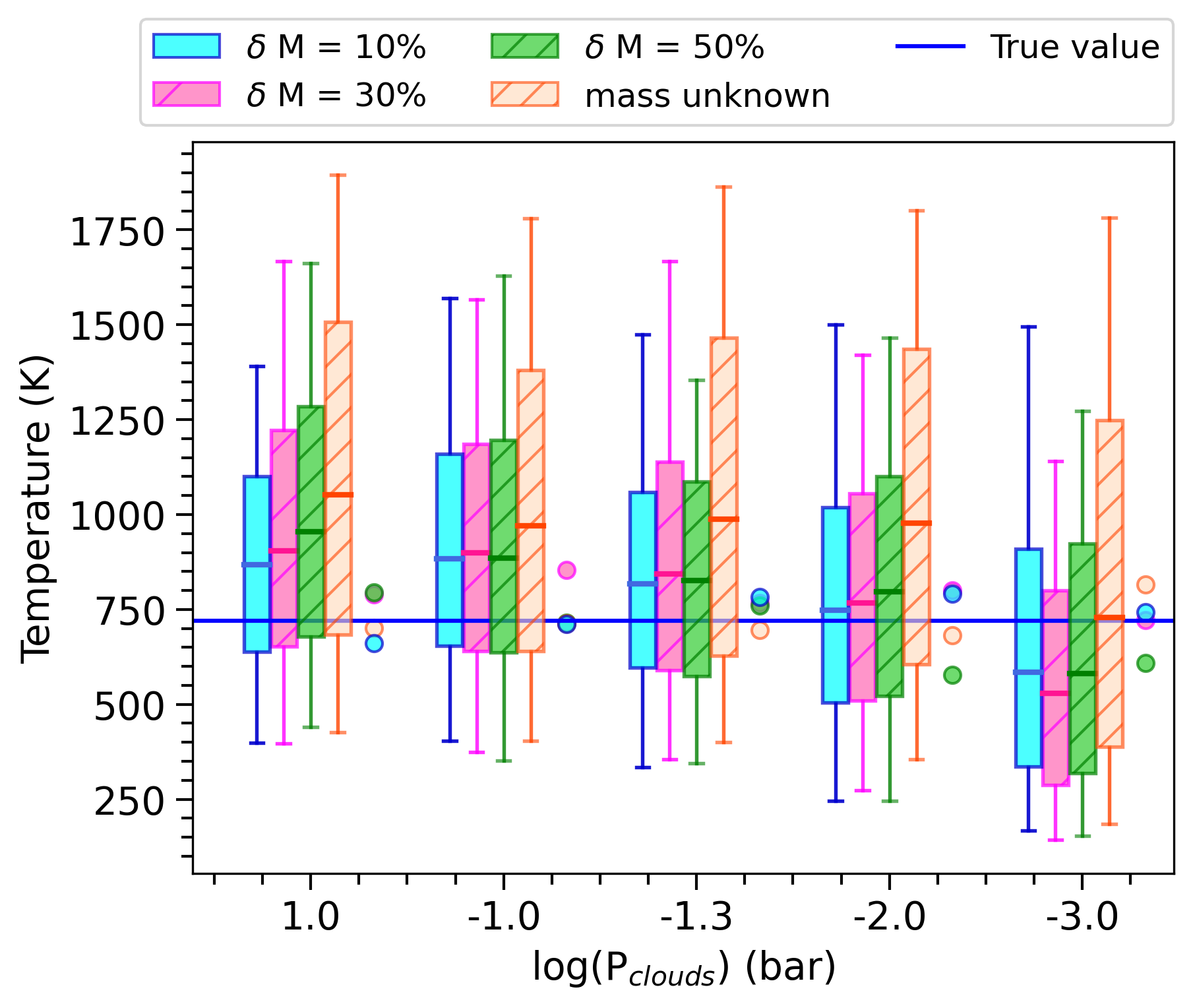}}
        \subfloat[Mass\label{fig:Mass_H2Odom11Secondary}]{\includegraphics[scale=0.42]{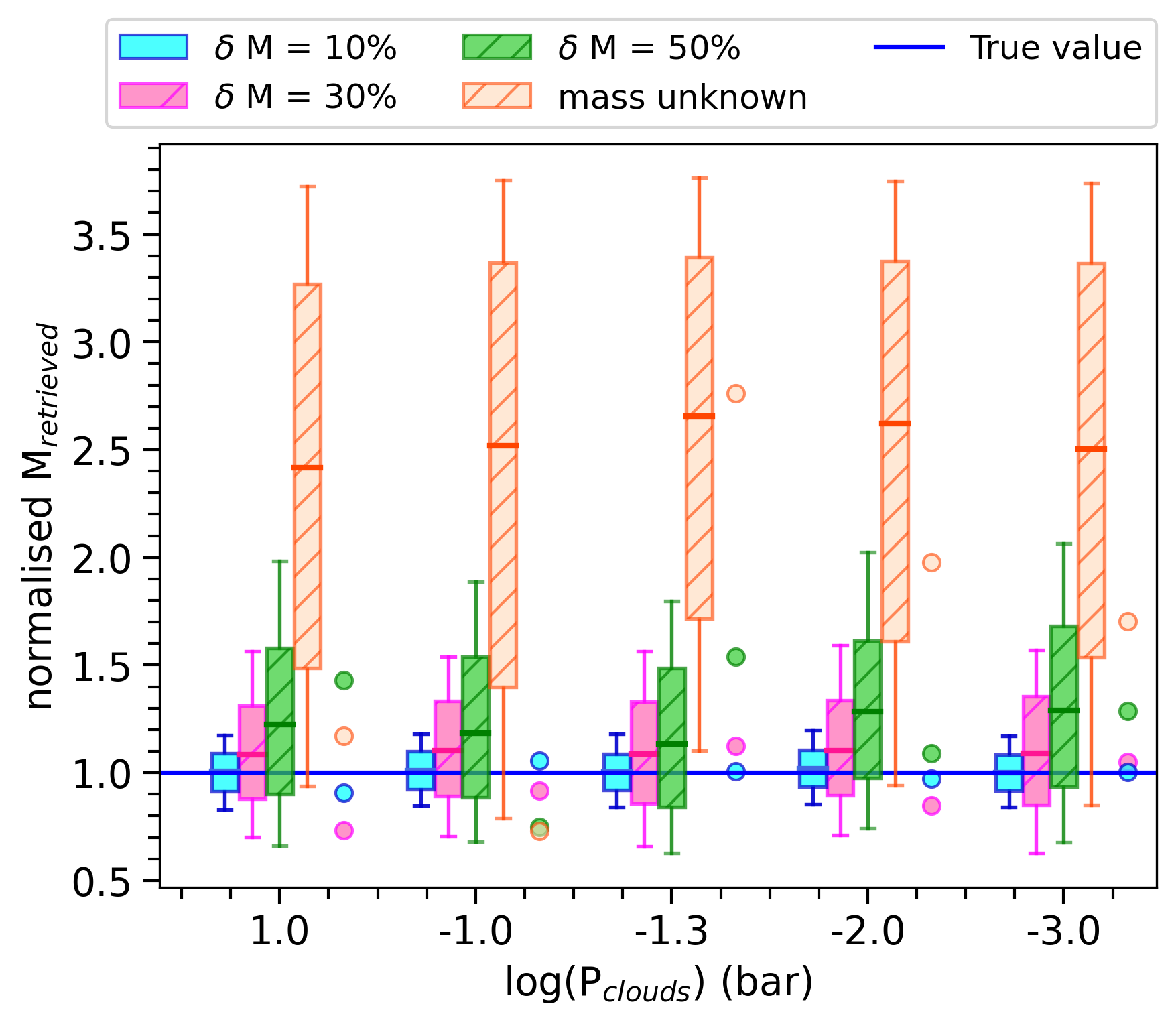}} 
        \\
        \subfloat[Clouds\label{fig:Clouds_H2Odom11Secondary}]{\includegraphics[scale=0.42]{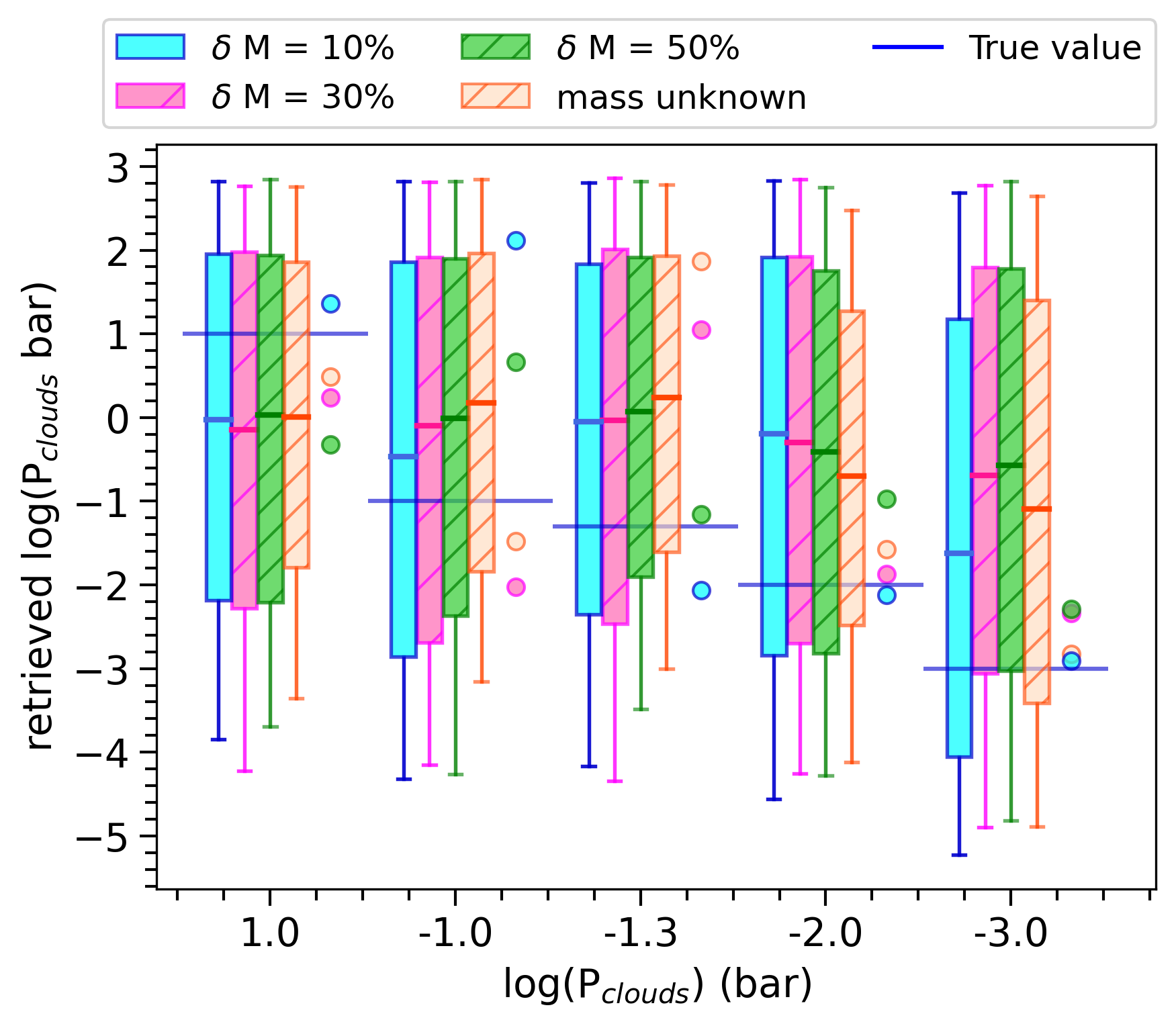}}
        \subfloat[$\mu$\label{fig:mu_H2Odom11Secondary}]{\includegraphics[scale=0.42]{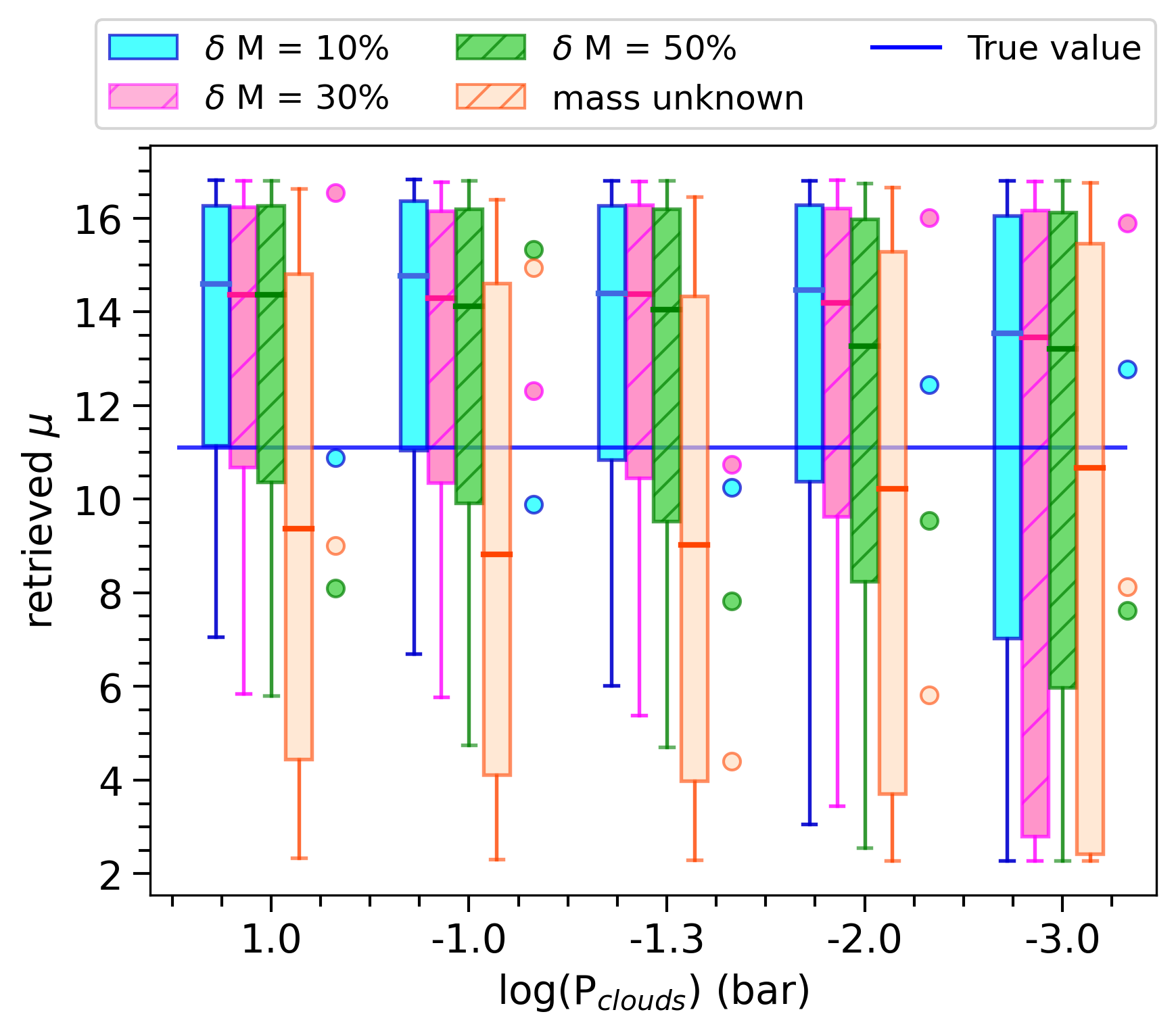}} 
        \subfloat[CH$_4$ mixing ratio\label{fig:CH4_H2Odom11Secondary}]{\includegraphics[scale=0.42]{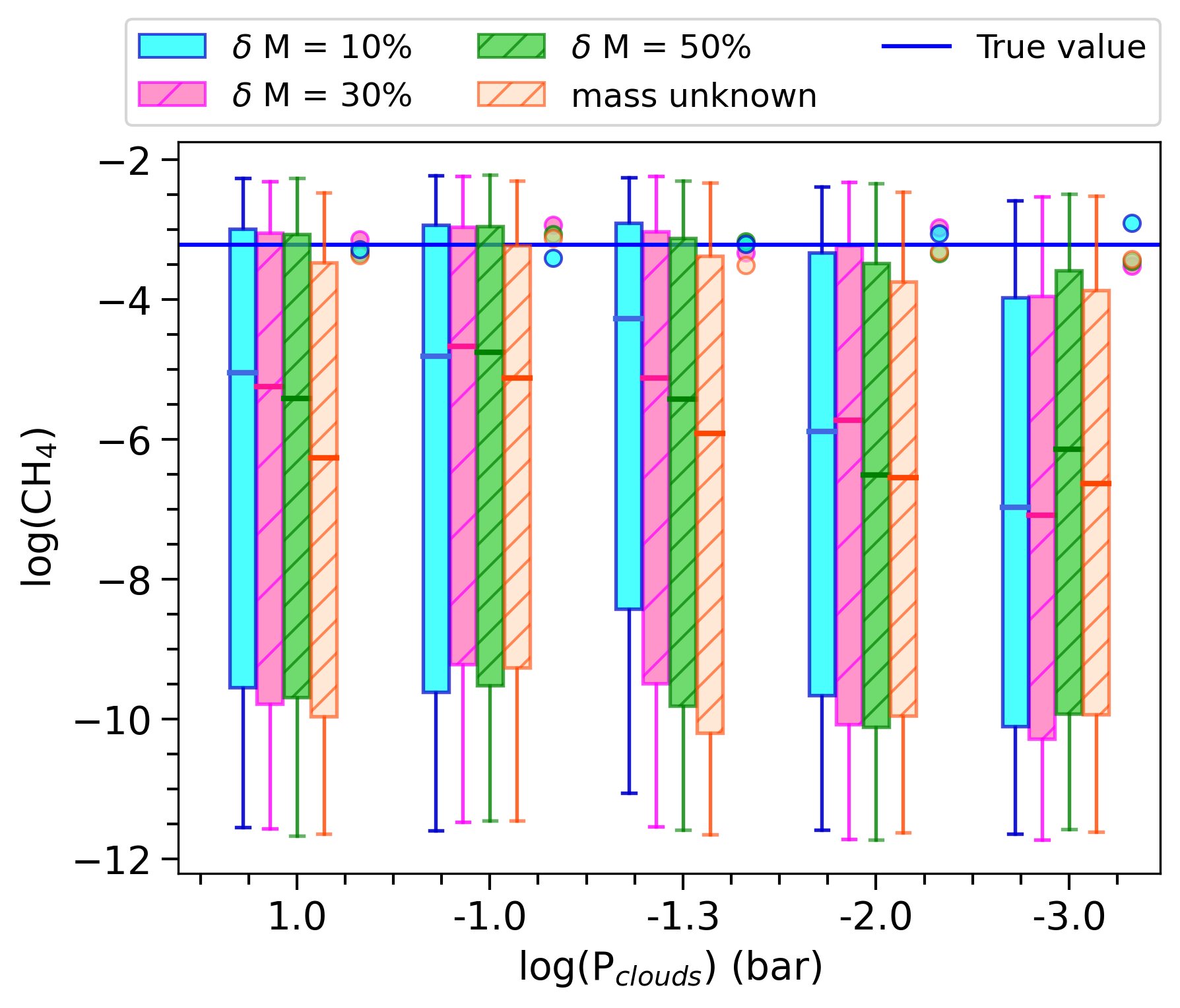}}
        \\
        \subfloat[H$_2$O/He \label{fig:H2OHe_H2Odom11Secondary}]{\includegraphics[scale=0.42]{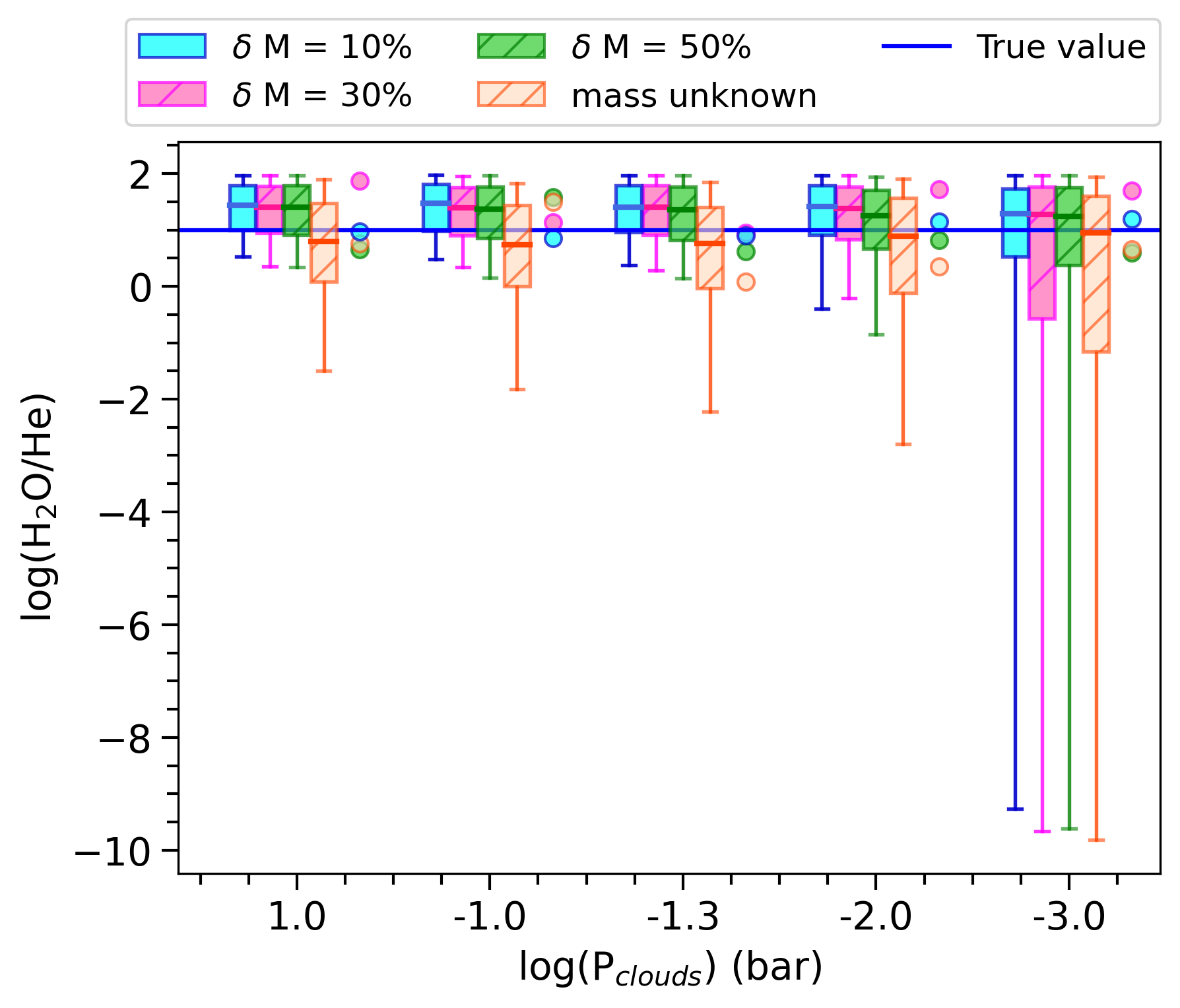}}
    }
    \caption{Retrieval results obtained from different scenarios of cloudy secondary H$_2$O-dominated atmospheres in the case of $\mu$=11.1. Differently coloured boxes represent the different mass uncertainties. Blue line highlights the true value. Points alongside the boxes highlight the MAP  parameters obtained for each analysed case. The size of the box and the error bar represent the points within 1$\sigma$ and 2$\sigma$ of the median of the distribution (highlighted with
solid lines), respectively.}
    \label{fig:H2O11_Secondary}
 \end{figure*}

  \begin{figure*}[h!]
 \centering
        {\subfloat[Radius\label{fig:Radius_COdom11Secondary}]{\includegraphics[scale=0.42]{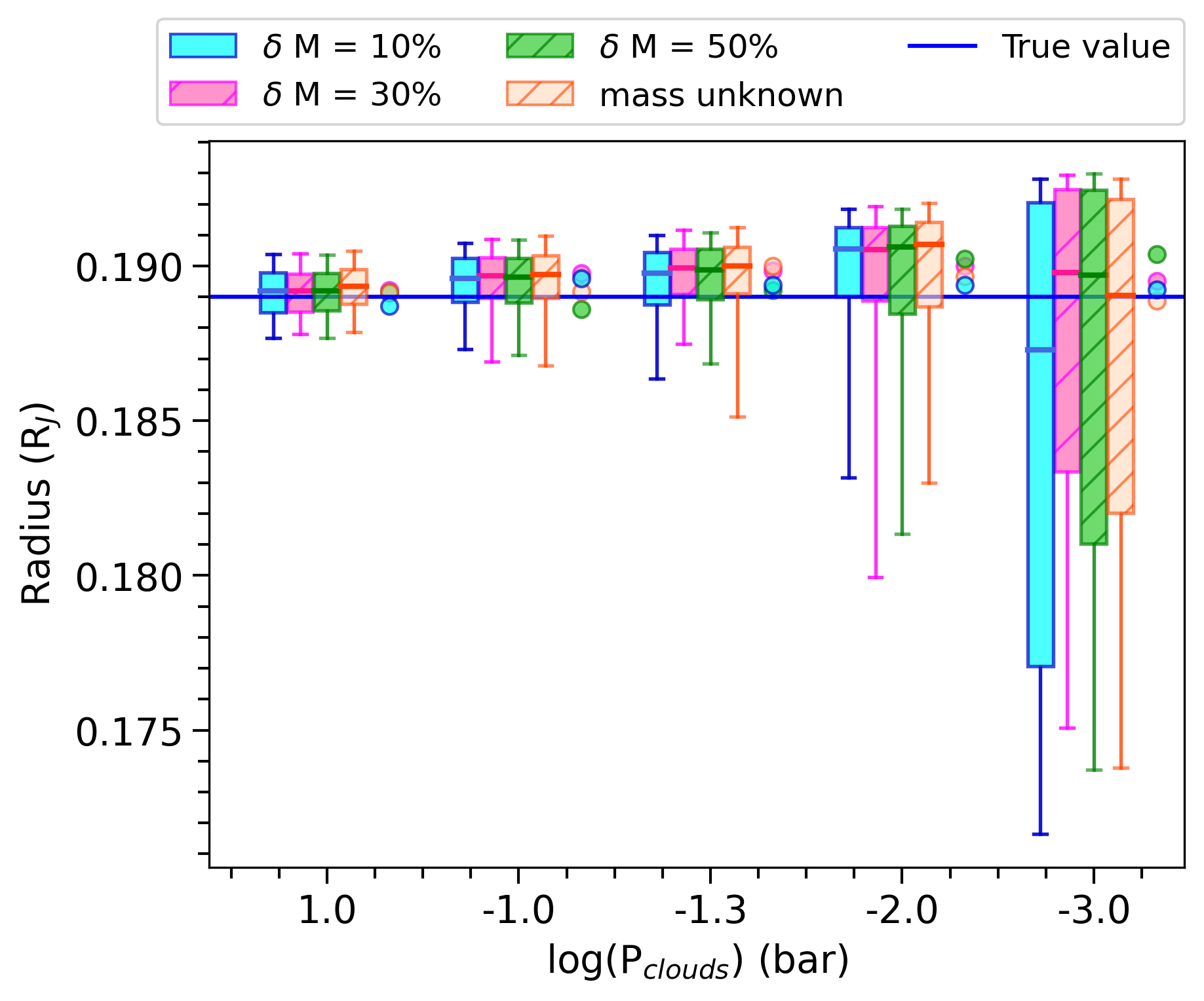}}
        \subfloat[Temperature\label{fig:Temp_COdom11Secondary}]{\includegraphics[scale=0.42]{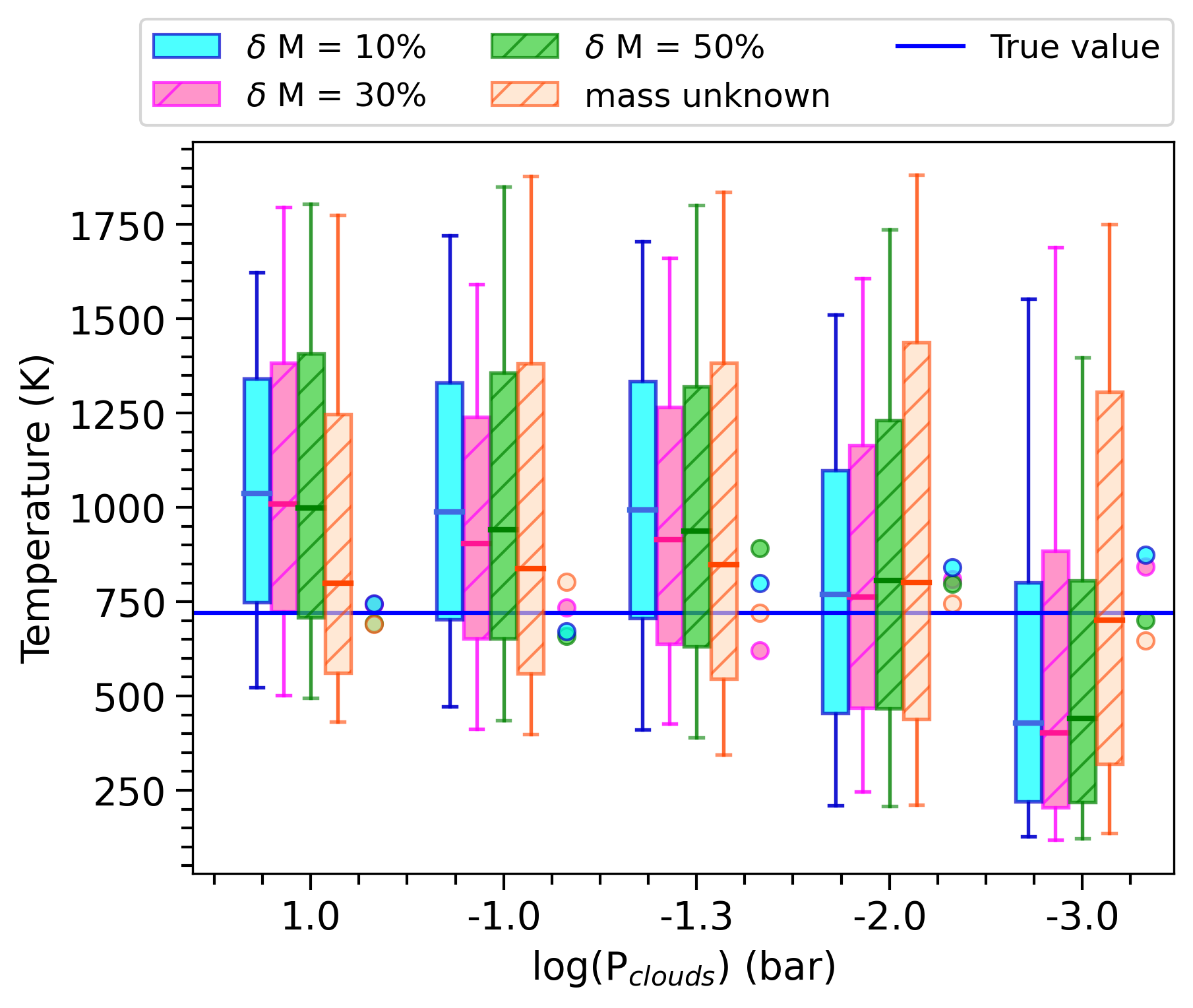}}
        \subfloat[Mass\label{fig:Mass_COdom11Secondary}]{\includegraphics[scale=0.42]{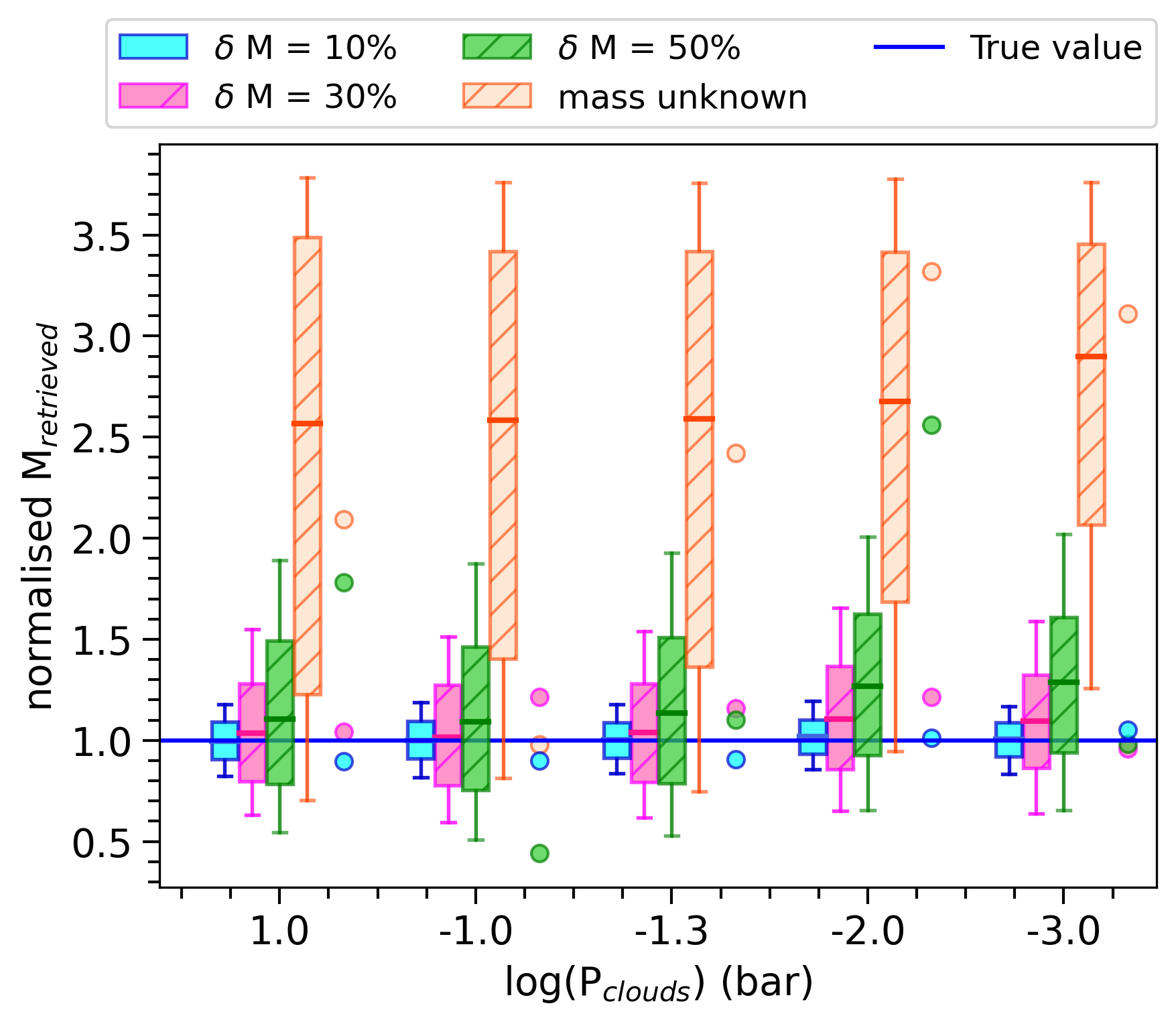}} 
        \\
        \subfloat[Clouds\label{fig:Clouds_COdom11Secondary}]{\includegraphics[scale=0.42]{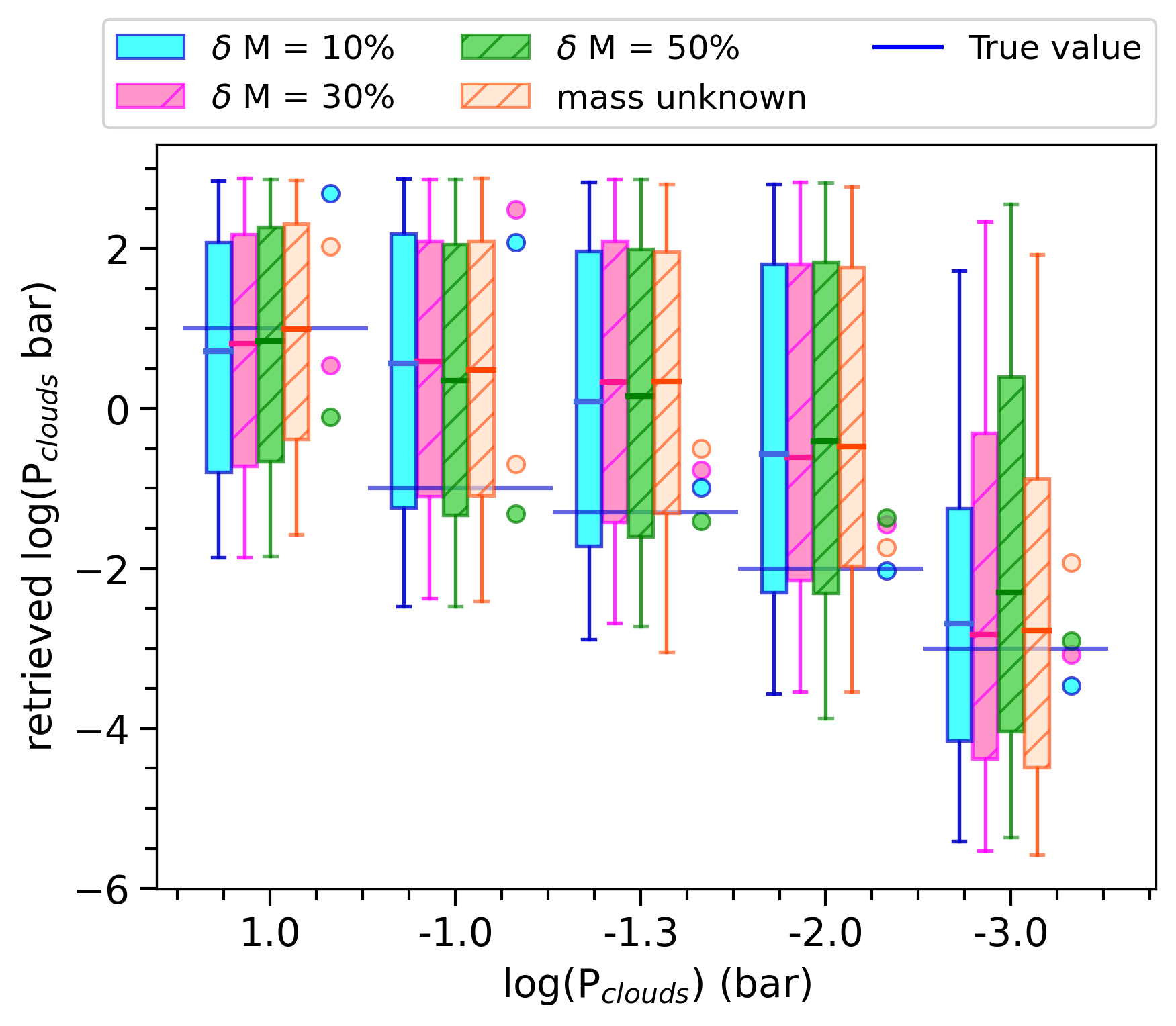}}
        \subfloat[$\mu$\label{fig:mu_COdom11Secondary}]{\includegraphics[scale=0.42]{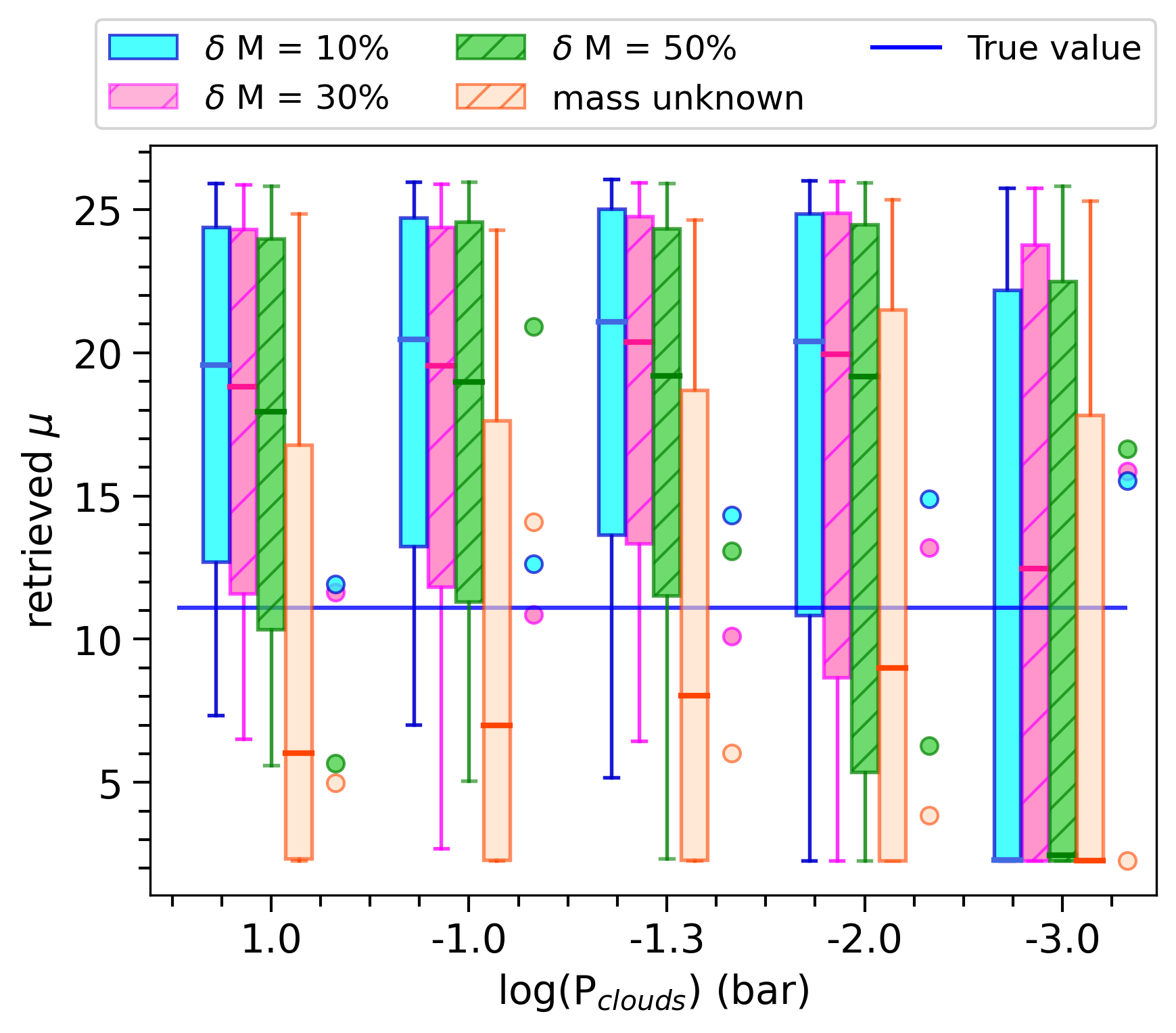}} 
        \subfloat[H$_2$O mixing ratio\label{fig:H2O_COdom11Secondary}]{\includegraphics[scale=0.42]{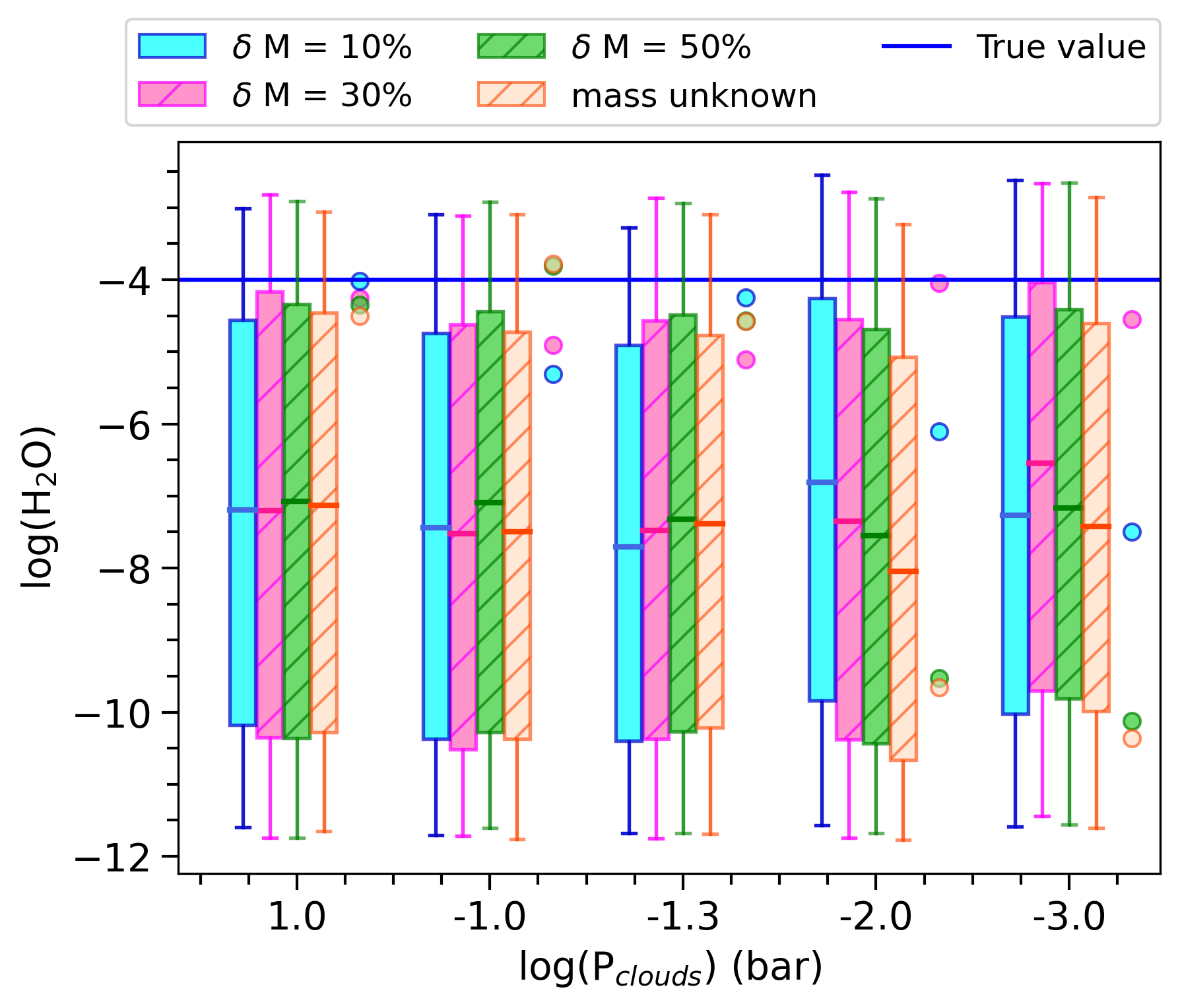}}
        \\
        \subfloat[CH$_4$ mixing ratio\label{fig:CH4_COdom11Secondary}]{\includegraphics[scale=0.42]{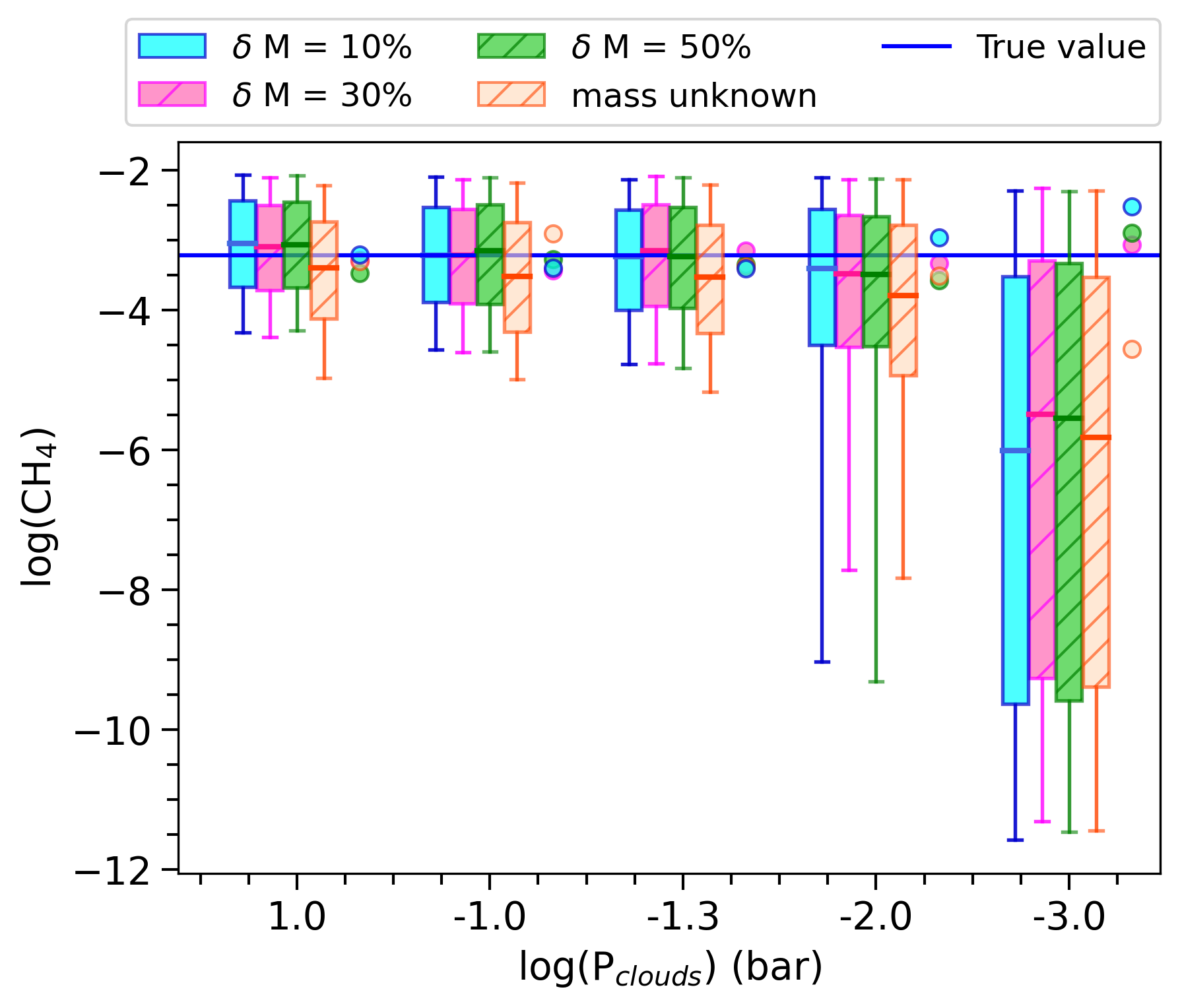}}
        \subfloat[CO/He \label{fig:COHe_COdom11Secondary}]{\includegraphics[scale=0.42]{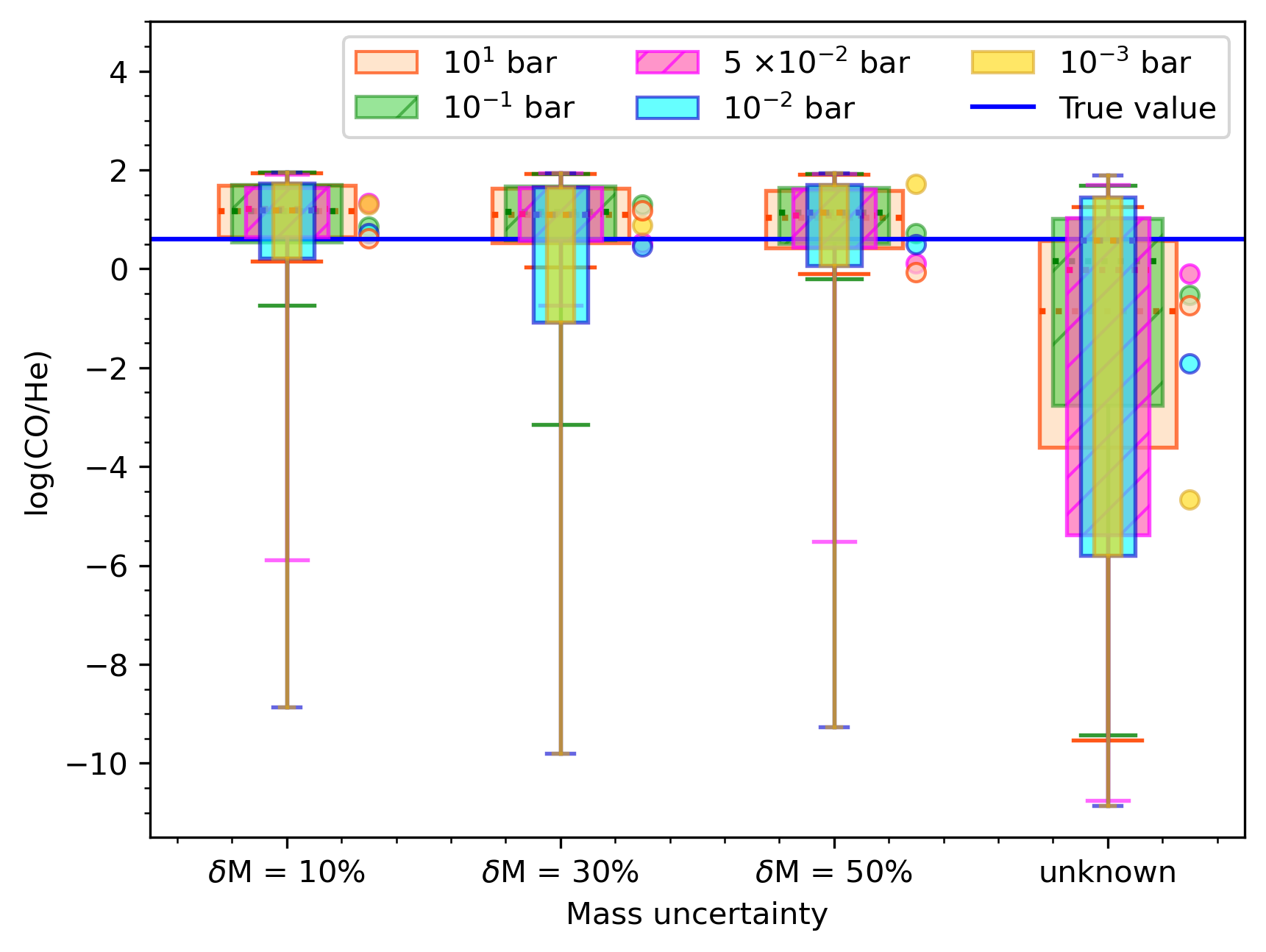}}
    }
    \caption{Retrieval results obtained from different scenarios of cloudy secondary CO-dominated atmospheres in the case of $\mu$=11.1. Differently coloured boxes represent the different mass uncertainties. Blue line highlights the true value. Points alongside the boxes highlight the MAP parameters obtained for each analysed case. The size of the box and the error bar represent the points within 1$\sigma$ and 2$\sigma$ of the median of the distribution (highlighted with
solid lines), respectively.}
    \label{fig:CO11_Secondary}
 \end{figure*}

   \clearpage
   \section{Tables with the results of the analysis of Sect. 2}\label{app:tabelle}
   \begin{table}[h!] \footnotesize
        \caption{Results from the retrieval performed for a hot Jupiter around a G star when we the mass is totally unknown and when we known it with an uncertainty of about 40\% and 10\%.}   
        \label{table:resumejupiter} 
        \centering             
\begin{tabular}{cccc|ccc|ccc}   
            \toprule[0.05cm]
             \multicolumn{10}{c}{\textbf{G star with a hot Jupiter}}\\
            \toprule
              & \multicolumn{3}{c}{\textbf{Mass totally unknown}}& \multicolumn{3}{c}{\textbf{$\delta$ M $\simeq$ 40\%}} & \multicolumn{3}{c}{\textbf{$\delta$ M $\simeq$ 10\%}} \\
            \cmidrule(lr){2-10}
            & {Clear Sky} & \multicolumn{2}{c}{Clouds} & {Clear Sky} & \multicolumn{2}{c}{Clouds}&  {Clear Sky} & \multicolumn{2}{c}{Clouds}  \\
            Parameters &  & Low & High & & Low & High &  & Low & High  \\
            \midrule
            Mass &   $<$5\% & $<$ 10\%& $>$50\% & $<$ 2 \%  & $<$ 10\% & $<$ 30\% & $<$ 2\% & $<$ 10\%& $\simeq$ 10\% \\
            Radius & \cellcolor{green!70} $\surd$& \cellcolor{green!70} $\surd$& \cellcolor{red!70} X& \cellcolor{green!70} $\surd$ & \cellcolor{green!70} $\surd$ & \cellcolor{green!70} $\sim$ & \cellcolor{green!70}$\surd$&  \cellcolor{green!70}$\surd$& \cellcolor{green!70}$\surd$\\
            Temperature & \cellcolor{green!70} $\surd$& \cellcolor{green!70} $\surd$& \cellcolor{green!70} $\surd$& \cellcolor{green!70} $\surd$ & \cellcolor{green!70} $\surd$ &  \cellcolor{green!70} $\surd$ &\cellcolor{green!70} $\surd$& \cellcolor{green!70} $\surd$&\cellcolor{green!70} $\surd$  \\
            H$_2$O & \cellcolor{green!70} $\surd$& \cellcolor{green!70} $\surd$& \cellcolor{green!70} $\surd$& \cellcolor{green!70} $\surd$& \cellcolor{green!70} $\surd$&  \cellcolor{green!70} $\surd$ &\cellcolor{green!70} $\surd$&\cellcolor{green!70} $\surd$&\cellcolor{green!70} $\surd$ \\
\end{tabular}
             \tablefoot{Green: well-retrieved values;  red: values within 2$\sigma$ or more of
the true value.}
            
\end{table}

\begin{table}[h!] \footnotesize
        \caption{Results from the retrieval performed for a neptunian planet around a G star when we know the mass with an uncertainty of about 40\% and 10\%.}   
        \label{table:target} 
        \centering             
        \begin{tabular}{cccc|ccc}   
            \toprule[0.05cm]
             \multicolumn{7}{c}{\textbf{G star with a Neptunian planet}}\\
            \toprule
             & \multicolumn{3}{c}{\textbf{$\delta$ M $\simeq$ 40\%}} & \multicolumn{3}{c}{\textbf{$\delta$ M $\simeq$ 10\%}} \\
            \cmidrule(lr){2-7}
            &  {Clear Sky} & \multicolumn{2}{c}{Clouds}&  {Clear Sky} & \multicolumn{2}{c}{Clouds}  \\
            Parameters & & Low & High &  & Low & High  \\
            \midrule
            Mass &   $<$ 2 \%  & $<$ 20\% & $<$ 30\% & $<$ 2\% & $<$ 10\%& $\simeq$ 10\% \\
            Radius & \cellcolor{green!70} $\surd$ & \cellcolor{green!70} $\surd$ & \cellcolor{green!70} $\surd$ & \cellcolor{green!70}$\surd$&  \cellcolor{green!70}$\surd$& \cellcolor{green!70}$\surd$\\
            Temperature &  \cellcolor{green!70} $\surd$ & \cellcolor{green!70} $\surd$ &  \cellcolor{green!70} $\sim$ &\cellcolor{green!70} $\surd$& \cellcolor{green!70} $\surd$&\cellcolor{green!70} $\sim$  \\
            H$_2$O &  \cellcolor{green!70} $\surd$& \cellcolor{green!70} $\surd$&  \cellcolor{green!70} $\surd$ &\cellcolor{green!70} $\surd$&\cellcolor{green!70} $\surd$&\cellcolor{green!70} $\surd$ \\
            CH$_4$ &  \cellcolor{green!70} $\surd$ & \cellcolor{green!70} $\surd$ & \cellcolor{green!70} $\sim$&\cellcolor{green!70} $\surd$ &\cellcolor{green!70} $\surd$&\cellcolor{green!70} $\sim$ \\
            CO     & \cellcolor{green!70} $\surd$ & \cellcolor{green!70} $\surd$ & \cellcolor{red!70} X&\cellcolor{green!70} $\surd$&\cellcolor{green!70} $\surd$&\cellcolor{red!70} X \\
            Clouds & \cellcolor{green!70} $\surd$ & \cellcolor{green!70} $\surd$ & \cellcolor{green!70} $\surd$&\cellcolor{green!70} $\surd$ &\cellcolor{green!70} $\surd$&\cellcolor{green!70} $\surd$\\
            \bottomrule[0.05cm]                
            \label{tab:neptunian}
            \end{tabular}
             \tablefoot{Green: well-retrieved values; red: values within 2$\sigma$ or more of
the true value.}
\end{table}

\begin{table}[h!] \footnotesize
        \caption{Results from the retrieval performed for a hot Jupiter around an 8th magnitude G star and a 10.5th magnitude G star.}   
        \label{table:resumemagnitude} 
        \centering             
        \begin{tabular}{cccc|ccc}   
            \toprule[0.05cm]
             \multicolumn{7}{c}{\textbf{Hot Jupiter around G stars of different magnitudes (mass uncertainty of 40\%)}}\\
            \toprule
             & \multicolumn{3}{c}{\textbf{m$_v$ $\simeq$ 8}} & \multicolumn{3}{c}{\textbf{m$_v$ $\simeq$ 10.5}} \\
            \cmidrule(lr){2-7}
            &  {Clear Sky} & \multicolumn{2}{c}{Clouds}&  {Clear Sky} & \multicolumn{2}{c}{Clouds}  \\
            Parameters & & Low & High &  & Low & High  \\
            \midrule
            Mass &   $<$ 2 \%  & $<$ 10\% & $<$ 30\% & $<$ 10\% & $\simeq$ 20\%& $\simeq$ 30\% \\
            Radius & \cellcolor{green!70} $\surd$ & \cellcolor{green!70} $\surd$ & \cellcolor{green!70} $\sim$ & \cellcolor{green!70}$\surd$&  \cellcolor{green!70}$\surd$& \cellcolor{green!70}$\surd$\\
            Temperature &  \cellcolor{green!70} $\surd$ & \cellcolor{green!70} $\surd$ &  \cellcolor{green!70} $\surd$ &\cellcolor{green!70} $\surd$& \cellcolor{green!70} $\surd$&\cellcolor{green!70} $\surd$  \\
            H$_2$O &  \cellcolor{green!70} $\surd$& \cellcolor{green!70} $\surd$&  \cellcolor{green!70} $\surd$ &\cellcolor{green!70} $\surd$&\cellcolor{green!70} $\surd$&\cellcolor{green!70} $\sim$ \\
            CH$_4$ &  \cellcolor{green!70} $\surd$ & \cellcolor{green!70} $\surd$ & \cellcolor{green!70} $\surd$&\cellcolor{green!70} $\surd$ &\cellcolor{green!70} $\surd$&\cellcolor{green!70} $\sim$ \\
            CO     & \cellcolor{green!70} $\surd$ & \cellcolor{green!70} $\surd$ & \cellcolor{green!70} $\surd$&\cellcolor{green!70} $\surd$&\cellcolor{green!70} $\surd$&\cellcolor{red!70} X \\
            Clouds & \cellcolor{green!70} $\surd$ & \cellcolor{green!70} $\surd$ & \cellcolor{green!70} $\surd$&\cellcolor{green!70} $\surd$ &\cellcolor{green!70} $\surd$&\cellcolor{green!70} $\sim$\\
            \bottomrule[0.05cm]                
            \label{tab:mv8mv10}
            \end{tabular}
             \tablefoot{Green: well-retrieved values;  red: values within 2$\sigma$ or more of
the true value.}
\end{table}

\begin{table*}\small
        \caption{Retrieved values obtained from the analyses performed on a N$_2$-dominated secondary atmosphere of a super-Earth around an 8th magnitude M star in the different  scenarios considered in this study.}   
        \label{table:Secondary} 
        \centering      
        \begin{tabular}{c@{\hskip 0.5cm}c@{\hskip 0.5cm}|c@{\hskip 0.5cm}c@{\hskip 0.5cm}c@{\hskip 0.5cm}c@{\hskip 0.5cm}}   
            \toprule[0.05cm]
            \multicolumn{6}{c}{\textbf{N$_2$-dominated secondary atmosphere  of a super-Earth around an 8th magnitude M star}}\\
            \toprule
             \multirow{1}*{Parameter} & True value & \multicolumn{4}{c}{$\mu$ = 2.3} \\
            \cmidrule(lr){1-6}
            \multirow{1}*{$\delta$M} & & 10\% & 30\% & 50\% & Unknown\\
            \midrule
            Mass        &       0.026   &       0.026   $^{+    0.001   }_{-    0.001   } \ [     0.026   ]$&     0.026   $^{+    0.002   }_{-    0.002   } \ [   0.026   ]$&     0.026   $^{+    0.002   }_{-    0.002   } \ [     0.025   ]$&     0.026   $^{+    0.002   }_{-    0.002   } \ [   0.027   ]$ \\
Radius  &       0.189   &       0.189   $^{+    0.001   }_{-    0.001   } \ [     0.189   ]$&     0.189   $^{+    0.001   }_{-    0.001   } \ [   0.189   ]$&     0.189   $^{+    0.001   }_{-    0.001   } \ [     0.189   ]$&     0.189   $^{+    0.001   }_{-    0.001   } \ [   0.189   ]$ \\
Temperature     &       720.33  &       725     $^{+    34      }_{-    31      } \ [     722     ]$&     728     $^{+    40      }_{-    41      } \ [   734     ]$&     728     $^{+    42      }_{-    38      } \ [     726     ]$&     730     $^{+    40      }_{-    39      } \ [   732     ]$ \\
log(H$_2$O)     &       -4      &       -4      $^{+    0.3     }_{-    0.3     } \ [     -4      ]$&     -4.1    $^{+    0.3     }_{-    0.3     } \ [   -4.0    ]$&     -4.1    $^{+    0.3     }_{-    0.3     } \ [     -4.0    ]$&     -4.0    $^{+    0.3     }_{-    0.3     } \ [   -4.1    ]$ \\
\end{tabular}
             \tablefoot{ We report the median values with its 1$\sigma$ error and the MAP values in square brackets.}
\end{table*}

\begin{table*}\small
        \caption{Retrieved values obtained from the analyses performed on a N$_2$-dominated secondary atmosphere of a super-Earth around a 10.5th magnitude M star in the different  scenarios considered here. }   
        \label{table:Secondarymv10} 
        \centering      
        \begin{tabular}{c@{\hskip 0.5cm}c@{\hskip 0.5cm}|c@{\hskip 0.5cm}c@{\hskip 0.5cm}c@{\hskip 0.5cm}}   
            \toprule[0.05cm]
            \multicolumn{5}{c}{\textbf{N$_2$-dominated secondary atmosphere  of a super-Earth around a 10.5th magnitude M star}}\\
            \toprule
             \multirow{1}*{Parameter} & True value & \multicolumn{3}{c}{$\mu$ = 2.3} \\
            \cmidrule(lr){1-5}
            \multirow{1}*{$\delta$M} & & 10\% & 30\% & 50\% \\
            \midrule
            Mass        &       0.026   &       0.026$^{+0.002}_{-0.002}$ [0.025] &       0.028$^{+0.005}_{-0.005}$ [0.027]       &       0.030$^{+0.006}_{-0.005}$ [0.026] \\
Radius  &       0.189   &       0.190$^{+0.002}_{-0.002}$ [0.188]       &       0.190$^{+0.002}_{-0.002}$ [0.190] &       0.190$^{+0.002}_{-0.002}$ [0.189]       \\
Temperature     &       720.33  &       740$^{+86}_{-73}$ [722] &       772$^{+105}_{-99}$ [737]   &       803$^{+138}_{-100}$ [693]       \\
log(H$_2$O)     &       -4      &       -5.0$^{+1.1}_{-4.1}$[-3.9]      &       -5.0$^{+1.2}_{-3.6}$[-4.3]      &       -5.7$^{+1.6}_{-3.9}$[-3.7]      \\
log(CH$_4$)     &       -3.22   &       -3.2$^{+0.5}_{-0.5}$ [-3.3]     &       -3.2$^{+0.5}_{-0.5}$ [-3.43] &       -3.2$^{+0.5}_{-0.5}$ [-3.2]     \\
\end{tabular}
            \tablefoot{ We report the median values with its 1$\sigma$ error and the MAP values in square brackets.}
\end{table*}

\begin{table*}\small
        \caption{Retrieved values obtained from the analyses performed on a N$_2$-dominated cloudy secondary atmosphere (P$_{clouds} = 10^{-3}$ bar) of a super-Earth around an 8th magnitude M star in the different scenarios considered here.}   
        \label{table:SecondaryN2Cloud1e2} 
        \centering      
        \begin{tabular}{c@{\hskip 0.5cm}c@{\hskip 0.5cm}|c@{\hskip 0.5cm}c@{\hskip 0.5cm}c@{\hskip 0.5cm}c@{\hskip 0.5cm}}   
            \toprule[0.05cm]
            \multicolumn{6}{c}{\textbf{N$_2$-dominated cloudy secondary atmosphere  of a super-Earth around an 8th magnitude M star (P\textsubscript{clouds} = 10\textsuperscript{-3} bar)}}\\
            \toprule
             \multirow{1}*{Parameter} & True value & \multicolumn{4}{c}{$\mu$ = 2.3} \\
            \cmidrule(lr){1-6}
            \multirow{1}*{$\delta$M} & & 10\% & 30\% & 50\% & Unknown\\
            \midrule
            Mass        &       0.026   &       0.026   $^{+    0.002   }_{-    0.002   } \ [     0.026   ]$&     0.029   $^{+    0.005   }_{-    0.004   } \ [   0.028   ]$&     0.031   $^{+    0.007   }_{-    0.006   } \ [     0.025   ]$&     0.038   $^{+    0.009   }_{-    0.006   } \ [   0.028   ]$\\
Radius  &       0.189   &       0.191   $^{+    0.004   }_{-    0.006   } \ [     0.187   ]$&     0.193   $^{+    0.005   }_{-    0.005   } \ [   0.196   ]$&     0.193   $^{+    0.004   }_{-    0.006   } \ [     0.188   ]$&     0.198   $^{+    0.003   }_{-    0.004   } \ [   0.191   ]$\\
Temperature     &       720.33  &       727     $^{+    61      }_{-    57      } \ [     743     ]$&     747     $^{+    72      }_{-    68      } \ [   738     ]$&     768     $^{+    78      }_{-    74      } \ [     722     ]$&     813     $^{+    111     }_{-    77      } \ [   743     ]$\\
log(H$_2$O)     &       -4      &       -6      $^{+    2       }_{-    4       } \ [     -4      ]$&     -5.8    $^{+    1.6     }_{-    3.9     } \ [   -5.3    ]$&     -5.7    $^{+    1.6     }_{-    3.5     } \ [     -4.0    ]$&     -7      $^{+    2       }_{-    3       } \ [   -4      ]$\\
\end{tabular}
            \tablefoot{ We report the median values with its 1$\sigma$ error and the MAP values in square brackets.}
\end{table*}

\begin{table*}\small
        \caption{Retrieved values obtained from the analyses performed on a N$_2$-dominated cloudy secondary atmosphere (P$_{clouds} = 10^{-1}$ bar) of a super-Earth around an 8th magnitude M star in the different  scenarios considered here. }   
        \label{table:SecondaryN2Cloud1e4} 
        \centering      
        \begin{tabular}{c@{\hskip 0.5cm}c@{\hskip 0.5cm}|c@{\hskip 0.5cm}c@{\hskip 0.5cm}c@{\hskip 0.5cm}c@{\hskip 0.5cm}}   
            \toprule[0.05cm]
            \multicolumn{6}{c}{\textbf{N$_2$-dominated cloudy secondary atmosphere  of a super-Earth around an 8th magnitude M star (P\textsubscript{clouds} = 10\textsuperscript{-1} bar)}}\\
            \toprule
             \multirow{1}*{Parameter} & True value & \multicolumn{4}{c}{$\mu$ = 2.3} \\
            \cmidrule(lr){1-6}
            \multirow{1}*{$\delta$M} & & 10\% & 30\% & 50\% & Unknown\\
            \midrule
            Mass        &       0.026   &       0.026   $^{+    0.002   }_{-    0.002   } \ [     0.025   ]$&     0.026   $^{+    0.002   }_{-    0.002   } \ [   0.026   ]$&     0.026   $^{+    0.002   }_{-    0.002   } \ [     0.026   ]$&     0.026   $^{+    0.002   }_{-    0.002   } \ [   0.026   ]$\\
Radius  &       0.189   &       0.189   $^{+    0.002   }_{-    0.003   } \ [     0.189   ]$&     0.189   $^{+    0.002   }_{-    0.003   } \ [   0.187   ]$&     0.189   $^{+    0.002   }_{-    0.003   } \ [     0.188   ]$&     0.189   $^{+    0.002   }_{-    0.003   } \ [   0.189   ]$\\
Temperature     &       720.33  &       727     $^{+    37      }_{-    35      } \ [     712     ]$&     729     $^{+    41      }_{-    40      } \ [   721     ]$&     728     $^{+    44      }_{-    41      } \ [     715     ]$&     730     $^{+    44      }_{-    43      } \ [   718     ]$\\
log(H$_2$O)     &       -4      &       -3.9    $^{+    0.7     }_{-    0.5     } \ [     -4.2    ]$&     -4.0    $^{+    0.7     }_{-    0.6     } \ [   -3.6    ]$&     -4.0    $^{+    0.7     }_{-    0.6     } \ [     -3.8    ]$&     -4.0    $^{+    0.7     }_{-    0.6     } \ [   -4.0    ]$\\
\end{tabular}
            \tablefoot{ We report the median values with its 1$\sigma$ error and the MAP values in square brackets.}
\end{table*}

\begin{table*}\small
        \caption{Retrieved values obtained from the analyses performed on a H$_2$O-dominated cloudy secondary atmosphere (P$_{clouds} = 10^{-3}$ bar) of a super-Earth around an 8th magnitude M star in the different  scenarios considered here. }   
        \label{table:SecondaryH2OCloud1e2} 
        \centering      
        \begin{tabular}{c@{\hskip 0.5cm}c@{\hskip 0.5cm}|c@{\hskip 0.5cm}c@{\hskip 0.5cm}c@{\hskip 0.5cm}c@{\hskip 0.5cm}}   
            \toprule[0.05cm]
            \multicolumn{6}{c}{\textbf{H$_2$O-dominated  cloudy secondary atmosphere  of a super-Earth around an 8th magnitude M star (P\textsubscript{clouds} = 10\textsuperscript{-3} bar)}}\\
            \toprule
             \multirow{1}*{Parameter} & True value & \multicolumn{4}{c}{$\mu$ = 2.3} \\
            \cmidrule(lr){1-6}
            \multirow{1}*{$\delta$M} & & 10\% & 30\% & 50\% & Unknown\\
            \midrule
            Mass        &       0.026   &       0.026   $^{+    0.002   }_{-    0.002   } \ [     0.024   ]$&     0.029   $^{+    0.005   }_{-    0.005   } \ [   0.024   ]$&     0.032   $^{+    0.006   }_{-    0.005   } \ [     0.024   ]$&     0.034   $^{+    0.011   }_{-    0.007   } \ [   0.029   ]$\\
Radius  &       0.189   &       0.191   $^{+    0.003   }_{-    0.004   } \ [     0.188   ]$&     0.192   $^{+    0.004   }_{-    0.004   } \ [   0.187   ]$&     0.194   $^{+    0.003   }_{-    0.004   } \ [     0.187   ]$&     0.194   $^{+    0.004   }_{-    0.004   } \ [   0.191   ]$\\
Temperature     &       720.33  &       711     $^{+    63      }_{-    53      } \ [     715     ]$&     736     $^{+    72      }_{-    68      } \ [   715     ]$&     758     $^{+    80      }_{-    74      } \ [     721     ]$&     774     $^{+    101     }_{-    82      } \ [   737     ]$\\
log(CH$_4$)     &       -3.22   &       -3.7    $^{+    0.5     }_{-    0.7     } \ [     -3.0    ]$&     -3.7    $^{+    0.5     }_{-    0.7     } \ [   -3      ]$&     -3.8    $^{+    0.5     }_{-    0.7     } \ [     -3.0    ]$&     -3.7    $^{+    0.6     }_{-    0.6     } \ [   -3.4    ]$\\
\end{tabular}
            \tablefoot{ We report the median values with its 1$\sigma$ error and the MAP values in square brackets.}
\end{table*}

\begin{table*}\small
        \caption{Retrieved values obtained from the analyses performed on a H$_2$O-dominated cloudy secondary atmosphere (P$_{clouds} = 10^{-1}$ bar) of a super-Earth around an 8th magnitude M star in the different considered scenarios.}   
        \label{table:SecondaryH2OCloud1e4} 
        \centering      
        \begin{tabular}{c@{\hskip 0.5cm}c@{\hskip 0.5cm}|c@{\hskip 0.5cm}c@{\hskip 0.5cm}c@{\hskip 0.5cm}c@{\hskip 0.5cm}}   
            \toprule[0.05cm]
            \multicolumn{6}{c}{\textbf{H$_2$O-dominated cloudy secondary atmosphere  of a super-Earth around an 8th magnitude M star (P\textsubscript{clouds} = 10\textsuperscript{-1} bar)}}\\
            \toprule
             \multirow{1}*{Parameter} & True value & \multicolumn{4}{c}{$\mu$ = 2.3} \\
            \cmidrule(lr){1-6}
            \multirow{1}*{$\delta$M} & & 10\% & 30\% & 50\% & Unknown\\
            \midrule
            Mass        &       0.026   &       0.027   $^{+    0.002   }_{-    0.002   } \ [     0.026   ]$&     0.027   $^{+    0.003   }_{-    0.002   } \ [   0.027   ]$&     0.027   $^{+    0.003   }_{-    0.003   } \ [     0.027   ]$&     0.027   $^{+    0.003   }_{-    0.003   } \ [   0.027   ]$\\
Radius  &       0.189   &       0.191   $^{+    0.001   }_{-    0.001   } \ [     0.190   ]$&     0.191   $^{+    0.001   }_{-    0.001   } \ [   0.189   ]$&     0.191   $^{+    0.001   }_{-    0.002   } \ [     0.190   ]$&     0.191   $^{+    0.001   }_{-    0.001   } \ [   0.19    ]$\\
Temperature     &       720.33  &       704     $^{+    39      }_{-    35      } \ [     703     ]$&     715     $^{+    49      }_{-    45      } \ [   724     ]$&     717     $^{+    53      }_{-    46      } \ [     725     ]$&     717.815 $^{+    52.961  }_{-    49.599  } \ [   713.822 ]$\\
log(CH$_4$)     &       -3.22   &       -3.5    $^{+    0.3     }_{-    0.2     } \ [     -3.3    ]$&     -3.5    $^{+    0.3     }_{-    0.2     } \ [   -3.2    ]$&     -3.5    $^{+    0.3     }_{-    0.2     } \ [     -3.4    ]$&     -3.502  $^{+    0.276   }_{-    0.24    } \ [   -3.411  ]$\\
\end{tabular}
            \tablefoot{ We report the median values with its 1$\sigma$ error and the MAP\ values in square brackets.}
\end{table*}

\begin{table*}\small
        \caption{Retrieved values obtained from the analyses performed on a CO-dominated cloudy secondary atmosphere (P$_{clouds} = 10^{-3}$ bar) of a super-Earth around an 8th magnitude M star in the different considered scenarios. }   
        \label{table:SecondaryCOCloud1e2} 
        \centering      
        \begin{tabular}{c@{\hskip 0.5cm}c@{\hskip 0.5cm}|c@{\hskip 0.5cm}c@{\hskip 0.5cm}c@{\hskip 0.5cm}c@{\hskip 0.5cm}}   
            \toprule[0.05cm]
            \multicolumn{6}{c}{\textbf{CO-dominated cloudy secondary atmosphere  of a super-Earth around an 8th magnitude M star (P\textsubscript{clouds} = 10\textsuperscript{-3} bar)}}\\
            \toprule
             \multirow{1}*{Parameter} & True value & \multicolumn{4}{c}{$\mu$ = 2.3} \\
            \cmidrule(lr){1-6}
            \multirow{1}*{$\delta$M} & & 10\% & 30\% & 50\% & Unknown\\
            \midrule
            Mass        &       0.026   &       0.026   $^{+    0.002   }_{-    0.002   } \ [     0.026   ]$&     0.029   $^{+    0.005   }_{-    0.005   } \ [   0.028   ]$&     0.031   $^{+    0.007   }_{-    0.006   } \ [     0.026   ]$&     0.037   $^{+    0.009   }_{-    0.007   } \ [   0.027   ]$\\
Radius  &       0.189   &       0.191   $^{+    0.005   }_{-    0.005   } \ [     0.191   ]$&     0.192   $^{+    0.005   }_{-    0.006   } \ [   0.191   ]$&     0.193   $^{+    0.005   }_{-    0.006   } \ [     0.189   ]$&     0.196   $^{+    0.004   }_{-    0.004   } \ [   0.191   ]$\\
Temperature     &       720.33  &       727     $^{+    62      }_{-    58      } \ [     715     ]$&     751     $^{+    66      }_{-    67      } \ [   721     ]$&     762     $^{+    83      }_{-    73      } \ [     722     ]$&     813     $^{+    105     }_{-    87      } \ [   742     ]$\\
log(H$_2$O)     &       -4      &       -5.4    $^{+    1.5     }_{-    3.5     } \ [     -4.6    ]$&     -6      $^{+    2       }_{-    3       } \ [   -4      ]$&     -6      $^{+    2       }_{-    3       } \ [     -4      ]$&     -6      $^{+    2       }_{-    4       } \ [   -4      ]$\\
\end{tabular}
            \tablefoot{ We report the median values with its 1$\sigma$ error and the MAP values in square brackets.}
\end{table*}

\begin{table*}\small
        \caption{Retrieved values obtained from the analyses performed on a CO-dominated cloudy secondary atmosphere (P$_{clouds} = 10^{-1}$ bar) of a super-Earth around an 8th magnitude M star in the different considered scenarios.}   
        \label{table:SecondaryCOCloud1e4} 
        \centering      
        \begin{tabular}{c@{\hskip 0.5cm}c@{\hskip 0.5cm}|c@{\hskip 0.5cm}c@{\hskip 0.5cm}c@{\hskip 0.5cm}c@{\hskip 0.5cm}}   
            \toprule[0.05cm]
            \multicolumn{6}{c}{\textbf{CO-dominated cloudy secondary atmosphere  of a super-Earth around an 8th magnitude M star (P\textsubscript{clouds} = 10\textsuperscript{-1} bar)}}\\
            \toprule
             \multirow{1}*{Parameter} & True value & \multicolumn{4}{c}{$\mu$ = 2.3} \\
            \cmidrule(lr){1-6}
            \multirow{1}*{$\delta$M} & & 10\% & 30\% & 50\% & Unknown\\
            \midrule
            Mass        &       0.026   &       0.026   $^{+    0.002   }_{-    0.002   } \ [     0.026   ]$&     0.026   $^{+    0.002   }_{-    0.002   } \ [   0.026   ]$&     0.026   $^{+    0.002   }_{-    0.002   } \ [     0.026   ]$&     0.026   $^{+    0.002   }_{-    0.002   } \ [   0.027   ]$\\
Radius  &       0.189   &       0.188   $^{+    0.002   }_{-    0.003   } \ [     0.189   ]$&     0.189   $^{+    0.002   }_{-    0.003   } \ [   0.189   ]$&     0.188   $^{+    0.002   }_{-    0.003   } \ [     0.189   ]$&     0.189   $^{+    0.002   }_{-    0.003   } \ [   0.190   ]$\\
Temperature     &       720.33  &       728     $^{+    36      }_{-    34      } \ [     714     ]$&     731     $^{+    44      }_{-    42      } \ [   718     ]$&     732     $^{+    43      }_{-    42      } \ [     727     ]$&     729     $^{+    45      }_{-    39      } \ [   735     ]$\\
log(H$_2$O)     &       -4      &       -3.8    $^{+    0.7     }_{-    0.6     } \ [     -4.0    ]$&     -3.9    $^{+    0.7     }_{-    0.6     } \ [   -4.0    ]$&     -3.9    $^{+    0.7     }_{-    0.6     } \ [     -4.0    ]$&     -3.9    $^{+    0.7     }_{-    0.6     } \ [   -4.3    ]$\\
\end{tabular}
            \tablefoot{ We report the median values with its 1$\sigma$ error and the MAP values in square brackets.}
\end{table*}

  \end{appendix}
\end{document}